\documentclass[fleqn,usenatbib]{mnras}

\usepackage{newtxtext,newtxmath}

\usepackage[T1]{fontenc}

\DeclareRobustCommand{\VAN}[3]{#2}
\let\VANthebibliography\thebibliography
\def\thebibliography{\DeclareRobustCommand{\VAN}[3]{##3}\VANthebibliography}

\usepackage{graphicx}	
\usepackage{amsmath}	
\usepackage{shortcuts}
\usepackage{ulem}

\defcitealias{moeMindYourPs2017}{MD17}

\title[Compact mergers from varying initial conditions]{Compact object populations over cosmic time II. Compact object merger rates and masses over redshift from varying initial conditions}

\author[L. M. de S\'a et al.]{
Lucas M. de S\'a,$^{1}$\thanks{E-mail: lucasmdesa@usp.br}
L\'ivia S. Rocha,$^{2}$
Ant\^onio Bernardo,$^{1}$
Riis R. A. Bachega,$^{1}$
Jorge E. Horvath,$^{1}$
\\
$^{1}$Instituto de Astronomia, Geof\'isica e Ci\^encias Atmosf\'ericas, Universidade de S\~ao Paulo, Rua do Mat\~ao, 1228, Butant\~a, S\~ao Paulo, BR \\
$^{2}$Instituto de F\'isica, Universidade of S\~ao Paulo, Rua do Mat\~ao, 1371, Butant\~a, S\~ao Paulo, BR
}

\date{Accepted XXX. Received YYY; in original form ZZZ}

\pubyear{2024}

\begin{document}
\label{firstpage}
\pagerange{\pageref{firstpage}--\pageref{lastpage}}
\maketitle

\begin{abstract}
We perform a first study of the impact of varying two components of the initial conditions in binary population synthesis of compact binary mergers — the initial mass function, which is made metallicity- and star formation rate-dependent, and the orbital parameter (orbital period, mass ratio and eccentricity) distributions, which are assumed to
be correlated — within a larger grid of initial condition models also including alternatives for the primary mass-dependent binary fraction and the metallicity-specific cosmic star formation history. We generate the initial populations with the sampling code BOSSA and evolve them with the rapid population synthesis code COMPAS. We find strong suggestions that the main role of initial conditions models is to set the relative weights of key features defined by the evolution models. In the two models we compare, black hole-black hole (BHBH) mergers are the most strongly affected, which we connect to a shift from the common envelope to the stable Roche lobe overflow formation channels with decreasing redshift. We also characterize variations in the black hole-neutron star (BHNS) and neutron star-neutron star (NSNS) final parameter distributions. We obtain the merger rate evolution for BHBH, BHNS and NSNS mergers up to $z=10$, and find a variation by a factor of $\sim50\en60$ in the local BHBH and BHNS merger rates, suggesting a more important contribution from initial conditions than previously thought, and calling for a complete exploration of the initial conditions model permutations.
\end{abstract}




\section{Introduction}
\label{sec:1intro}

The build-up of new gravitational-wave (GW) observations of compact objects (CO) mergers \citep{ligoscientificcollaborationandvirgocollaborationGWTC1GravitationalWaveTransient2019,ligoscientificcollaborationandvirgocollaborationGWTC2CompactBinary2021,ligoscientificcollaborationGWTC3CompactBinary2023,theligoscientificcollaborationandthevirgocollaborationGWTC2DeepExtended2024} has allowed for an increasingly refined attempt to picture the properties of distant populations of merging CO binaries \citep{abbottBinaryBlackHole2019,abbottPopulationPropertiesCompact2021a,ligoscientificcollaborationPopulationMergingCompact2023,ligoscientificcollaborationGWTC3CompactBinary2023}. The most recent Gravitational Wave Transient Catalog, GWTC-3, has included compact binary mergers (CBMs) placed at redshifts up to $1.18^{+0.73}_{-0.53}$, and constrained the presence of overdensities in the primary mass distribution of black hole-black hole (BHBH) mergers, around $\sim10\,\Msun$ and $\sim35\,\Msun$, to within a few solar masses, the origins of which remain undetermined \citep{ligoscientificcollaborationPopulationMergingCompact2023}. Black hole-neutron star (BHNS) and neutron star-neutron star (NSNS) mergers still constitute a small part of the observed population, but the same level of detail can be expected in the future, both with continued improvements to the current generation of detectors \citep[e.g.,][]{adhikariCryogenicSiliconInterferometer2020,akutsuOverviewKAGRAKAGRA2021,paceStatusAdvancedVirgo2021} and the construction of new, third generation, detectors to gather extensive new data, such as Cosmic Explorer \citep{hallCosmicExplorerNextGeneration2022,evansCosmicExplorerSubmission2023} and the Einstein Telescope \citep{maggioreScienceCaseEinstein2020}; as well as of the Laser Interferometer Space Antenna \citep{amaro-seoaneLaserInterferometerSpace2017a, amaro-seoaneAstrophysicsLaserInterferometer2023a, breivikLISARoleTimedomain2023}.

Much work has been dedicated to placing constraints on the contribution from different formation channels, in an effort to discern the origins of the merging binaries. Isolated binary evolution typically relies on episodes of mass transfer, either stable Roche lobe overflow or Common Envelope evolution, to drive binaries towards small enough separations so that they will merge within a Hubble time after the second supernova \citep[e.g.,][]{belczynskiCompactBinaryMerger2016, klenckiImpactIntercorrelatedInitial2018,marchantRoleMassTransfer2021,baveraImpactMasstransferPhysics2021,gallegos-garciaBinaryBlackHole2021,sonRedshiftEvolutionBinary2022,piccoFormingMergingDouble2024}. Chemically homogeneously evolving \citep{mandelMergingBinaryBlack2016,deminkChemicallyHomogeneousEvolutionary2016,marchant2016aNewRouteTowards,dubuissonCosmicRatesBlack2020,sharpeInvestigatingChemicallyHomogeneous2024} and Population III stars \citep[e.g.,][]{belczynskiLikelihoodDetectingGravitational2017,liuGravitationalWavesRemnants2021,liuCorrectionGravitationalWaves2023,2023santoliquidoPopIII,tanikawaContributionPopulationIII2024a} have also been considered as pathways for isolated formation of BHBH mergers. External factors too might play an important role, such as through dynamical interactions in dense environments \citep[e.g.,][]{antoniniMergerRateBlack2020,santoliquidoCosmicMergerRate2020,mapelliMassRateHierarchical2021,traniCompactObjectMergers2022}; or the effect of evolution in an AGN disk \citep[e.g.,][]{yangHierarchicalBlackHole2019,arcaseddaQuiescentActiveGalactic2023,gauthambhaskarEnhancedBlackHole2023}. Finally, mergers should not hail only from binaries, but also from higher-order multiple systems \citep[e.g.,][]{hamersDoubleNeutronStar2019,toonenEvolutionStellarTriples2020,fragioneMergingBlackHoles2020a,vynatheyaHowImportantSecular2022}, which might become even more common than binaries for very massive progenitors \citep[][Figure 1, and discussion therein]{offnerOriginEvolutionMultiple2023}. Given this variety of channels, and the variety of physical assumptions that can be made within each of them, it is no surprise that estimates of the local BHBH merger rate, for example, have varied over six orders of magnitude \citep{mandelRatesCompactObject2022}. From GWTC-3, on the other hand, this rate has been inferred to be within $16\en61\,\Gpc^{-3}\,\yr^{-1}$ with $90\%$ credibility \citep{ligoscientificcollaborationPopulationMergingCompact2023}, suggesting that many particular combinations of binary evolution models could be ruled out at this point. Increasingly strong constraints on CBM populations could thus translate into increasingly strong constraints on binary evolution.

Evolution models, however, do not fully characterize a CBM population. Stellar/binary formation sets the stage for evolution to take place, and thus the properties of newborn, zero-age main sequence (ZAMS), populations can also be expected to add new dimensions to the parameter space to be explored when attempting to extract information about the life of binary stars from GW observations. The distributions followed by the parameters of ZAMS binaries, for example --- their constituent masses, orbital period and eccentricity ---, should affect the relative fraction of progenitors following one or another evolutionary pathway towards a particular type of CBM. These distributions benefit from being based on the observation of stars in the field, and young populations in particular, and have tended to be relatively stable over the years. The initial mass function (IMF) by \citet{salpeterLuminosityFunctionStellar1955} and \"Opik's law for orbital periods \citep{opikStatisticalStudiesDouble1924,abtNormalAbnormalBinary1983} are exemplary cases; some of the most frequent employed IMFs today, besides the Salpeter IMF itself, are still refinements of it \citep[e.g.,][]{millerInitialMassFunction1979,kroupaVariationInitialMass2001,chabrierGalacticStellarSubstellar2003}. More recently, the mass ratio distribution and binary fraction constraints by \citet{sanaBinaryInteractionDominates2012} set the standard for binary population synthesis (BPS). CBMs are predicted to reach coalescence times --- the time between the second supernova and merger --- of the order of $10\,\Gyr$. This implies that binaries with a wide range of ages and metallicities --- a key factor in their evolution --- can all contribute to relatively nearby mergers, a contribution which should become increasingly important as the number of detections is expanded and their range extended. Considering this variety of contributions further requires normalizing synthetic populations to an appropriate metallicity-specific cosmic star formation history (cSFH), yet another important ingredient of the initial conditions.

The study of the variation of star formation conditions over time is thus particularly well-matched to GW astronomy constraints. If variations of observed CBMs at different redshift can be related to the contribution of progenitors of different ages, then population synthesis and GW astronomy might together prove to be an important tool in probing the young Universe.

While initial conditions variations have been less widely explored in BPS than those of binary evolution models, the degree to which they could affect local merger rates has been evaluated before. The correlated orbital parameter distributions from \citet{moeMindYourPs2017} were implemented, alongside a metallicity-dependent IMF based on the model by \citet{marksEvidenceTopheavyStellar2012}, by \citet{klenckiImpactIntercorrelatedInitial2018} for CBM population synthesis; the authors found that both model variations only affected predicted local merger rates by a factor $\sim2$. A similarly small impact had been earlier found by \citet{deminkMergerRatesDouble2015} when implementing the initial orbital period distribution from \citet{sanaBinaryInteractionDominates2012}, which more strongly favors close initial orbits than \citet{opikStatisticalStudiesDouble1924}, although they found that uncertainties related to the IMF affected local rates by up to a factor of $\sim6$. In contrast, \citet{kruckowProgenitorsGravitationalWave2018} obtained local rate variations of about one order of magnitude by varying the IMF slope, while agreeing that the \citet{moeMindYourPs2017} distributions only affected these rates by a factor of $\sim2-3$. Later yet, \citet{santoliquidoCosmicMergerRate2021} found a negligible impact of varying the high-mass IMF slope on merger rates over cosmic time. The impact of a star formation rate (SFR)-dependent model of the IMF on CBM populations, however, has not so far been studied. 

Beyond local and time-dependent merger rates, the shape of final parameter distributions under different initial condition assumptions is of interest. Recent BPS work has increasingly constrained the formation channels of each kind of merger and particular features of the final parameter space associate to them under different evolution models \citep[e.g.,][]{kruckowProgenitorsGravitationalWave2018,chruslinskaDoubleNeutronStars2018,broekgaardenImpactMassiveBinary2022,iorioCompactObjectMergers2023a} and metallicity-specific cSFHs \citep[e.g.,][]{luFormationGW1908142021,santoliquidoCosmicMergerRate2021,bocoEvolutionGalaxyStar2021,broekgaardenImpactMassiveBinary2022,sonLocationsFeaturesMass2023,chruslinskaChemicalEvolutionUniverse2024} assumptions, but a similar exploration of the full range of uncertainties in initial condition models is yet to be performed.

In de S\'a et al. (submitted) (Paper I hereon), we presented BOSSA\footnote{Binary Object environment-Sensitive Sampling Algorithm}, an initial sampling algorithm for BPS which accounts for the above mentioned issues surrounding stellar/binary formation, in particular with regard to the IMF, its environment-dependence, consistent sampling when dealing with binaries or higher-order multiples, and the implications of stochasticity and self-regulation in star formation. BOSSA accounts for the metallicity- and SFR-dependent IMF from \citep{jerabkovaImpactMetallicityStar2018}; the correlated orbital parameters and mass-dependent multiplicity from \citep{moeMindYourPs2017}; and the collection of cSFH models by \citet{chruslinskaMetallicityStarsFormed2019,chruslinskaEffectEnvironmentdependentIMF2020}, for a total of 192 possible initial conditions.

In this work, we employ BOSSA to generate initial populations which are then evolved up to merger with the open-source rapid BPS code COMPAS \citep{stevensonFormationFirstThree2017a,vigna-gomezFormationHistoryGalactic2018a,compasRapidStellarBinary2022} \footnote{\url{https://compas.science/}} for two selected representative model permutations within our 192 set, and analyze and discuss the range of variations between them for BHBH, BHNS and NSNS mergers from the isolated binary evolution channel. Whenever possible, we also suggest explanations to understand how different initial conditions affect the relative weights of different formation channels. In Sec. \ref{sec:2models} we summarize the models included in BOSSA and COMPAS, and define the two model permutations that will be compared. In Sec. \ref{sec:3computation} we describe some of the problems in computing quantities continuously distributed over time, such as merger rates, from a population distributed over a discrete set of ages, as well as steps taken to mitigate them. In Sec. \ref{sec:4results} we discuss our results for each model and explore how different initial distributions impact the final parameters in detail. We summarize our conclusions in Sec. \ref{sec:5conclusions}.

We adopt a flat cosmology with $\Omega_\mathrm{M}=0.287$, $\Omega_\Lambda=0.713$ and $H_0=69.32\,\mathrm{km}\,\mathrm{s}^{-1}\,\mathrm{Mpc}^{-1}$ \citep{hinshawNINEYEARWILKINSONMICROWAVE2013} as a baseline hypothesis.
\section{Initial conditions and evolution models}
\label{sec:2models}

We generate composite binary populations of ZAMS binaries with BOSSA, wherein each simple stellar population (SSP) corresponds to a metallicity-SFR pair. We draw SSPs through scatterless sampling, which does not account for a physical spread around the mass-metallicity relation (MZR) and star formation-mass relation (SFMR); this allows for a one-to-one association between SFR and redshift which is assumed when computing merger rates over redshift (Section \ref{sec3sub:mrates}). We weight SSP sampling by the galaxy mass density as a function of galaxy mass and redshift, as set by a redshift-dependent galaxy stellar mass function (GSMF). Sampling is performed for 10 values of redshift and 10 metallicity per redshift, for $z=0-10$ and galaxy stellar masses $10^6-10^{12}\,\Msun$.\footnote{While the version of BOSSA released alongside Paper I performs SFRD-weighted sampling \textit{with} scatter by default, the simulations in this paper were run with an earlier version based on scatterless mass density-weighted sampling. Our conclusions remain the same under SFRD-weighted sampling as trends with metallicity and redshift are independent of this.}

 For a given redshift and galaxy stellar mass, metallicity and star formation rate are set by the MZR and SFMR, respectively. Although we do not explicitly employ a cSFH model, one is implicitly defined by the GSMF, MZR and SFMR. Redshifts $0.01$ and $10$ are manually added to the sample, with another 10 mass density-weighted metallicities drawn for each, as "boundary conditions"; $z=0.01$ is used instead of $z=0$ as the latter will generate no mergers, but we would still like to have a reference for the frequency of very young merging binaries. Each initial sample is thus a composite population consisting of 120 distinct SSP. We choose a sample size of $\sim10^6$ binaries for each SSP (due to the mass sampling method in BOSSA, the exact size of each sample varies; see Section \ref{sec2sub:imf}), yielding a composite population of $\sim10^8$ binaries. The "binaries" in the sample may include exclusively physical binaries, or also the inner binaries of higher-order multiples. This is briefly discussed below. The entire sampling process is described in detail in Section 3 of Paper I.

Each SSP can then be evolved with COMPAS, for which the metallicity is a given parameter, but not the redshift. We seek to compensate for the "degeneracy" that would result from individually evolving all $120$ SSPs by cross-matching them to a pre-evolved set of populations at different metallicities; this is discussed in Sec. \ref{sec:3computation}. Here, we briefly go over all models considered for initial conditions (Sec. \ref{sec2sub:initial_conditions}) and for key stages of stellar/binary evolution (Sec. \ref{sec2sub:evolution}). The initial sampling method is also summarized in Sec. \ref{sec2sub:initial_conditions}. We refer the reader to Paper I for a full discussion of the initial conditions models, their physical implications and implementation; as well as discussion and consistency tests of the initial sampling method.

\subsection{Initial conditions}
\label{sec2sub:initial_conditions}

Each system in the initial population is defined by eight parameters at zero-age main sequence (ZAMS): the mass of the primary (most massive) star, $m_1$; its number of companions, $\ncp$; the orbital period, $P$, mass ratio, $q$, and eccentricity, $e$, of each companion; and the metallicity, $\Z$, star formation rate (SFR) and redshift, $z$, characterizing the environment in which it was formed. The SFR figures as a parameter for the binary because our varying initial mass function (IMF) model is SFR-dependent; thus, for fixed redshift and metallicity, the SFR is still necessary to set the IMF from which the component masses were drawn. In turn, we discuss in the initial mass function (Sec. \ref{sec2sub:imf}), the matter of multiplicity (Sec. \ref{sec2sub:multiplicity}), the orbital parameter distributions (Sec. \ref{sec2sub:orbital_parameters}), the environmental conditions (Sec. \ref{sec2sub:environment}), and our sampling method (Sec. \ref{sec2sub:sampling}).

\subsubsection{Initial mass function}
\label{sec2sub:imf}

We consider two models for the IMF, termed "Invariant" and "Varying". As the Invariant IMF we employ the \citet{kroupaVariationInitialMass2001}, which is equal to the \citet{salpeterLuminosityFunctionStellar1955} IMF for $>0.5\,\Msun$ and widely used in BPS. The Invariant model reflects the assumption that the IMF of any stellar system is the same as the Milky Way IMF; thus, other Salpeter "descendants", such as the \citet{millerInitialMassFunction1979} and \citet{chabrierGalacticStellarSubstellar2003} IMFs would serve the same purpose. As the "Varying" IMF, we employ the \citet{jerabkovaImpactMetallicityStar2018} IMF \citep[but see also][]{yanOptimallySampledGalaxywide2017}, computed within the integrated galaxy-wide IMF theory \citep[or IGIMF, originally by][]{kroupaGalacticFieldInitialMass2003}, based on the empirical fits by \citet{marksEvidenceTopheavyStellar2012}, for its dependence on $\FeHinline$ as a metallicity measure\footnote{Defined with relation to solar abundances, $\FeHinline=\log\lrp{\frac{N_\mathrm{Fe}}{N_\mathrm{H}}}-\log\lrp{\frac{N_{\mathrm{Fe}}}{N_{\mathrm{O}}}}_\odot$.}; and \citet{gunawardhanaGalaxyMassAssembly2011}, for its SFR-dependence. The \citet{jerabkovaImpactMetallicityStar2018} IMF \citep[usually argued for on the basis of the Jeans' mass temperature-dependance; see, e.g.,][]{larsonEarlyStarFormation1998,larsonThermalPhysicsCloud2005,bateDependenceInitialMass2005,bonnellJeansMassOrigin2006}; reproduces the long-expected environment-sensitivity of the IMF \cite[see, e.g.][]{larsonEarlyStarFormation1998,larsonThermalPhysicsCloud2005,bateDependenceInitialMass2005,bonnellJeansMassOrigin2006}; in particular, it shows that the low- and high-mass slopes of the IMF can vary independently. With decreasing metallicity, the IMF becomes primarily bottom-light; and with increasing SFR, top-heavy. In either case, we restrict sampled masses to $0.8-150\,\Msun$ for the primary component, where $0.8\,\Msun$ is the lower limit of our orbital parameter distributions (Section \ref{sec2sub:orbital_parameters}), and $150\,\Msun$ we take to be the upper limit for star formation \citep{figerUpperLimitMasses2005,oeyStatisticalConfirmationStellar2005,koenUpperLimitStellar2006,maizapellanizPismis241Stellar2007}. For companions we allow $0.08-150~\Msun$, where $0.08\,\Msun$ is simply the hydrogen-burning limit. We treat the IMF as a probability density function for simplicity, but point out that this might not accurately represent actual individual stellar populations \citep[see Paper I, and also][]{kroupaStellarSubStellarInitial2013,kroupaInitialMassFunction2021a}.

In IGIMF theory, each individual cluster within a galaxy is allowed to possess a different stellar IMF (sIMF), while cluster (gas) masses themselves are distributed according to an \textit{embedded cluster} IMF (eIMF), which is allowed to vary from galaxy to galaxy, within the constraints of empirical data. The \textit{galaxy-wide} IMF (gwIMF) is then the result of integrating over the sIMF of all of a galaxy's clusters. In the Varying model, for each metallicity-SFR (or metallicity-redshift) pair, we sample masses directly from the corresponding gwIMF, instead of sampling from individual clusters. We note that, in a strictly IGIMF theory approach, masses should be sampled directly from the clusters through \textit{optimal sampling}, which always yields the same mass sample for a given cluster; for statistical studies, however, it has been noted  that random sampling, which we adopt here, is adequate \citep[see Paper I, and also][]{kroupaStellarSubStellarInitial2013,kroupaInitialMassFunction2021a}.

\subsubsection{Orbital parameters}
\label{sec2sub:orbital_parameters}

We collectively refer to $P$, $q$ and $e$ as \textit{orbital parameters}; if triples and higher-order multiples are allowed, then this triad of parameters must be specified for each companion. We always take $q=m_1/\mcp\leq1$, where $\mcp$ is the companion mass. For the orbital parameters we again defined a Invariant and a Varying model. For the "Invariant" case we choose common, uncorrelated distributions: \citet{opikStatisticalStudiesDouble1924}'s law for the orbital period, a log-uniform distribution between $10^{0.4}$ and $10^{3}\,\mathrm{d}$; a uniform distribution for $q$, from \citet{sanaBinaryInteractionDominates2012}; and $e=0$. The most significant feature of these distributions is that they are \textit{invariant}, and describe only \textit{physical binaries}, not higher-order multiples. For the Varying case we adopt the distributions from \citet{moeMindYourPs2017}, which are empirical power-law/log-linear function series, describing the correlation of $P$ with $m_1$; and of both $q$ and $e$ with $P$ and $m_1$. As discussed briefly in Paper I, and thoroughly in \citet{moeMindYourPs2017}, these correlations are thought to emerge from the pre-ZAMS evolution of multiple systems. Besides the fact that they are correlated, it is also essential to note that the sample upon which the fits were performed includes both binaries and higher-order multiples. Therefore, strictly speaking, these distributions are not \textit{binary} orbital parameter distributions. They do, however, contain information about multiplicity that allows differentiating between different-order multiples if certain assumptions are made; we summarize this process in the next Section. Otherwise, they are simply employed as probability density functions.

\Citet{moeMindYourPs2017} fitted the Varying distributions within $0.8\leq m_1/\,\Msun\leq40\,\Msun$, $10^{0.2}\leq P/\mathrm{d} \leq 8$, $0.1\leq q\leq1$ and $0\leq e <e_\mathrm{max}\lrp{P}$. The maximum eccentricity is defined by a "no interaction at ZAMS" criterion. We keep the orbital period and mass ratio ranges, but extrapolate their distributions up to $m_1=150\,\Msun$. In the case of $q$ and $e$, their distributions were already found to be invariant above $\sim10\,\Msun$, thus are also invariant for $m_1>40\,\Msun$. For the orbital period, however, our extrapolation leads to a growing excess of short-period orbits with primary mass; the implications of this choice, as well as an alternative, are discussed in Paper I (see Figures 4 and 14 therein). Additionally, the $e$ distribution is not well-defined for $P>10^6\,\mathrm{d}$, and so we must extrapolate it up to $P=10^8\,\mathrm{d}$. \citet{moeMindYourPs2017} discuss the source of this limitation and issues with extrapolation; Paper I discusses why the issues with extrapolation are lessened within BPS specifically. For the sake of comparison, we also extrapolate the Invariant distributions, so that $0.8\leq m_1/\,\Msun\leq150\,\Msun$, $10^{0.2}\leq P/\mathrm{d} \leq 8$, $0.1\leq q\leq1$ and $0\leq e <e_\mathrm{max}\lrp{P}$.

Possible errors stemming from setting these limits are discussed in Paper I, where they are found to not introduce significant deviations between the sampled population and the empirical distributions. We do not take into account the spread around these fits \citep[also modeled by][]{moeMindYourPs2017} at this time.

The effect of metallicity on ZAMS orbital parameters has been so far most clearly constrained from measurements of the binary fraction at different separations, in connection with the orbital period distribution. When it comes to the close binary fraction ($a\lesssim10\,\mathrm{AU}$), \citet{moeCloseBinaryFraction2019} reviewed the available literature and sought to reconcile often conflicting observed trends, finding that the close binary fraction of solar-type stars is strongly anticorrelated with metallicity --- in other words, the orbital period distribution trends towards closer orbits with decreasing metallicity. That of massive stars, on the other hand, has been found not to vary significantly with metallicity \citep{moeCLOSEBINARYPROPERTIES2013,dunstallVLTFLAMESTarantulaSurvey2015,almeidaTarantulaMassiveBinary2017}.

As further discussed in Paper I, orbital parameter correlations are thought to emerge from pre-ZAMS evolution, for which the initial orbital period is a key factor. Binary formation is typically recognized to occur through one of two channels: turbulent molecular cloud core fragmentation or fragmentation of a gravitationally unstable disk, generally responsible for the formation of wide and close binaries, respectively. Within this picture, \citet{moeCloseBinaryFraction2019} proposed that the metallicity-dependence of the close binary fraction stems from an increasing fraction of fragmenting disks as metallicity decreases down to $0.1\Z_\odot$. As they suggested, disk fragmentation should always occur for massive stars, such that no close binary fraction metallicity-dependence would emerge from this mechanism. \citet{el-badryWideBinaryFraction2019} similarly found that the binary fraction of solar-type stars is constant for separations $\gtrsim250\,\mathrm{AU}$, but is significantly anticorrelated with metallicity within $50-100\,\mathrm{AU}$, suggesting that that the $100-200\,\mathrm{AU}$ separation range could be a transition region between core fragmentation- and disk fragmentation-dominated regimes of binary formation.

Insofar as mass ratio and eccentricity distributions for massive binaries have been constrained under varying metallicities, no variations have been found so far \citep[][Section 2.3]{offnerOriginEvolutionMultiple2023}. Given current evidence, and our focus on CO progenitors, we do not include any metallicity-dependence in the orbital parameter distributions.

\subsubsection{Multiplicity}
\label{sec2sub:multiplicity}

In addition to finding a uniform mass ratio distribution, \citet{sanaBinaryInteractionDominates2012} also determined a constant binary fraction of $0.69\pm0.09$ for O-type stars. Based on that, the majority of BPS works interested in CO progenitors assumes a constant binary fraction of $0.7$ for $m_1\geq5\,\Msun$ primaries, or even $1$. The \textit{companion frequency} -- which is the number of companions per primary of mass $m_1$, per decade of orbital period -- fitted by \citet{moeMindYourPs2017}, however, leads to an average number of companions, $\ncp$, that increases with $m_1$, and becomes greater than $2$ even below $m_1=40\,\Msun$; in Paper I, it was found that the extrapolation up to $m_1=150\,\Msun$ leads to average companion numbers greater than $3$ starting at $m_1\approx60\,\Msun$. 

Paper I followed the suggestion by \citet{moeMindYourPs2017} and recovered the multiplicity fractions from $\ncp=0$ (isolated fraction) up to $\ncp=4$ (quintuple fraction) by assuming that $\ncp$ follows a Poissonian distribution with expected value constrained by the companion frequency. This is an extrapolation from the behavior found for solar-type primaries by \citet{krausMAPPINGSHORESBROWN2011}, from a survey of the Taurus-Auriga star forming region, which, as a sparse association, shows a greater wide companion frequency than clusters or field stars, and is thus thought to more accurately represent a primordial population. Hence, we deal with two alternatives for the treatment of multiplicity.

In the \textit{All Multiples} (AM) case, any primary can have up to $\ncp=4$ companions, according to the binary fractions recovered in this way, and we are able to distinguish between physical binaries and inner binaries of higher-order multiples, and chose to evolve both or the former only. In the \textit{Only Binaries} (OB) case, we sum all multiple fractions into a single binary fraction, which monotonically increases with $m_1$, up to $\sim1$ at $150\,\Msun$, but has a mass-weighted average of $\approx0.74$ in $\lrs{5\,\Msun,150\,\Msun}$, remaining compatible with \citet{sanaBinaryInteractionDominates2012}. 

Even when not evolving inner binaries of higher-order multiples, allowing for them in the AM case decreases the total \textit{star-forming mass} corresponding to a given binary sample. This is important for computing merger rates, and further discussed in Sec. \ref{sec:3computation}. Paper I explores differences between the OB binaries, AM physical binaries and AM inner binaries with regard to initial conditions. In order to facilitate comparison with other population synthesis work, here we always use the OB model for multiplicity.

While evidence for a slightly lower binary fraction in low-metallicity environments has recently emerged \citep{bodensteinerYoungMassiveSMC2021a, neugentRedSupergiantBinary2021}, these are still early observations which could be attributable to binary evolution or selection effects \citep[][Section 2.3]{offnerOriginEvolutionMultiple2023}. We thus do not include any metallicity-dependence in our multiple fractions.

\subsubsection{Environmental conditions}
\label{sec2sub:environment}

For the environmental conditions (redshift, metallicity and SFR), we employ the metallicity-specific cSFH by \citet{chruslinskaMetallicityStarsFormed2019,chruslinskaEffectEnvironmentdependentIMF2020}, computed for a collection of 24 different permutations of the redshift-dependent GSMF, which gives the number density of star-forming galaxies as a function of their stellar mass; the MZR, which connects stellar mass to metallicity; and the SFMR, which connects stellar mass to the SFR. For the GSMF model, \citet{chruslinskaInfluenceDistributionCosmic2019} rely on 13 previous fits of the GSMF, which employed varying definitions of star-forming galaxies. Based on the SFMR by \citet{renziniOBJECTIVEDEFINITIONMAIN2015}, who did not impose a SFR-cut on their sample, the authors verify that not accounting for star formation in quiescent galaxies induces an error of $\lesssim1\%$ in the star-forming mass. 

 All distributions are either fitted or extrapolated up to $z=10$, adopted as the redshift of the beginning of star formation. The SFMR fits collected by \citet{chruslinskaMetallicityStarsFormed2019} relied on SFR measurements which assumed a \citet{kroupaVariationInitialMass2001} IMF.  \citet{chruslinskaEffectEnvironmentdependentIMF2020} employed the spectral synthesis code \texttt{P\'EGASE} to compute publicly-available corrections to the SFMR for the \citet{jerabkovaImpactMetallicityStar2018} IMF\footnote{\url{https://ftp.science.ru.nl/astro/mchruslinska/}}. We are thus able to use their collected GSMF, MZR and SFMR in both the Invariant and Varying IMF models. While galaxy mass estimates are also affected by the IMF assumption \citep[by a factor of $\sim2$, see also][]{haslbauer2024environmentDependentsIMF}, we do not include any corrections to the GSMF, as here masses are only used to connect galaxy metallicities, SFRs and redshifts, and do not feature in our calculations and results \citep[see also Section 2.4 of ][]{chruslinskaEffectEnvironmentdependentIMF2020}.

A further source of uncertainty is introduced into our models by these relations, since MZR is based on $\ZOH=12+\log\lrp{\mathrm{O/H}}$ measurements, whereas \citet{jerabkovaImpactMetallicityStar2018} parameterize the Varying IMF as a function of $\FeHinline$. As there is no straightforward conversion between iron and oxygen abundances, we adopt the simple relation suggested by \citet{chruslinskaEffectEnvironmentdependentIMF2020} when working with the same Varying IMF, which connects $\FeHinline$ and $[\mathrm{O/H}]$ through a two-part linear function, such that oxygen is always more abundant than iron in environments with subsolar metallicity.

Out of the 24 possible model permutations, \citet{chruslinskaMetallicityStarsFormed2019} determine a baseline permutation between the two possible metallicity extremes, which is not necessarily the most accurate but facilitates comparison between all permutations. This baseline permutation, which we refer to as the "Moderate Metallicity" model, adopts the  "Fixed Slope" GSMF from \citet{chruslinskaMetallicityStarsFormed2019}; the MZR calibration by \citet{maiolinoAMAZEEvolutionMass2008}, refined by \citet{mannucciLSDLymanbreakGalaxies2009}; and the "Moderate Flattening" SFMR model, based on \citet{boogaardMUSEHubbleUltra2018} and \citet{speagleHIGHLYCONSISTENTFRAMEWORK2014}.

\subsubsection{Sampling}
\label{sec2sub:sampling}

Instead of directly sampling $m_1$ from the IMF and $q$ from its own distribution, thus defining each companion mass $\mcp$, as is commonly done, we sample \textit{all} component masses from the IMF and pair them according to the mass ratio distribution. This is done by means of performing an initial sampling of a \textit{mass pool} from the IMF, from which masses are then paired according to the mass ratio distribution. In Paper I this process is described and motivated in detail, and we verify that it reproduces successfully both the mass ratio distribution and the IMF when \textit{all} component masses are considered. Individual component masses ($m_1$, $m_2$, $m_3$...), on the other hand, deviate from the IMF according to the tendency of multiplicity to increase with primary mass. This process is the same regardless of the chosen IMF and orbital parameter distributions. In the OB multiplicity model, $m_2$ follows a distribution slightly flatter than the assumed IMF for masses $\lesssim10\,\Msun$,  while $m_1$ deviates little from the IMF. For masses $\gtrsim10\,\Msun$, $m_1$ follows a slightly flatter distribution, while $m_2$ becomes significantly steeper. A detailed analysis of these variations can be found in Paper I.

\subsection{Binary evolution}
\label{sec2sub:evolution}

Because we do not seek to constrain evolution models, we keep all of the default models available within the COMPAS code. Below, we summarize the modeling of some of the key stages for our further discussion, and refer the reader to \citet{compasRapidStellarBinary2022} for a full breakdown.

COMPAS models isolated binary evolution, with single stellar evolution based on fits from \citet{hurleyComprehensiveAnalyticFormulae2000,hurleyEvolutionBinaryStars2002a} to the models of \citet{polsStellarEvolutionModels1998} for non-rotating stars with masses between $0.1-50\,\Msun$ and metallicity between $Z=10^{-4}$ and $Z=0.03$. While metallicities are kept to the original range, masses are extrapolated smoothly up to $150\,\Msun$ \citep{compasRapidStellarBinary2022}. With decreasing metallicity, stars generally become hotter and evolve more quickly, with a nuclear burning timescale that is shorter by a factor of $\sim2$ for $Z=10^{-4}$ compared to $Z=0.02$. As a consequence, they leave the main sequence and expand at an earlier age.

An important feature of massive stars evolution is wind mass loss, for which the prescription by \citet{belczynskiMAXIMUMMASSLAR2010} is adopted. This relies on results by \citet{vinkNewTheoreticalMassloss2000,vinkMasslossPredictionsStars2001} for the mass loss rates of $T>12,500\,\mathrm{K}$ stars and, for cooler $T<12,500\,\mathrm{K}$ stars, on the model by \citet{hurleyComprehensiveAnalyticFormulae2000} based on the prescriptions of \citet{nieuwenhuijzenParametrizationStellarRates1990}, corrected by the works of \citet{kudritzkiAbsoluteScaleMassloss1978,kudritzkiRadiationdrivenWindsHot1989} and \citet{ vassiliadisEvolutionLowIntermediateMass1993}, depending on the stellar life stage. Helium star mass-loss rates are set to the prescription by \citet{belczynskiMAXIMUMMASSLAR2010}, based on \citet{hamannSpectrumFormationClumped1998} and \citet{vinkMetallicityDependenceWolfRayet2005}. Stars that reach the Humphreys-Davidson limit for luminosity \citep{humphreysLuminousBlueVariables1994} are classified as Luminous Blue Variables (LBV) and are set to the prescription by \citet{belczynskiMAXIMUMMASSLAR2010}, with the standard calibration factor $f_\mathrm{LBV}=1.5$. Wind accretion is not considered in our work. While on the main sequence, wind mass loss rates increase with metallicity, as $\propto \Z^{1/2}$, when $T<12,500\,\mathrm{K}$; and $\propto\log\Z$ otherwise.

Mass transfer (MT) occurs when one star fills its Roche lobe (Roche-lobe overflow, RLOF) according to the Roche lobe radius approximation by \citet{eggletonAproximationsRadiiRoche1983}. COMPAS employs an estimate of the response of the donor star's radius to the mass loss, $\zeta_\ast=\d\ln R_\ast/\d\ln M_\ast$, relative to the response of the Roche-lobe radius itself \citep{paczynskiEvolutionCloseBinaries1972,hjellmingThresholdsRapidMass1987,sobermanStabilityCriteriaMass1997}, to determine whether MT is stable or unstable. For main sequence (MS) and helium main sequence (HeMS) stars, $\zeta_\ast=2$ is assumed, and for Hertzsprung Gap (HG) stars, $\zeta_\ast=6.5$, as in \citet{vigna-gomezFormationHistoryGalactic2018a}, based on \citet{geAdiabaticMassLoss2015}. For later stages, $\zeta_\ast$ is computed from the prescription by \citet{sobermanStabilityCriteriaMass1997}. MT from stripped post-helium-burning stars is assumed to be always stable, as suggested in  \citet{taurisULTRASTRIPPEDTYPEIc2013,taurisUltrastrippedSupernovaeProgenitors2015}. For stable MT episodes (stable RLOF), the entire envelope is assumed to be lost by the donor, if it has a clear/core envelope structure (assumed to be the case for MS and HeMS stars). Otherwise, it is assumed that the donor loses the minimal mass necessary to fit within its Roche lobe (HG and later stages). The fraction of mass effectively accreted by the companion, $\beta$, is set by limiting the accretion rate to $10$ times its thermal rate, $M_\mathrm{a}/\tau_{\mathrm{KH},\mathrm{a}}$, \citep{paczynskiEvolutionCloseBinaries1972,neoEffectRapidMass1977,hurleyEvolutionBinaryStars2002a,schneiderEVOLUTIONMASSFUNCTIONS2015}, or Eddington-limited if the accretor is a compact object. Mass that is not accreted carries away the specific angular momentum of the accretor \citep[see, e.g.,][]{sobermanStabilityCriteriaMass1997}. 

Unstable MT always leads to a common envelope (CE) phase which causes the binary to inspiral \citep{paczynskiCommonEnvelopeBinaries1976,podsiadlowskiCommonEnvelopeEvolutionStellar2001,ivanovaCommonEnvelopeEvolution2013}. COMPAS follows \citet{webbinkDoubleWhiteDwarfs1984} and \citet{dekoolCommonEnvelopeEvolution1990} in computing the post-CE separation: a parameter $\alpha_\mathrm{CE}$, set to $1$, expresses the fraction of the lost orbital energy that goes into unbinding the envelope, while a structural parameter $\lambda$ determines the inner boundary of the envelope. The structural parameter is calculated in the "Nanjing lambda" fits by \citet{xuBindingEnergyParameter2010,xuERRATUMBindingEnergy2010}, including improvements made in \texttt{StarTrack} \citep{belczynskiCompactObjectModeling2008} by \citet{dominikDOUBLECOMPACTOBJECTS2012}. HG donors are always assumed to not survive a CE phase \citep["pessimistic" model from][]{dominikDOUBLECOMPACTOBJECTS2012}. No mass is accreted during a CE phase.

Stars with helium core masses within $1.6\en2.25\,\Msun$ at the base of the Asymptotic Giant Branch (AGB), and a carbon-oxygen core that reaches $1.38\,\Msun$, are assumed to undergo an electron-capture supernova (ECSN). The remnants of ECSNe are always assumed to be NSs with mass $1.26\,\Msun$. For core-collapse supernovae (CCSNe), we adopt the "delayed" prescription from \citet{fryerCOMPACTREMNANTMASS2012}, which does not produce an empty "lower mass gap" between NSs and BHs. The neutron star maximum mass is set to $2.5\,\Msun$ (\citeauthor{yeInferringNeutronStar2022}, \citeyear{yeInferringNeutronStar2022}; \citeauthor{aiWhatConstraintsCan2023}, \citeyear{aiWhatConstraintsCan2023}; \citeauthor{rochaMassDistributionMaximum2024}, \citeyear{rochaMassDistributionMaximum2024}; cf. \citeauthor{alsingEvidenceMaximumMass2018}, \citeyear{alsingEvidenceMaximumMass2018}; \citeauthor{shaoMaximumMassCutoff2020}, \citeyear{shaoMaximumMassCutoff2020}; \citeauthor{fanMaximumGravitationalMass2024}, \citeyear{fanMaximumGravitationalMass2024}). CCSNe are reached by stars with a helium core $>2.25\,\Msun$ at the base of the asymptotic giant branch, but the corresponding limit in terms of progenitor mass depends on the details of stellar evolution. To account for this, \citet{fryerCOMPACTREMNANTMASS2012} assume a metallicity-dependent minimum progenitor mass for NS formation based on \citet{poelarendsSupernovaChannelSuperAGB2008}, such that $M_\mathrm{NS}^\mathrm{lower}\propto\log\Z$ above $\Z=10^{-3}\Z_\odot$, with $M_\mathrm{NS}^\mathrm{lower}=9\,\Msun$ at solar metallicity; and $M_\mathrm{NS}^\mathrm{lower}=6.3\,\Msun$ at $Z=10^{-3}\Z_\odot$.

Stars with $35-60\,\Msun$ helium cores are assumed to undergo pulsational pair-instability supernovae (PPISNe), leading to severe mass loss that does not disrupt the star, for which we adopt the prescription by \citet{marchantPulsationalPairinstabilitySupernovae2019}, as implemented in \citet{stevensonImpactPairinstabilityMass2019}. In this model, helium cores within $60\en135\,\Msun$ lead to pair-instability supernovae (PISNe), leaving no compact remnants. Supernova kicks are assumed to be isotropic, and their magnitudes are drawn from a Maxwellian distribution with $\sigma_\mathrm{CCSN}=265\,\mathrm{km}\,\mathrm{s}^{-1}$ \citep{hobbsStatisticalStudy2332005} or $\sigma_\mathrm{ECSN}=30\,\mathrm{km}\,\mathrm{s}^{-1}$ \citep{vigna-gomezFormationHistoryGalactic2018a}. BH natal kicks are scaled down according to the fraction of matter falling back onto the proto-NS \citep{fryerCOMPACTREMNANTMASS2012}.

\subsection{Model definitions}
\label{sec2sub:model_definitions}

The full set of initial condition models results in a grid of 192 possible permutations. While evaluating this entire grid is of great interest (and will be necessary, as discussed further in Sec. \ref{sec:5conclusions}), this first work is dedicated to a deeper analysis of only two permutations. We always use the default evolution models within COMPAS, the Moderate Metallicity cSFH and the OB multiplicity model. The two model permutations we test, Invariant and Varying, each employ the respectively named IMF and orbital parameter distributions, with the SFMR normalized to their specific IMF. In order to facilitate comparison between the two, we sample the 120 SSPs from the Varying IMF version of the cSFH, but normalization through the star-forming mass is always performed with the appropriate cSFH. This SSP sample is shown in Figure 7 of Paper I.   
\section{Computing time-evolving properties}
\label{sec:3computation}

When computing time-evolving quantities, it is important that we make a clear distinction between the two kinds of redshift that are tracked in our sample: the redshift at ZAMS, $\zzams$, and the redshift at merger, $\zmerger$. The first, $\zzams$, is sampled from the GSMF, as described in the previous section, and is not an observable, but instead a proxy for the total system age. The second, $\zmerger$, is an observable: it is exactly the redshift at which we would observe the binary merge.

An SSP formed at $\zzams$ generates a "continuous" sequence of mergers for all $\zmerger<\zzams$, but $\zzams$ is always a discrete variable due to the nature of the initial sampling. While we would like to study both how mergers vary with time (i.e., as a function of $\zmerger$) and how populations of different ages contribute to the cosmic population (i.e., as a function of $\zzams$), the latter suffers from a far greater limitation of resolution than the former.

Any properties that depend on an integration over $\zzams$ --- namely, merger rates --- thus present the challenge of how to best choose a set of $\zzams$ in order to minimize the resolution problem, and how to avoid biasing lower $\zmerger$ due to the greater number of $\zzams$ samples contributing to them. In Sec. \ref{sec3sub:caveats} we address this issue and how it was considered in settling on the 120 SSP configuration described in Sec. \ref{sec:2models}. In Sec. \ref{sec3sub:crossmatching} we lay out the \textit{cross-matching} step which was adopted to allow for an increased resolution on redshift/SFR, as well as metallicity, at the cost of resolution on mass and orbital parameters. We describe the computation of merger rates, which relies on interpolating between the initial set of $\zzams$, in Sec. \ref{sec3sub:mrates}.

\subsection{Caveats of the sample setup}
\label{sec3sub:caveats}

The main limiting factor on the quality of our results, as far as the continuous time-evolution of CBM populations is concerned, is the fact that, due to the Varying IMF, we start by sampling on a two- instead of one-dimensional space: the SFR-metallicity plane, or alternatively, the redshift-metallicity plane. Previous works that has characterized the time and/or metallicity evolution of CBMs has typically relied on a grid of around 30 metallicities within $\FeHinline\sim-2\en0.1$ \citep[e.g][]{belczynskiFirstGravitationalwaveSource2016,chruslinskaInfluenceDistributionCosmic2019,klenckiImpactIntercorrelatedInitial2018,neijsselEffectMetallicityspecificStar2019,broekgaardenImpactMassiveBinary2022}, although larger \citep[e.g., 53 metallicities in][]{broekgaardenImpactMassiveBinary2021a} grids, or individual sampling per binary \citep[as in][]{sonRedshiftEvolutionBinary2022} have also been implemented. Taking $30$ values per variable as a reference resolution might lead us to a currently impractical $\sim900$ SSPs per choice of model permutation, instead of only $\sim30$ as has been common.

We settled on a standard $120$ SSP, consisting of 12 SFR--redshift pairs and 10 metallicities per redshift, after attempting different configurations, as a compromise between resolution and practical constraints. Because this resolution is still relatively low, it was important to perform the galaxy stellar mass-weighted sampling of SFR/redshift and metallicity described in Sec. \ref{sec:2models} in order to obtain the best representation of the SFH. 

With regard to the metallicity, the greatest caveat is that COMPAS is limited to the $\FeHinline=-2.1\en0.3$ range, whereas the Moderate Metallicity cSFH leads to metallicities below this range (see Figure 7 in Paper I) for 16 out of the 120 SSP: all of the $\zzams=10$ populations, and the low-metallicity extreme population for the next 6 $\zzams$ (namely, $\zzams=4.64,3.05,2.49,2.09,1.80$ and $1.55$). We opted to approximate the evolution of any sample below $\FeHinline=-2.1$ to its evolution at $\FeHinline=-2.1$. We consider that this introduces relatively little error for the full population, as, this metallicity measure being in log-scale, galaxies quickly enrich to $\left[\mathrm{Z}\right]\sim-4$, within $\sim10\,\mathrm{Myr}$; and $\left[\mathrm{Z}\right]\sim-2$, within $\sim100\,\mathrm{Myr}$. For our configuration in particular, the error is lessened for $\zzams\lesssim6$ populations, as they are concentrated around higher metallicities. Even for higher redshifts we might expect a relatively small induced error, as CBM populations might already be less sensitive to metallicity below $\FeHinline=-1$ (see discussion in Section \ref{sec4sub:formation_efficiency}).

Low-resolution is otherwise partially compensated for by the quantile-based sampling of metallicity and $\zzams$ (see Section 3.1 in Paper I), which ensures that both quantities are more densely sampled where the distribution is steeper. Whenever we need to interpolate from the sampled metallicity and $\zzams$ to arbitrary values, this allows for the best approximation at the given resolution. 

\subsection{Minimizing the number of BPS runs}
\label{sec3sub:crossmatching}

While we initially sample over SFR/redshift and metallicity, only metallicity is an evolution parameter, such that is not necessary to run COMPAS multiple times for different SFRs with similar metallicities. To avoid this, we build a pre-evolved grid of binary populations at 33 metallicites sampled uniformly between $\FeHinline=-2.1\en0.3$. Instead of evolving each individual SSP in the "working sample", we perform a \textit{cross-matching} step between the working and the pre-evolved sample with the nearest metallicity, wherein each individual binary in the SSP is matched to the one in the pre-evolved sample with the closest initial parameters, and assigned its final parameters. 

Since this in principle leads to a degree of error in the evolution, we take steps to control to what degree the individual binary parameters are allowed to differ between the working and the pre-evolved samples. Because the Varying IMF fluctuates around the \citet{salpeterLuminosityFunctionStellar1955} IMF, we build a list of $200$ values of $m_1$ as the median of $200$ quantiles of the \citet{salpeterLuminosityFunctionStellar1955} IMF. From the chosen orbital parameter distributions, we then again employ quantile medians to select $100$ values of $P$ per $m_1$; and $100$ values of $q$ and $10$ of $e$ per $\lrp{m_1,P}$, totaling $1000$ $\lrp{q,e}$. This standard sample of $2\times10^7$ binaries is run for each of the $33$ metallicities, totaling $6.6\times10^8$ binaries in the pre-run grid, whith the same order as the working sample. A different such pre-evolved sample is generated for each choice of orbital parameter distribution. We find that this resolution is sufficient to allow the differentiation of population components and formation channels, with only two instances of visible artifacts in the final parameters during our analysis of the results in Sec. \ref{sec:4results}.

\subsection{Computing merger rates over time}
\label{sec3sub:mrates}

For each merger in our sample, the time of merger, $t_\mathrm{m}$, and merger redshift, $\zmerger$, can be found from its $\zzams$ and delay time, $t_\mathrm{d}$. Once all $t_\mathrm{m}$ are known, binning over $t_\mathrm{m}$ allows the merger rate per star-forming mass for a given type of merger, to be estimated as a function of metallicity, $\zzams$ and merger time, as

\begin{equation}
    \label{eq:sample_rate}
    \mathcal{R}_\mathrm{sp}(t_\mathrm{m}, \zzams,\Z) = \frac{\d^2 N_\mathrm{merger}}{\d t_\mathrm{m} \d M_\mathrm{sf}},
\end{equation}

\noindent where $M_\mathrm{sf}$ is the total star-forming mass corresponding to the SSP with the given $\zzams,\Z$; $V_\mathrm{c}$ is the comoving volume; and $N_\mathrm{merger}$ is the number of mergers at $t_\mathrm{m}$ from the same SSP. Because SSPs are defined by their $\FeHinline$ and the cSFH is fitted over $\ZOH$, through this section we use $\Z$ to indicate metallicity regardless of the specific quantity explicitly used in the calculations. Whenever necessary, these quantities are converted as described in Section \ref{sec2sub:environment}.

The star-forming mass, $M_\mathrm{sf}$, accounts for the initial mass of the entire stellar population of which the binary population we evolve is a part of, including systems with $<0.8\,\Msun$ primaries, which we do not evolve, and all isolated stars (as well as higher-order multiples had we used the AM model instead). BOSSA tracks the mass formed in isolated stars, and outputs the star-forming mass for all systems with $\geq0.8\,\Msun$ primaries. Because we assume that the IMF describes the ZAMS mass distribution of both primaries and companions (see Section \ref{sec2sub:imf}), $M_\mathrm{sf}$ can be found for each SSP by normalizing its IMF to the $\geq0.8\,\Msun$ star-forming mass in that mass range (from BOSSA), and then integrating $m\times\mathrm{IMF}(m)$ over $0.08-150\,\Msun$.

The rate in equation \eqref{eq:sample_rate} should be understood as the contribution from the SSP corresponding to a given $\zzams$ and $\Z$ to mergers at a later time $t_\mathrm{m}$. For a fixed $\zzams$, the contribution from each metallicity can be estimated by interpolating from the 10 sampled $\Z$. Integrating over metallicity then yields the total contribution from a given $\zzams$ to the merger rate density at $t_\mathrm{m}$. Analogously, interpolating from the $12$ sampled $\zzams$ and integrating yields the total rate at $t_\mathrm{m}$.

The \textit{physical} contribution to the total merger rate \textit{density} from stars formed at a given $\zzams$ with a given $\Z$ is found from the corresponding $\mathcal{R}_\mathrm{sp}$ as

\begin{align}
    \nonumber
    \label{eq:physical_rate_contribution}
    \mathcal{R}_\mathrm{ph}(t_\mathrm{m},\zzams,\Z) =& \mathcal{R}_\mathrm{sp}(t_\mathrm{m},\zzams,\Z) \times\\
    &\frac{\d^3 M_\mathrm{sf}}{\d V_\mathrm{c} \d\Z\d\zzams}(\zzams,\Z),
\end{align}

\noindent where $V_\mathrm{c}$ is the comoving volume; the second term describes the total star-forming mass density per metallicity and redshift bin from an assumed cSFH. Integrating over metallicity then yields the total contribution from a given $\zzams$ to the merger rate density at $t_\mathrm{m}$. Analogously, interpolating from the $12$ sampled $\zzams$ and integrating yields the total rate at $t_\mathrm{m}$. Finally, the total merger rate at $t_\mathrm{m}$ is found as

\begin{align}
    \label{sec3eq:mrate_integral}
    \nonumber
    \mathcal{R}_\mathrm{m}\lrp{t_\mathrm{m}} =& \int_{10}^{\zmerger}\int_{10^{-5}\Zsun}^{10^{0.5}\Zsun}\,\mathcal{R}_\mathrm{ph}(t_\mathrm{m},\zzams,\Z)\,\d\Z\,\d\zzams\\ 
    \nonumber
    =&\int_{10}^{\zmerger}\int_{10^{-5}\Zsun}^{10^{0.5}\Zsun}\,\frac{\d^2N_\mathrm{merger}}{\d t_\mathrm{m}\d M_\mathrm{sf}}\lrp{t_\mathrm{m},\zzams,\Z}\times \\
    & \frac{\d^3 M_\mathrm{sf}}{\d V_\mathrm{c}\d\Z\d\zzams}\lrp{\zzams,\Z}\,\d\Z\,\d\zzams,
\end{align}

\noindent where we set the metallicity integration limits to the approximate minimum and maximum metallicities reached by the metallicity-specific cSFH from \citet{chruslinskaEffectEnvironmentdependentIMF2020} for galaxy stellar masses within $10^6-10^{12}\,\Msun$. We fix $\mathcal{R}_\mathrm{sp}$ to zero at the metallicity boundaries.

The integral in equation \ref{sec3eq:mrate_integral} is in practice approximated by a Riemann sum over metallicity and redshift bins, as has been typically done \citep[e.g.,][]{dominikDoubleCompactObjects2015,mandelMergingBinaryBlack2016,belczynskiCompactBinaryMerger2016,neijsselEffectMetallicityspecificStar2019,baveraImpactMasstransferPhysics2021,broekgaardenImpactMassiveBinary2022}. We do not integrate explicitly over delay time, since each binary in our sample has both a well defined time of formation and merger encoded in $\zzams$ and $\zmerger$. 

\section{Results}
\label{sec:4results}

\begin{figure*}
    \centering
    \includegraphics[width=\textwidth]{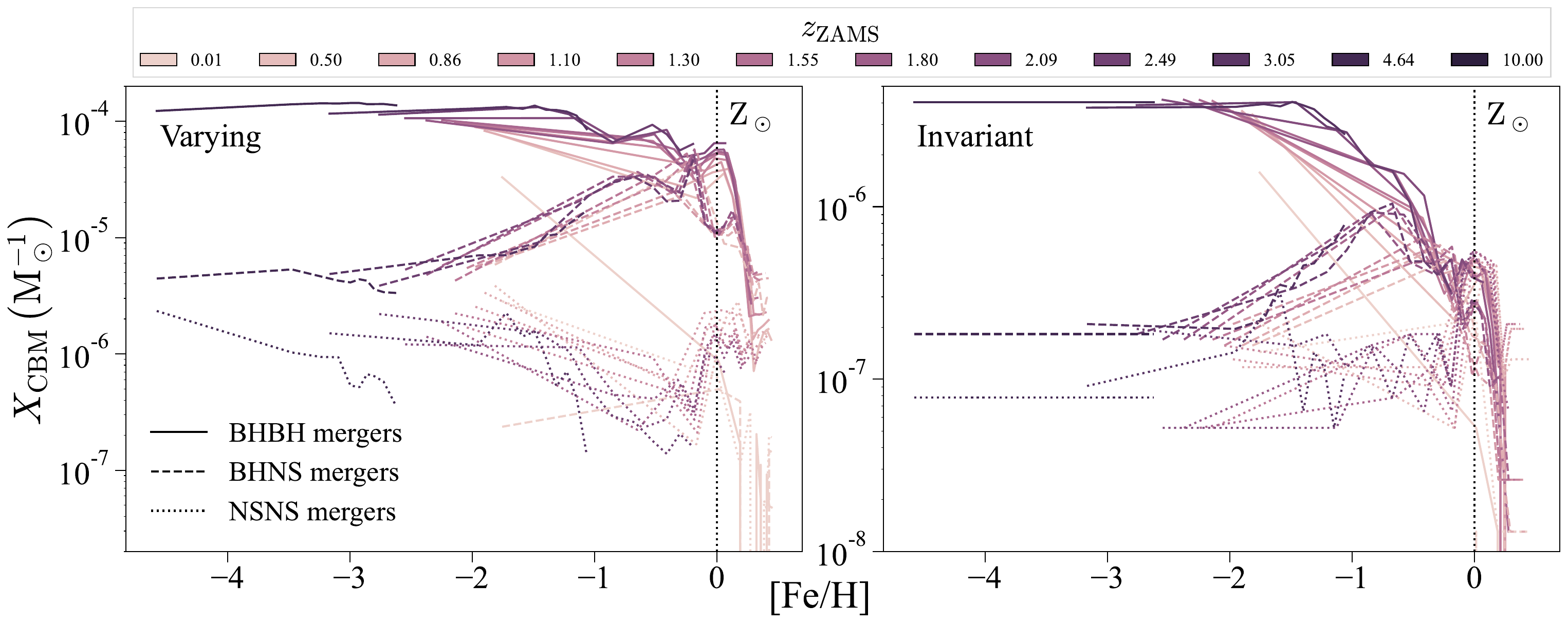}
    \caption{Formation efficiency of BHBH (solid lines), BHNS (dashed lines) and NSNS (dotted lines) merger progenitors, at different $\zzams$ (color code), as a function of metallicity, for both Varying (left) and Invariant (right) models. No distinction is made between BHNSs and NSBHSs. Compact object formation is generally more common in the Varying model. The Invariant model show effects of evolution metallicity-dependence: BH formation becomes increasingly difficult with growing metallicity. In the Varying model this competes with the effect of the variation of the IMF, which becomes top-heavy at high SFRs, which are on average associated with higher metallicities (see Sec. \ref{sec2sub:imf}), but also with increasing $\zzams$. BHBH formation is sharply cut off at solar metallicity, with BHNS formation also being suppressed, albeit to a lesser degree; the NSNS fraction is amplified as the most massive stars drop into the NS progenitor range due to wind mass loss, becoming dominant in the Invariant model, or sitting just below BHNSs in the Varying model. Below $\FeHinline=-2.1$, the formation efficiency can only vary due to variations of the IMF or due to a coalescence time-cutoff, but this latter effect is not significant at these metallicities (see text for discussion). This makes the formation efficiency constant in that region in the Invariant model.}
    \label{sec4fig:metal_fractions}
\end{figure*}

\begin{figure}
    \centering
    \includegraphics[width=\columnwidth]{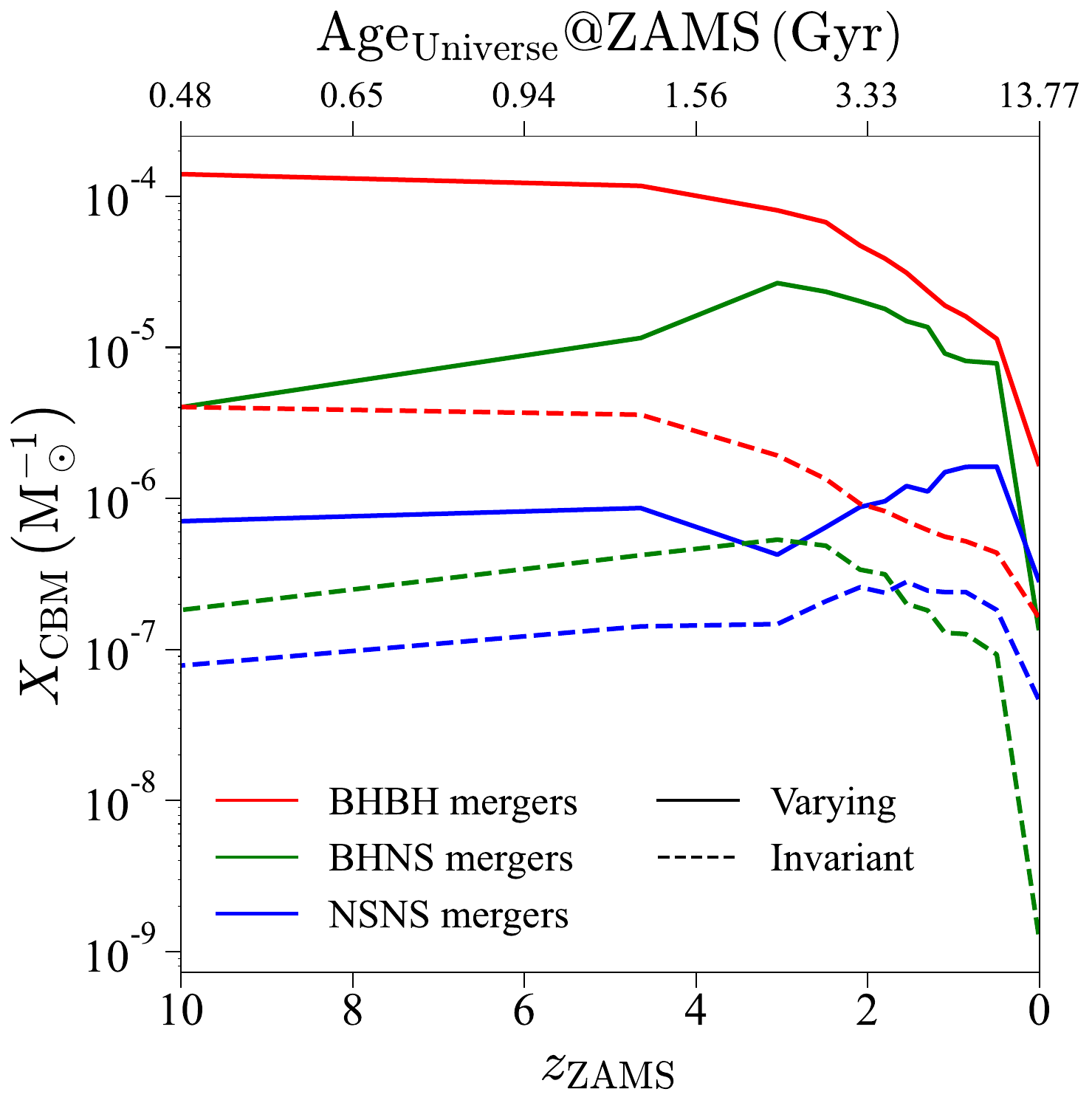}
    \caption{Formation efficiency of BHBH (red lines), BHNS (green lines) and NSNS (blue lines) progenitors as a function of $\zzams$, for both the Varying (solid lines) and Invariant (dashed lines) models. No distinction is made between BHNSs and NSBHSs. The Varying model increases compact object formation efficiency overall, but it most strongly affects BH progenitors. CBM progenitor formation generally starts to be suppressed below $\zzams\sim2$. In both models BHBH merger progenitors are always dominant but drop monotonically with decreasing redshift. BHNS formation efficiency rises in response down to $\zzams\sim3$, and also falls for lower redshift. The NSNS formation efficiency increases monotonically with decreasing redshift. The last point suffers from undersampling (see text and Fig. \ref{sec4fig:metal_fractions}) and is thus not reliable.}
    \label{sec4fig:redshift_frac}
\end{figure}

In this section we examine the formation efficiencies as a function of $\zzams$, and the CBM properties (primary mass, mass ratio and coalescence times) as function of both $\zmerger$ and $\zzams$, the latter as a proxy of progenitor age and mean metallicity, for BHBH, BHNS and NSNS mergers. In some cases we reference the delay time of binaries, which is equal to the coalescence time minus the time between ZAMS and the second supernova. Whenever relevant we make a distinction within BHNS mergers between those where the first CO to form was the BH (BHNS) or the NS (NSBH). Otherwise we use BHNS to refer to both classes simultaneously. Whether or not the distinction is made is explicitly stated at the beginning of each Section. Some similarly-named variables are defined in Table \ref{sec4tab:variables}.

Of the $\sim10^8$ initial binaries, only $\lesssim10^5$ evolve up to CBMs within their available time for merger (lookback time at the respective $\zzams$). In the Invariant model, the final population contains $\sim10^4$ BHBHs, $\sim10^3$ BHNSs and NSNSs each, and $\sim10^2$ NSBHs. In the Varying model, the final population holds $\sim10^5$ BHBHs, $\sim10^4$ BHNSs, $\sim10^3$ NSNSs and $\sim10^2$ NSBHs.

Because our $\zzams$ sampling is coarse, tracking the $\zmerger$-evolution of the populations requires us to define relatively broad intervals over which to collect CBMs, which must be chosen in such a way as to not introduce an artificial towards one $\zmerger$ bin or another. Were these intervals defined with a regular redshift-width of $1$, for example, the $\lrs{9,10}$ interval would be undersampled in relation to $\lrs{0,1}$, and the entire $\lrs{5,9}$ interval would be populated solely by the relatively few mergers in the tail of the $\zzams=10$ population. We thus define 4 $\zmerger$ intervals of varying width, such that each contains three of the twelve $\zzams$ and begins at the earliest of the three.

Finally, the $\zzams=0.01$ population must be treated with particular care. The lookback time at that redshift is $\approx0.14\,\Gyr$, and, because many BCOs have coalescence times of order $\sim1\en10\,\Gyr$, this population suffers from a particularly strong coalescence time-cutoff, the exact degree of which depends on the typical coalescence times of each merger class. In some cases, the remaining population is too small to be appropriately compared to the others, but even when a significant portion of it survives, it will still have a greater bias toward short coalescence times than the rest. This means that, in terms of tracking the influence of evolving star/binary formation conditions and metallicity-dependent evolution, the $\zzams=0.01$ population cannot be relied upon in the same way as those of greater redshift. Therefore, unless this population is explicitly mentioned, our discussions of any $\zzams$-evolution are concentrated on the $\zzams=0.5\en10$ interval.

\begin{table}
 \caption{Summary of similarly-named variables.}
 \label{sec4tab:variables}
 \begin{tabular}{ll}
  \hline
  Variable & Definition \\
  \hline
  $z_\mathrm{ZAMS}$ & \multicolumn{1}{p{6.5cm}|}{Redshift corresponding to the age of the Universe at ZAMS for a given merging binary.} \\
   $z_\mathrm{merger}$ & \multicolumn{1}{p{6.5cm}|}{Observable redshift of merger.} \\
  $t_\mathrm{d}$ & \multicolumn{1}{p{6.5cm}|}{Delay time: time between ZAMS and merger.} \\
  $t_\mathrm{c}$ & \multicolumn{1}{p{6.5cm}|}{Coalescence time: time between second supernova and merger.} \\
  $t_\mathrm{m}$ & \multicolumn{1}{p{6.5cm}|}{Merger time: age of the Universe at merger.} \\
  $m_i$ & \multicolumn{1}{p{6.5cm}|}{Mass of the $i$th component of a multiple of arbitrary order, i.e., $m_1$ would be primary mass, $m_2$ secondary, $m_3$ tertiary...} \\
  $m_\mathrm{cp}$ & \multicolumn{1}{p{6.5cm}|}{Companion mass: refers to a companion star of arbitrary order.} \\
  \hline
 \end{tabular}
\end{table}

\subsection{Formation efficiency}
\label{sec4sub:formation_efficiency}

For a given population (an SSP or a union of SSPs over metallicity for fixed redshift), we define the formation efficiency of CBMs of a given type as

\begin{equation}
    \label{eq:formation_eff_def}
    X_\mathrm{CBM} = \frac{N_\mathrm{CBM}}{M_\mathrm{sf}},
\end{equation}

\noindent where $M_\mathrm{sf}$ is the star-forming mass corresponding to that population, and $N_\mathrm{CBM}$ is the number of CBMs of that type in the population that did merger, considering their individual $\zzams$. In Fig. \ref{sec4fig:metal_fractions}, we show the formation efficiency for BHBH, BHNS and NSBHs for each SSP, as a function of metallicity and $\zzams$, for both models. We do not distinguish between BHNSs and NSBHs in this section.

The $\zzams=0.01$ population, in addition to the time-based undersampling already discussed, is also affected by a metallicity-related undersampling, as the increasingly unfavorable conditions for CO formation at supersolar metallicities lead to some cases of no CBMs being formed at our initial sampling resolution, yielding the sharp drop and jagged pattern seen in this region, in Fig. \ref{sec4fig:metal_fractions}, for both models. Both models display the effect of metallicity-dependence in the evolutionary models, and of a coalescence time-cutoff for lower $\zzams$. In the Invariant model, this cutoff is the only effect of $\zzams$ for a fixed metallicity, but it is only significant for the lower $\zzams$ values (see also Section \ref{sec4sub:ctimes}), which are shifted to higher metallicities. As a result, variations of $X_\mathrm{CBM}$ with $\zzams$ are suppressed for $\FeHinline\lesssim-2.1$. The Varying model is subject to the same trends, but also displays the effect of the metallicity- and redshift (SFR)-dependence of the IMF.

Across both models, the common dominant feature is a sharp increase/cutoff around solar metallicity, which is a consequence of metallicity-dependent evolution, and the increasing wind mass loss efficiency with metallicity in particular. Below solar metallicity, there is a general dominance of BHBH progenitor formation, followed by BHNSs and NSNSs. The BH progenitor preference for low metallicities is well established \citep[e.g.,][]{dominikDOUBLECOMPACTOBJECTS2012,belczynskiFirstGravitationalwaveSource2016,eldridgeBPASSPredictionsBinary2016,mapelliCosmicMergerRate2017,giacobboProgenitorsCompactobjectBinaries2018,neijsselEffectMetallicityspecificStar2019,santoliquidoCosmicMergerRate2021,broekgaardenImpactMassiveBinary2022} and is linked to decreased wind mass loss efficiencies, which result in more massive progenitors and less intense supernova kicks, increasing the likelihood of the binary to remain bound and merge within the available time. In addition to winds, the sharp drop in BHBH and BHNS formation efficiency at solar metallicity is also connected to the greater stellar radius for a fixed mass at higher metallicities \citep{hurleyComprehensiveAnalyticFormulae2000}, which makes binaries more likely to engage in a CE phase earlier, while the donor is still in the HG, preventing them from forming CBMs under our "pessimistic" CE assumption \citep{belczynskiEffectMetallicityDetection2010,klenckiImpactIntercorrelatedInitial2018}.

We find NSNS mergers to be much less sensitive to metallicity, in agreement with previous work (\citeauthor{giacobboProgenitorsCompactobjectBinaries2018},\citeyear{giacobboImpactElectroncaptureSupernovae2019}; \citeauthor{klenckiImpactIntercorrelatedInitial2018}, \citeyear{klenckiImpactIntercorrelatedInitial2018}; \citeauthor{broekgaardenImpactMassiveBinary2022}, \citeyear{broekgaardenImpactMassiveBinary2022}; cf. \citeauthor{gallegos-garciaEvolutionaryOriginsBinary2023}, \citeyear{gallegos-garciaEvolutionaryOriginsBinary2023}). However, as both BHBH and BHNS fractions drop at and above solar metallicity, NSNS fractions slightly increase in response, which we suggest to be at least partially connected to a greater fraction of BH progenitors falling into the NS progenitor range due to wind mass loss. This trend is also suggested by the redshift-dependence of NSNS primary masses in Sec. \ref{sec4sub:nsns_masses}. 

Whether BHNS or NSNS progenitor formation becomes dominant at supersolar metallicities is model-dependent: the former is the case in the Varying model, and the latter in the Invariant model, although the degree to which one is dominant depends on redshift. For supersolar metallicities and lower redshifts, NSNS progenitor formation in some cases becomes comparable to or even more efficient than that of BHBH progenitors. Because we have assumed that evolution for metallicites below $\FeHinline=-2.1$ is as at $\FeHinline=-2.1$, the formation efficiency can only vary significantly in that region due to variations of the IMF, which affect both the progenitor mass distribution and the star-forming mass corresponding to a given SSP size. Consequently, the formation efficiency only varies within this region in the Varying model, where it is dominated by the tendency of the SFR to grow with metallicity, and which makes the IMF more top-heavy (Sec. \ref{sec2sub:imf}). A top-heavy IMF means both an increased fraction of BH progenitors relative to NS progenitors and an increase in $M_\mathrm{sf}$ for a fixed SSP size. The latter effect tends to decrease all formation efficiencies, whereas the former tends to increase that of BHBHs and decrease that of NSNSs, while BHNSs favor an intermediate range. The results in this metallicity range are monotonically increasing and decreasing BHBH and BHNS progenitor formation efficiencies, respectively, and approximately stable BHNS progenitor formation.

In the $-2.1<\FeHinline<0$ range, the Varying model sees the varying IMF compete with the metallicity-dependence of winds and radii discussed above. In the Invariant model, the latter effects increasingly move BHBH progenitors to the BHNS progenitor range, up to $\FeHinline\sim-1$. For even higher metallicities, however, the NSNS progenitor range becomes increasingly favored, up to the sharp shift at solar metallicity. This effect also sets the overall behavior of the formation efficiencies within $-2.1<\FeHinline<0$ in the Varying model, but the top-heavy IMF contributes to keep BHBH progenitor formation more common than BHNS progenitor formation up to $\FeHinline\sim-0.5$. The shift to NSNS progenitor dominance, as also mentioned before, never happens in the Varying model. Overall, Fig. \ref{sec4fig:metal_fractions} also shows  that there is little variation of the formation efficiencies below $\FeHinline=-1$, relative to the sharp features above. This suggests that a grid of evolved binaries with metallicites sampled uniformly over $\Z$, instead of log-uniformly on $\Z$, might improve the precision of our simulations in the future.

The $\zzams$-dependence of formation efficiencies is more clearly seen in Fig. \ref{sec4fig:redshift_frac}, where they are integrated over metallicity. While $\zzams=0.01$ is unreliable, the earlier part of the curve makes clear the trends present in each model. The BHBH evolution is similar in both, with a decrease over time, while it remains as the most common CBM progenitor produced at all $\zzams$. Again, CO progenitor production is overall increased in the Varying model in comparison with the Invariant one. All formation efficiencies are increased by at least one order of magnitude overall, with BHBHs being the most privileged and NSNSs the least. The models otherwise differ with regard to whether BHNS or NSNS merger progenitors are more commonly produced at low redshifts. In common, both models have the BHNS efficiency to monotonically increase down to $\zzams\sim3$ and then drop, while the NSNS efficiency increase monotonically with decreasing redshift (not considering $\zzams=0.01$). In contrast, the Varying model has the BHNS formation efficiency still one order of magnitude greater than that of NSNSs locally, while in the Invariant model they cross at $\zzams\sim2$, resulting in a NSNS progenitor formation efficiency about two times greater than that of BHNSs, locally. 

\subsection{Component masses}
\label{sec4sub:masses}

\begin{figure*}
    \centering
    \includegraphics[width=\textwidth]{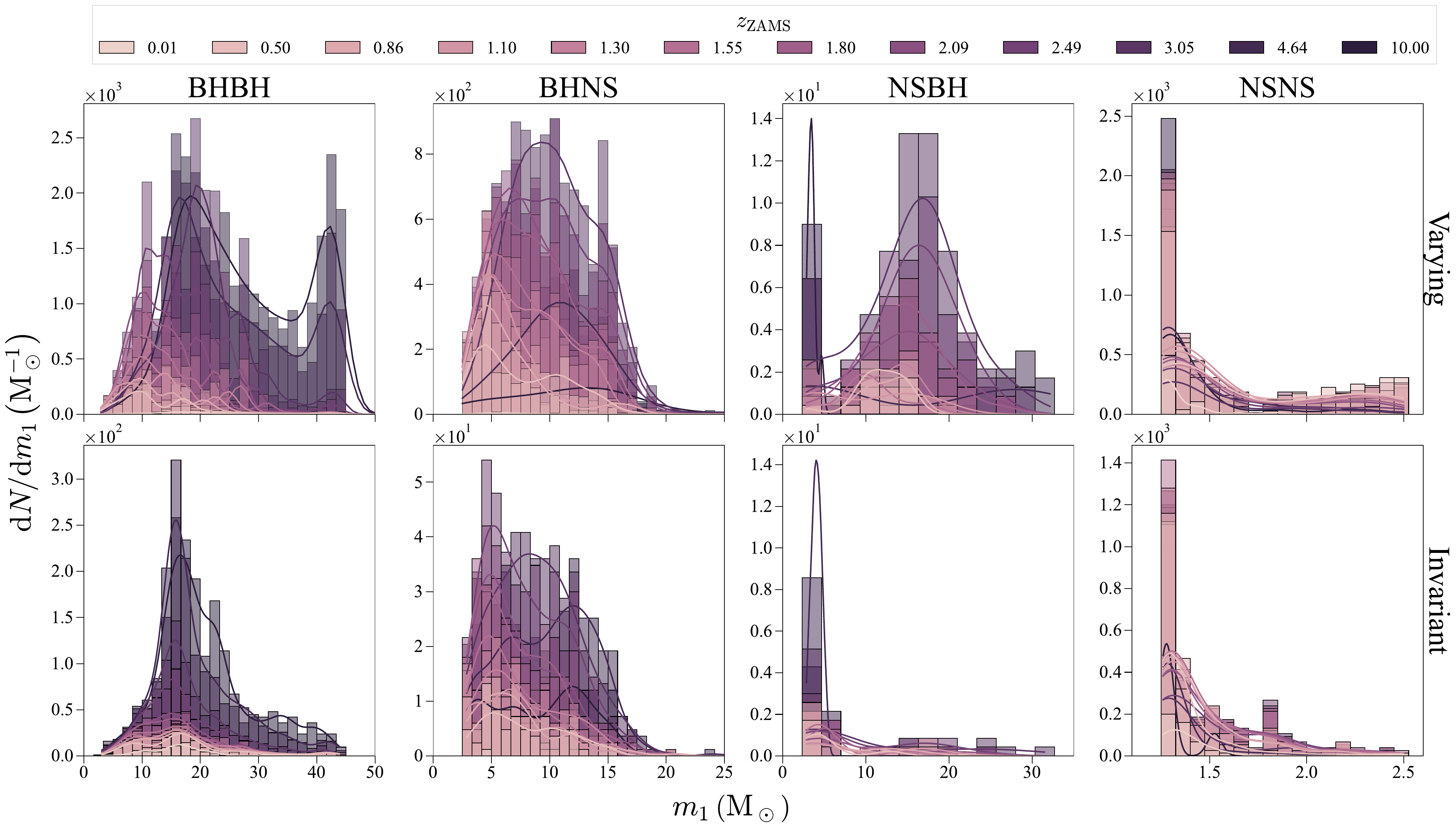}
    \caption{Primary mass ($m_1$) distributions of BHBH, BHNS, NSBH and NSNS mergers (columns from left to right, in that order) in the Varying (top row) and Invariant (bottom row) models, for different $\zzams$ (color code), as both histograms and kernel density estimates (KDEs, solid curves). KDE bandwidths are set according to sample sizes in order to provide the best visualization of the $\zzams$-evolution of the distribution. Due to a lookback time of $\approx0.14\,\mathrm{Gyr}$ at $\zzams=0.01$, those specific CBM samples suffer a strong coalescence time cutoff (see Sec. \ref{sec4sub:ctimes}) and do not accurately reflect the influence of evolving initial conditions and metallicity on binary formation/evolution. At $\zzams=0.01$, no NSBHs are formed in either model, and the BHNS samples in either model, as well as the Invariant BHBH sample, are too small to be distinguishable.}
    \label{sec4fig:zzams_m1}
\end{figure*}

In the following Sections we analyze the resulting CBM primary mass ($m_1$) and mass ratio ($q$) distributions, and their evolution both in terms of progenitor age ($\zzams$) and merger redshift ($\zmerger$). This Section differentiates between BHNSs and NSBHs, and uses the term "BH+NS" whenever both are considered simultaneously.

While we characterize the resulting distributions, our main concern is evaluating the differences between the Invariant and Varying models. This, however, requires understanding the typical formation channels that connect the initial to final parameter space, which are tracked by COMPAS, but which we have not kept in our synthesis output. We thus rely on previous work which has characterized CBM formation channels reproduced by COMPAS in its default settings to interpret some of our results. In our simulations, we note that each case results in certain robust features that are relatively independent of the initial condition models, and that are also present in the aforementioned previous work. Whenever these features have been previously linked to certain evolutionary channels, we tentatively associate the features we have obtained to the same channels. 

In practical terms, we adopt a picture within which formation channels, set by the chosen evolution models, drive the location of certain features, while the initial condition models set their relative weights. We note that a similar conclusion has been reached by \citet{sonLocationsFeaturesMass2023}, who verified that the key features of the BHBH $m_1$ distribution obtained with COMPAS for a fixed set of evolution models are robust against variations of the assumed metallicity-specific cSFH. This picture, if directly confirmed, is in itself a useful principle for studying model variations in BPS, which we come back to in Sec. \ref{sec:5conclusions}.

Some of the discussion relies on verifying the distribution of initial \textit{progenitor} parameters for a given set of CBMs. In keeping with notation, these quantities are always labeled as "ZAMS". Any quantities that do not receive this label refer to the CBM parameters themselves, i.e., at merger. Orbital period is always given in days, even if the unit is suppressed for simplicity of notation.

\subsubsection{Black hole+black hole mergers}
\label{sec4sub:bhbh_masses}

The BHBH $m_1$ distributions are shown in terms of $\zzams$ in the left panels of Fig. \ref{sec4fig:zzams_m1}, and are characterized by peaks at $\sim9$, $\sim16$ and $\sim45\,\Msun$. For both models, the distributions sharply stops at an upper mass gap beginning at $\sim45\,\Msun$, a result of the PPISNe model by \citet{marchantPulsationalPairinstabilitySupernovae2019}, as implemented by \citet{stevensonImpactPairinstabilityMass2019}, yielding remnant masses within $\sim30-45\,\Msun$ from helium core masses within $\sim35\en60\,\Msun$. In the Invariant model, the $\sim16\,\Msun$ peak is dominant at all $\zzams$, with a high-mass tail up to $\sim45\,\Msun$ increasingly significant with growing redshift. There is little evidence of a secondary peak, although we notice a slight flattening toward low masses at lower $\zzams$. In the Varying model, on the other hand, the distribution strongly varies with $\zzams$, shifting from the $\sim9$ to the $\sim16\,\Msun$ peak between $\zzams=0.5$ and $\zzams=2.49$. For even higher redshifts, the distributions continues to shift toward high masses, producing a prominent $\gtrsim20\,\Msun$ tail and a pileup at $\sim45\,\Msun$ which has been termed the "PPISNe pileup". The $\zzams=0.01$ sample is visible only in the Varying model, due to a shifting of the BHBHs to greater coalescence times, as discussed in Sec. \ref{sec4sub:bhbh_ctimes}.

\begin{figure*}
    \centering
    \includegraphics[width=\textwidth]{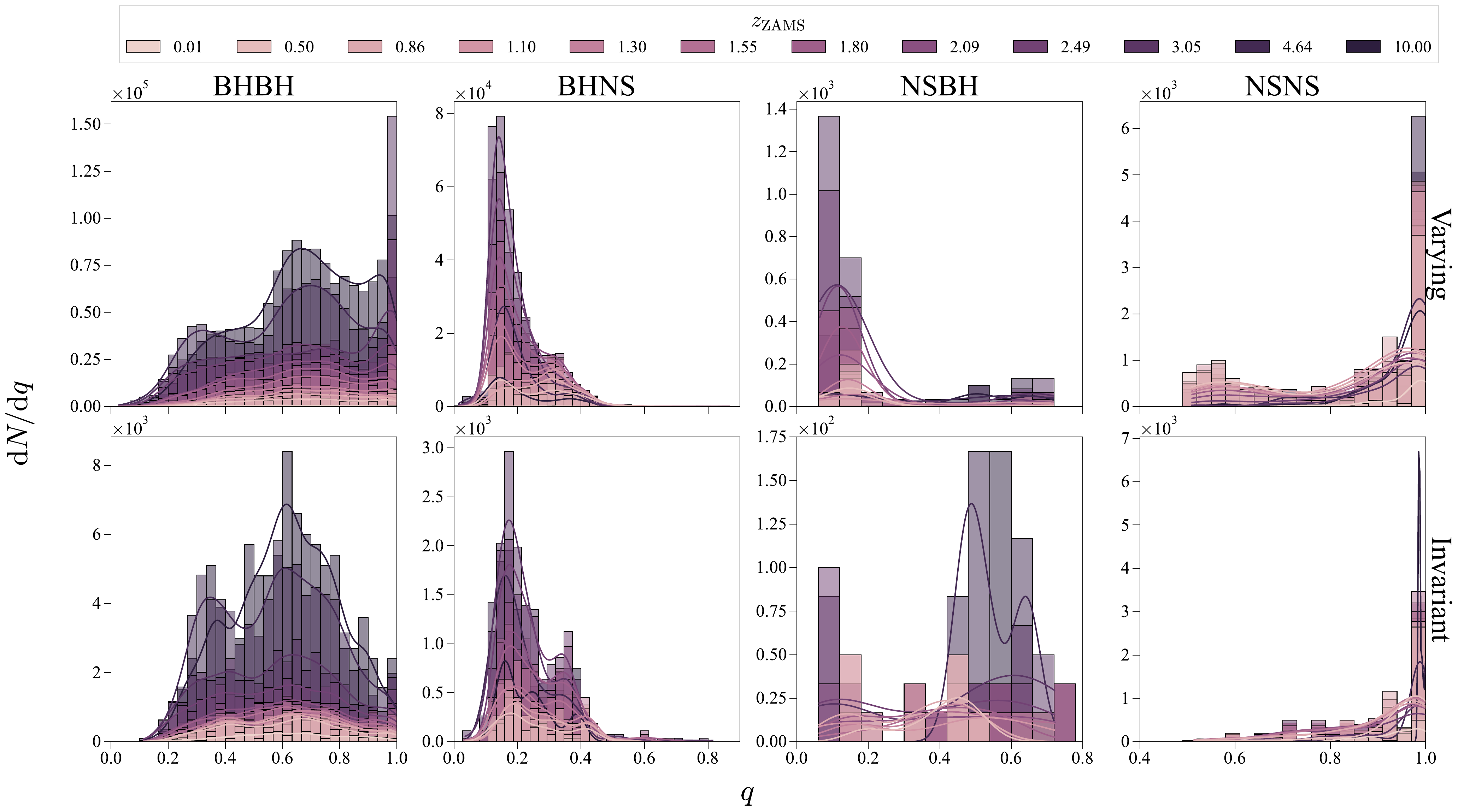}
    \caption{The same as Fig. \ref{sec4fig:zzams_m1}, but for the mass ratio ($q=m_2/m_1\leq1$). No NSBHs are produced for $\zzams=0.01$, and only the NSNS $\zzams=0.01$ distributions are large enough to be clearly visible.}
    \label{sec4fig:zzams_q}
\end{figure*}

The BHBH mass ratio over $\zzams$ distributions in Fig. \ref{sec4fig:zzams_q} are characterized by two common peaks, one about $q\sim0.3$ and a second about $q\sim0.6$, although they become increasingly flat toward low redshift (young progenitors). In the Invariant model, the two peaks are only clearly distinguished above $\zzams=3.05$, and with growing redshift the upper peak becomes increasingly dominant. In the Varying model the peaks are shifted slightly apart, but the overall behavior remains: the distribution is dominated by the upper peak at high redshifts, but becomes increasingly flatter for decreasing $\zzams$, and the two peaks become distinguishable above $\zzams=4.64$. Additionally, the Varying model is overall shifted to greater $q$ and exhibits a pileup of symmetric ($q=1$) systems not present in the Invariant model, and which is more significant for greater $\zzams$.

BHBH peaks around $m_1\sim9\,\Msun$ and $m_1\sim16\,\Msun$ have been characterized with COMPAS by \citet{sonRedshiftEvolutionBinary2022} and verified to be robust against metallicity-specific SFH variations by \citet{sonLocationsFeaturesMass2023}. The authors differentiated between the CE and stable RLOF channels, which are generally agreed to be the main sources of BHBH mergers \citep[while relative weights are still under debate, see, e.g.,][]{neijsselEffectMetallicityspecificStar2019,baveraImpactMasstransferPhysics2021,marchantRoleMassTransfer2021}. The CE channel includes systems which have undergone at least one CE phase, while the stable RLOF channel contains all systems that have undergone \textit{only} stable MT. Typically, both channels involve stable RLOF of the primary onto the secondary as the first MT episode, and are differentiated by a second episode from the secondary onto the collapsed primary. They find the CE channel to be characterized by a $\sim18\,\Msun$ peak and shorter delay times peaking at $t_\mathrm{d}\sim0.1-1\,\mathrm{Gyr}$ (as the CE phase is more efficient in shrinking the orbit), and to remain the overall dominant channel at all redshifts up to $10$. The stable RLOF channel, on the other hand, is found to be characterized by longer delay times of $\sim10\,\mathrm{Gyr}$ and to be the dominant formation channel for $m_1\gtrsim20\,\Msun$ BHBHs, particularly for low metallicities ($Z\leq\Zsun/10$), but its resulting mass distribution is also characterized by a bump around $m_1\sim9\,\Msun$ at high metallicities ($Z\geq\Zsun/5$). They find the stable RLOF channel to be always subdominant in relation to the entire population, but to have an increasing contribution with decreasing $\zmerger$ associated to its typically longer delay times, up to $\sim40\%$ of the BHBH merger detection rate at $z=0$. It is found to be the main formation channel for $m_1\gtrsim20\,\Msun$ BHBHs, but the $\sim9\,\Msun$ bump remains nonetheless subsumed by the CE channel within the full $m_1$ distribution, even at low redshift. They also find the CE channel to preferentially form asymmetrical binaries, with a peak at $q\approx0.3$, although the distribution is broad and covers the entire $0.2\lesssim q\lesssim1$ range. The stable RLOF channel, in contrast, is found to form binaries within $0.6\lesssim q\lesssim0.8$.

This balance between the CE and RLOF channels was determined in a model close to our Invariant model, and is comparable to the lower-left panels of Figs. \ref{sec4fig:zzams_m1} and \ref{sec4fig:zzams_q}. We leave a comparison of the delay/coalescence times to Sec. \ref{sec4sub:bhbh_ctimes}. In Fig. \ref{sec4fig:zzams_m1}, for the Invariant model, the primary mass distribution is dominated by the CE peak (which here appears closer to $\sim16\,\Msun$), with a lesser contribution at lower masses and lower $\zzams$ (higher metallicity) from the RLOF channel, and a growing high-mass tail at high redshift (lower metallicity) which is the main contribution from the RLOF channel, up to supplying a PPSINe pileup. In Fig. \ref{sec4fig:zzams_q}, the two peaks are generally comparable, but the upper peak becomes increasingly dominant as the distribution shifts to higher masses and accrues a greater contribution from the stable RLOF channel. We interpret the CE channel as contributing to both peaks, although favoring $q\sim0.3$, while the stable RLOF channel strongly favors $q\sim0.6$. In the Varying model, the stable RLOF channel becomes dominant within $\zzams\leq2.49$ instead (shift to $\sim9\,\Msun$). Because this is true even at the lowest redshifts, it is not simply explained by the varying IMF, and we look instead to the ZAMS orbital period distribution: while in the Invariant model it is sampled from a log-uniform distribution, it is double-peaked at $10^{0.2}$ and $10^{4}\,\mathrm{d}$ in the Varying model, with the lower peak being dominant above $m_1\approx40\,\Msun$ and the upper peak below (Sec. \ref{sec2sub:orbital_parameters}). 

\begin{figure}
    \centering
    \includegraphics[width=\columnwidth]{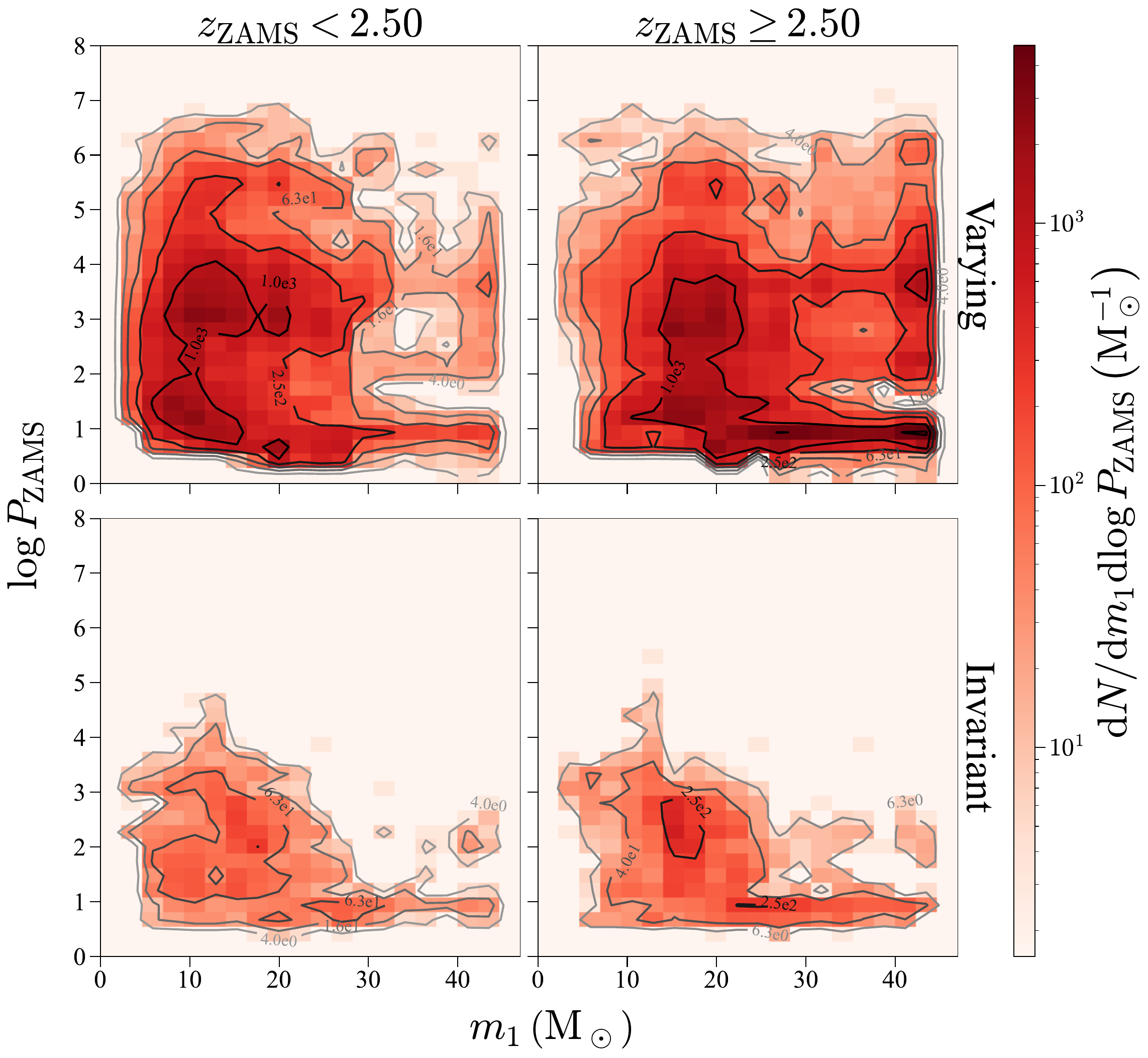}
    \caption{Primary mass ($m_1$)--initial (progenitor) orbital period ($P_\ZAMS$) distributions of BHBH CBMs as a two-dimensional histogram, for $\zzams<2.5$ (left column) and $\zzams\geq2.5$ (right column), from the Varying (top row) and Invariant (bottom row) models. Each two-dimensional bin is assigned the frequency of binaries per mass-orbital period decade bin ($\d N/\d m_1\d\log P_\ZAMS$), which is encoded both by the right color bar and the grayscale solid contours. The histogram is defined over $30$ uniform bins within $0\leq\log P_\ZAMS\leq8$, and $20$ uniform bins within $0\leq m_1/\Msun\leq47$. The lightly colored background is not populated. The plane can be divided around $\log P_\ZAMS\sim1$, with shorter initial periods generally preferring $m_1\gtrsim20\,\Msun$, and longer initial periods $m_1\sim10\en20\,\Msun$, although the choice of models heavily influences the detailed behavior.}
    \label{sec4fig:bhbh_mco1_pzams}
\end{figure}

We verify the interplay between $\zzams$, $P_\ZAMS$ and $m_1$ in Fig. \ref{sec4fig:bhbh_mco1_pzams}. Broadly, we note that $\log P_\ZAMS\sim1$ forms a threshold between two components of the BHBH population: $\log P_\ZAMS\gtrsim1$ binaries generally favor $m_1\sim10\en20\,\Msun$, while $\log P_\ZAMS\lesssim1$ binaries generally lead to primaries across the entire $2.5\leq m_1/\Msun\lesssim47$ range. In terms of mass distribution, the first component fits with the CE channel, while the second with the stable RLOF channel. Taking $\zzams=2.5$ as a threshold, as seen in Fig. \ref{sec4fig:zzams_m1}, the lower redshift range favors relatively less massive binaries than the upper range in both models, but it also favors lower $P_\ZAMS$, in particular for the component we associate with the stable RLOF channel. This $\zzams\en P_\ZAMS$ trend for $\log P_\ZAMS\lesssim1$ is particularly noticeable in the Varying model, and is also in accordance with the two-channel interpretation, as young CBMs would preferentially have shorter orbits, and the stable RLOF channel is less efficient at hardening the orbit. The overall shift of massive progenitors toward shorter initial orbits thus appears to contribute significantly to amplifying the stable RLOF channel, which manifests itself as an important contribution to the $\sim9\,\Msun$ peak at low redshift, and the amplified high-mass tail+PPISNe pileup at high redshift. It is important to note that our extrapolation of the companion frequency fit by \citet{moeMindYourPs2017} (Sec. \ref{sec2sub:orbital_parameters}) is an important factor in this result, and to consider alternative ways to extend the fit. We further discuss this in Sec. \ref{sec4sub:varying_bhbh_overestimate}.

We highlight that an amplification of the stable RLOF channel does not necessarily imply on a suppression of the CE channel. From Fig. \ref{sec4fig:bhbh_mco1_pzams}, the $\log P_\ZAMS\sim1$ threshold is at most a soft boundary between the two channels, and even for longer periods we see that the CE channel itself still contributes to $\sim9\,\Msun$ considerably in the Varying model, even at low redshift. Fig. \ref{sec4fig:bhbh_mco1_pzams} also highlights the presence of an extreme $\log P_\ZAMS>6$ population of BHBH mergers which would not ordinarily be expected to merge within a Hubble time. We verify that all binaries in this region are born as extreme eccentrics, with $e_\ZAMS>0.9$, close to the maximum limit set by a $70\%$ Roche lobe filling factor at periastron \citep[Equation (3) of][]{moeMindYourPs2017}. Binaries in this region have a mean primary mass of $\approx25\,\Msun$ and mean separation at periastron of $\approx10\,\mathrm{AU}$, which is enough to start an episode of RLOF, upon which COMPAS assumes the orbit to be circularized to periastron, i.e., an orbital period of $\approx1800\,\mathrm{d}$, which brings these systems back to the expected pre-interaction parameter space. This highlights the fact that it is the distribution of separations \textit{at periastron} which might more strongly affect the subsequent evolution of the population, as pointed out by \citet{deminkMergerRatesDouble2015} and \citet{klenckiImpactIntercorrelatedInitial2018}. The correlation between progenitor parameters, evolutionary pathways and merger parameters is complicated and remains to be investigated in full detail.

Within this two-channel picture, the relative shift to $q\sim0.6$ at high $\zzams$ relies on a significant contribution of the CE channel to this region, as it is still dominant at high redshift. This is consistent with the CE channel, while having a $q\sim0.3$ peak, significantly covering a broad range of mass ratios, which "dilutes" its contribution in relation to the more concentrated $q\sim0.6$ stable RLOF contribution. The $q=1$ pileup in the Varying model, as well a general filling of the $q\gtrsim0.8$ region, emerges primarily as a consequence of our extrapolation of the companion frequency distribution from \citet{moeMindYourPs2017} to $m_1>40\,\Msun$, strongly favoring close progenitors, which are both more symmetric initially and favor further symmetrization through mass transfer/loss. This leads to the $q=1$ pileup, amplified at higher redshifts due to the top-heaviness of the IMF. 

\begin{figure*}
    \centering
    \includegraphics[width=\textwidth]{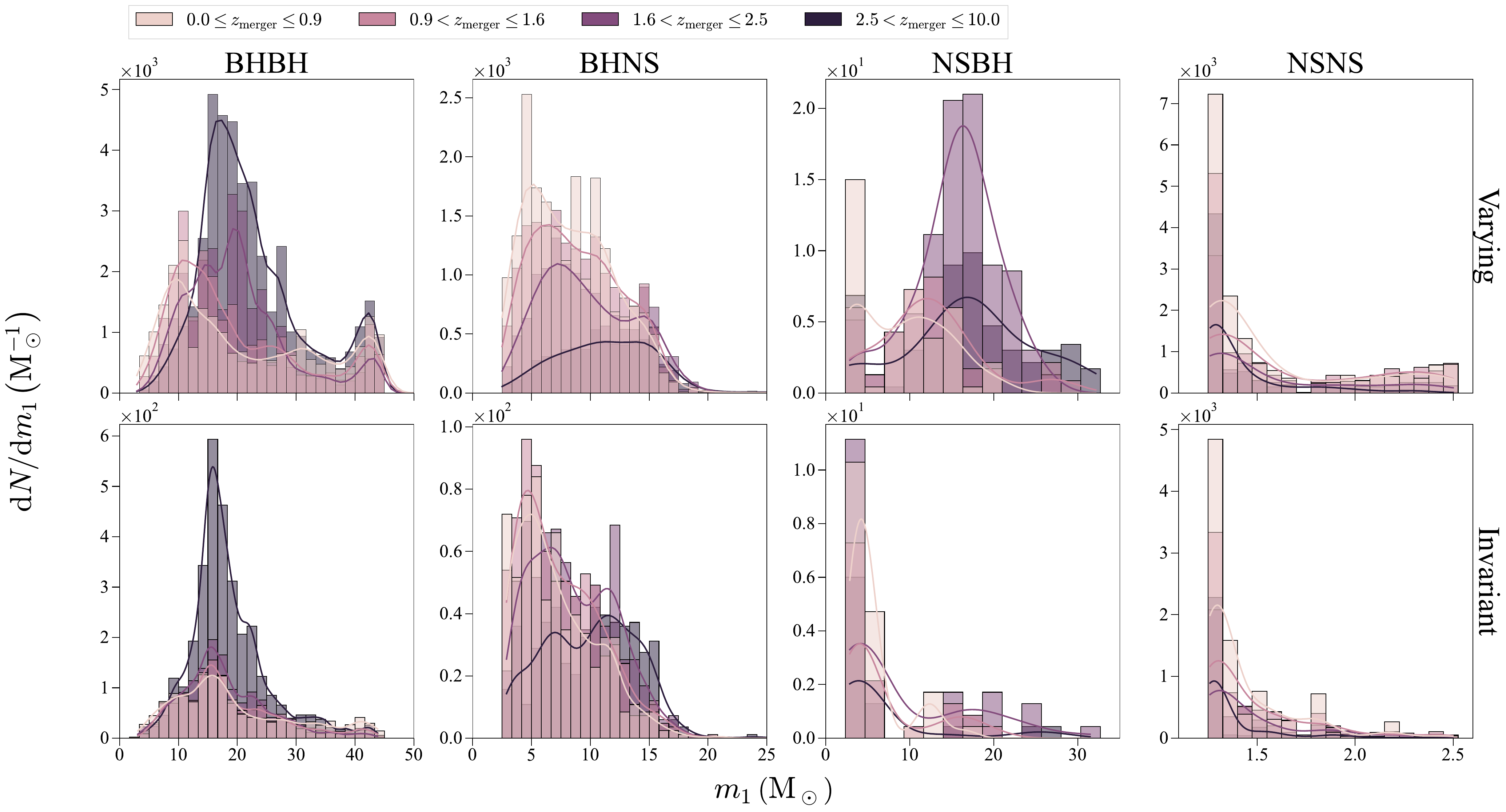}
    \caption{Primary mass ($m_1$) distribution of BHBH, BHNS, NSBH and NSNS mergers (columns from left to right, in that order) in the Varying (top row) and Invariant (bottom row) models, for different ranges of $\zmerger$ (color code), as both histograms and kernel density estimates (KDEs, solid curves). KDE bandwidths are set according to sample sizes in order to provide the best visualization of the $\zmerger$-evolution of the distribution. The $\zmerger$ ranges are defined so that each contains three of the twelve $\zzams$, and starts at the earliest of the three.}
    \label{sec4fig:zmerger_m1}
\end{figure*}

In terms of $\zmerger$, Fig. \ref{sec4fig:zmerger_m1} (first column left to right) shows that the variation of progenitor birth conditions leads directly to a strong correlation between $m_1$ and $\zmerger$ for BHBHs. The Invariant model results in a single-peaked distribution centered on the CE $\sim16\,\Msun$ peak, which nevertheless flattens and becomes shifted toward lower masses with decreasing redshift, particularly below $\zmerger=1.6$. However, the $m_1\gtrsim20\,\Msun$ tail also flattens and becomes relatively more important for $\zmerger<0.9$. We associate this to an increased contribution of the stable RLOF channel at low redshift, as a consequence of its typically longer coalescence times, which we verify in Sec. \ref{sec4sub:bhbh_ctimes}. This result is in line with the above mentioned $\zmerger$-evolution found by \cite{sonRedshiftEvolutionBinary2022}, where the RLOF contribution is minimal at high $\zmerger$ but grows to $\sim0.4$ of the total merger rate at $\zmerger=0$.

In the Varying model, the $\zmerger$-evolution of the mass distribution reflects the shift from the $\sim16\,\Msun$ to the $\sim9\,\Msun$ peak at low $\zzams$. However, the transition happens within $0.9<\zmerger\leq1.6$, whereas in terms of progenitor formation it happens within $2.49\leq\zzams\leq3.05$. This relative delay is a natural consequence of the relatively longer coalescence times characteristic of the stable RLOF channel, which also contribute to select against CBMs from this channel at high $\zmerger$, since at that point they would still not yet have had time to merge. The PPISNe pileup is featured at all $\zmerger$, being most prominent in the $2.5<\zmerger\leq10.0$ bin, followed by the $0.0\leq\zmerger\leq0.9$ bin. This is also connected to the characteristic coalescence times of each channel, a discussion left to Sec. \ref{sec4sub:bhbh_ctimes}. 

We find a $\sim35\,\Msun$ bump for $\zmerger<0.9$ in both models, but most prominently in the Varying model, and note that a $35^{+1.7}_{-2.9}\,\Msun$ overdensity has been found in the GWTC-3 BHBH $m_1$ distribution by \citet{ligoscientificcollaborationPopulationMergingCompact2023}, overlaid on a power-law distribution with another overdensity at $10^{+0.29}_{-0.59}\,\Msun$. This is remarkably similar to the shape obtained from the Varying model for $\zmerger<0.9$, which coincides with the approximate range of detections in GWTC-3, except for the PPISNe pileup, which is distinct from the $\sim35\,\Msun$ feature. The upper feature is also encountered by \citet{sonRedshiftEvolutionBinary2022}, who determine that it is an artifact from the transition between the prescription for remnant masses from core-collapse supernovae \citep{fryerCOMPACTREMNANTMASS2012} and those from PPISNe, although it is also not clear that the transition between them should be smooth. However, in that work the model by \citet{farmerMindGapLocation2019} is employed for PPISNe, instead of the model by \citet{marchantPulsationalPairinstabilitySupernovae2019} implemented here. It remains for us to verify why this $\sim35\,\Msun$ appears only at low $\zmerger$, and its robustness against model variations. The lower feature, which is well-matched to the peak in the observed distribution, has been characterized under variations of the evolution models in \citet{sonNoPeaksValleys2022}.

In the first column (left to right) of Fig. \ref{sec4fig:zmerger_q}, the mass ratio distributions retain their $\zzams$-evolution features: at $q\sim0.3$ and $q\sim0.6\en0.8$, common to both models, and a $q=1$ pileup in the Varying model. Similarly to Fig. \ref{sec4fig:zzams_q}, here a $q\sim0.3$ peak can be distinguished in the Invariant model, while in the Varying model its relative contribution is decreased such that it approaches a plateau within $q\sim0.3\en0.5$. Following $\zzams$, the upper feature is relatively more prominent for $\zmerger>2.5$. However, in both models, the peak within $q\sim0.6\en0.8$ shifts from the lower to the upper end of this interval with \textit{decreasing} $\zmerger$, a trend not present with regard to $\zzams$, and which is instead related to the coalescence times in these systems, discussed in Sec. \ref{sec4sub:bhbh_ctimes}. Finally, the $q=1$ pileup follows from the correlation between initial orbital period and progenitor mass, as discussed above, being more significant for $1.6<\zmerger\leq2.5$, while the production of $q=1$ binaries peaks at the highest redshifts (Fig. \ref{sec4fig:zzams_q}), suggesting that these binaries favor longer coalescence times. We also verify this correlation in a following section.

\subsubsection{Black hole+neutron star mergers}
\label{sec4sub:bhns_masses}

From the second column (left to right) in Fig. \ref{sec4fig:zzams_m1} we see that the BHNS distribution displays a similar surface-level $m_1\en\zzams$ trend to that of BHBHs, shifting from a lower, $\sim5\,\Msun$, to an upper, $\sim12\,\Msun$, peak with increasing $\zzams$. The $\zzams=0.01$ samples are too small to be visible (about $100$ times smaller than the others) because BHNSs in general have coalescence times $\gtrsim0.1\,\Gyr$ (Sec. \ref{sec4sub:bhns_ctimes}). In the Invariant model, the upper $\sim12\,\Msun$ feature manifests itself as a slight flattening even for low redshifts; and the lower, $\sim9\,\Msun$, feature remains present either as a secondary peak or a flattening even as the distribution shifts to the upper peak over $\zzams=3.05$. The Varying model, in contrast, results in a single-peaked distribution around $\sim12\,\Msun$ over $\zzams=4.64$, with little sign of the lower feature. Relative to the BHBH $m_1$ distribution, the key difference is that BHNSs rarely host BHs more massive than $\sim20\,\Msun$ (while we do not treat this range further here, its contribution to the BHNS merger rate is separately plotted in Sec. \ref{sec3sub:mrates}).

The mass ratio distributions in the second column (left to right) of Fig. \ref{sec4fig:zzams_q} reveal a common peak within $0.1\lesssim q\lesssim0.2$ in both models at all $\zzams$, as well as a secondary feature within $0.3\lesssim q\lesssim0.5$, beyond which the distribution is suppressed due to the relative lack of the $\lesssim5\,\Msun$ primaries that would host NS companions within that mass ratio interval. The relative shift to the lower peak with increasing $\zzams$ in both models is a direct consequence of the shift toward higher primary masses. Because the Varying model results in overall more massive primaries, its BHNSs trend toward slightly lower $q$, and their resulting mass ratio distribution is concentrated on a narrower $q\sim0.15$ peak. Even more extreme mass ratios are made unlikely by the combination of the BH mass distribution dropping below $\sim8\,\Msun$ (Fig. \ref{sec4fig:zzams_m1}) and the NS mass distribution below $\sim1.2\,\Msun$ (due to ECSNe and CCSNe models discussed in Sec. \ref{sec4sub:nsns_masses}).

Only $\sim6\%$ of all BH+NS mergers result in NSBH (the primary progenitor becomes the compact secondary) systems in the Invariant model, and an even lower $\sim1\%$ in the Varying model. Consequently, it is difficult to clearly identify the shape of the primary mass distribution of this class in Fig. \ref{sec4fig:zzams_m1} (third column left to right), in particular for the $\zzams=10$ and $0.01<\zzams<1.7$ samples, which, for both models, are similarly sized and contain each less than $10\%$ of the full NSBH sample. No NSBHs are produced at $\zzams=0.01$ in either model. Generally, however, in the Varying model NSBHs cluster within $m_1\sim10\en20\,\Msun$, and in the Invariant model within $m_1\sim5\en10\,\Msun$, in both cases with a slight tendency for the mass to grow with redshift up to $\zzams=3.05$. The $\zzams=10$ sample is the standout case, with \textit{all} BHs hosted by NSBHs falling into the lower mass gap, i.e., having $\lesssim5\,\Msun$, and even reaching into $\lesssim3\,\Msun$ in the Varying model, breaking the trend of increasing mass with redshift. In the Varying model the $\zzams=4.64$ sample also breaks with this trend, although it does not yield a single narrow peak.

For BHNSs, the difference between the resulting $m_1$ distributions from the Invariant and Varying models is small when compared to what is seen for BHBH progenitors. We may point out that, although the overall behavior is the same, the shift between the lower and upper features in the Varying model happens more gradually, starting from the lowest redshifts, whereas in the Invariant model it happens rather abruptly, in comparison, somewhere between $\zzams=2.49$ and $\zzams=4.64$. We attribute this to the varying IMF, which becomes smoothly top-heavy starting from the lowest redshifts, and with increasingly more massive progenitors induces more massive BHs as well. The broadly similar $m_1$ distributions lead to also similar $q$ distributions, largely defined by the constraint of the maximum NS mass (set to $2.5\,\Msun$).

In terms of formation channels, BHNSs may present a somewhat similar picture to that of BHBHs. With the default COMPAS settings (their fiducial "A" model), \citet{broekgaardenImpactMassiveBinary2021a} performed a comprehensive study of BHNS merger populations and found that their formation is under all conditions dominated by the same CE channel (or channel I; $\approx86\%$ of all \textit{detectable} BHNS mergers). Two minor channels follow: the stable RLOF channel (or channel II; $\approx4\%$ of all detectable BHNS mergers), and the single-core CE as \textit{first} MT channel (or channel III; $\approx4\%$ of all detectable BHNS mergers also). The relative inefficiency of channel II in producing BHNS mergers helps explain why BHNSs are so much less affected by the change in initial conditions when compared to BHBHs, which \textit{can} be efficiently produced by channel II. 
Channel I has also been characterized as the dominant formation channel for BHNS mergers by \citet{dominikDOUBLECOMPACTOBJECTS2012},\citet{kruckowProgenitorsGravitationalWave2018} and by \citet{iorioCompactObjectMergers2023a}, although the last find a relatively more expressive contribution from channels II and III due chiefly to different stellar evolution models. The main effect of the Varying model seems to be increasing the size of the population, as noted by the normalization in Fig. \ref{sec4fig:zzams_m1} and the formation efficiency in Fig. \ref{sec4fig:metal_fractions}.

In \citet{broekgaardenImpactMassiveBinary2021a}, both channels II and III are found to become more significant at low metallicities (relative to channel I), specially channel III. Both channels display significant overlap with channel I and between each other in terms of produced $m_1$, making it difficult to distinguish them in the BHNS $m_1$ distribution as we had done in the BHBH case. Channel III is found to contribute to the entire $m_1\sim3-20\,\Msun$ range, and both secondary channels partially produce $\sim5\,\Msun$ primaries. In Figure 4 of the aforementioned work, however, the three channels are shown to occupy distinct regions of the $q_\ZAMS\en a_\ZAMS$ plane ($a_\ZAMS$ the semi-major axis at ZAMS). Channel II (stable RLOF) predominantly covers the $q_\ZAMS\sim0.3\en0.6$, $a_\ZAMS\sim0.1\en0.5\,\mathrm{AU}$ region, while channel I (CE) covers $q_\ZAMS\sim0.2\en0.6$, $a_\ZAMS\sim0.5\en15\,\mathrm{AU}$, and channel III (single-core CE as first MT episode) the $q_\ZAMS\sim0.4\en0.9$, $a_\ZAMS\sim5\en30\,\mathrm{AU}$. They also determine that Channel I retreats to $a_\ZAMS\lesssim5\,\mathrm{AU}$ for low metallicites ($\Z\leq\Zsun/10$). 

\begin{figure*}
    \centering
    \includegraphics[width=\textwidth]{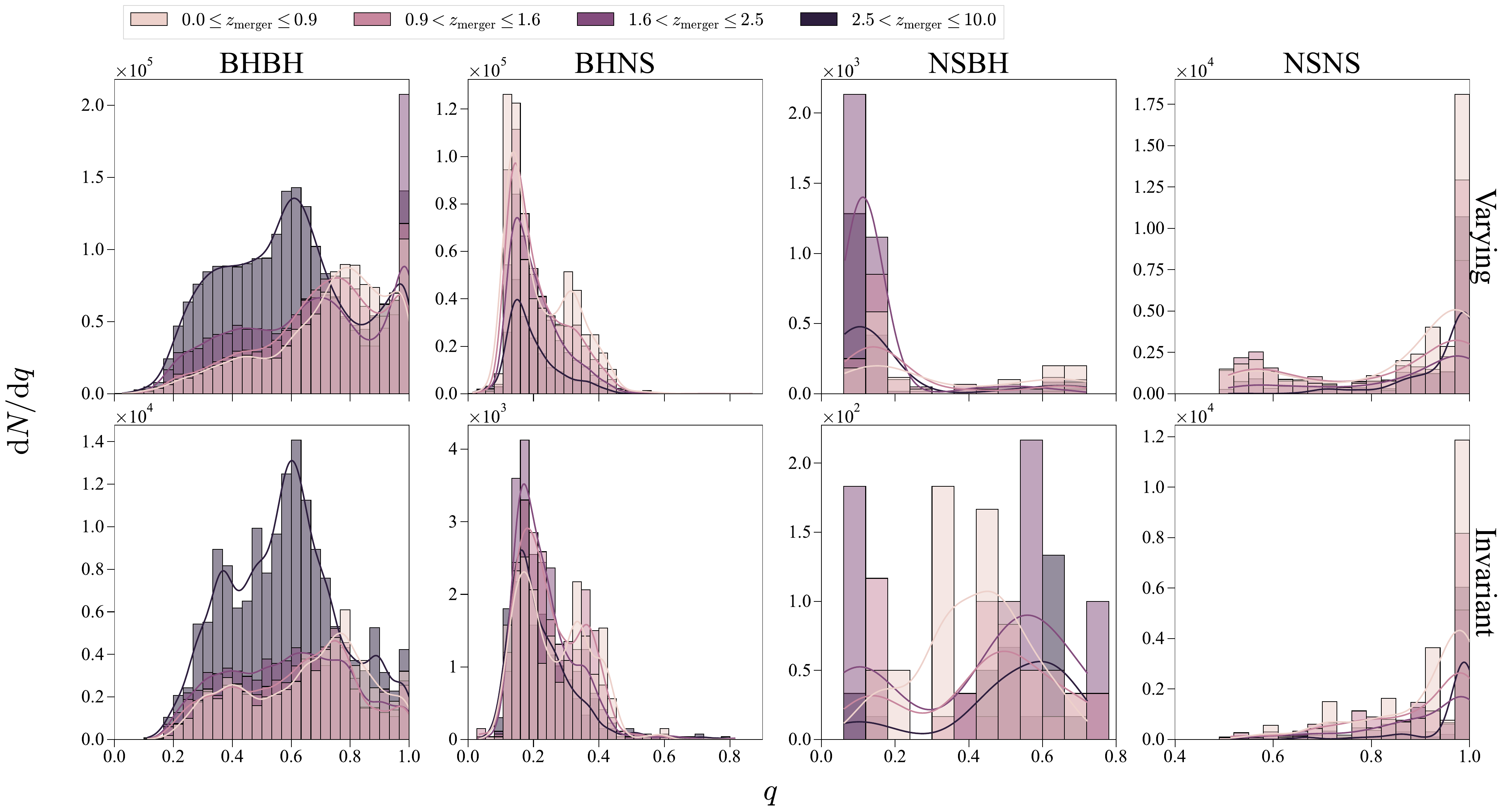}
    \caption{The same as Fig. \ref{sec4fig:zmerger_m1}, but for the mass ratio ($q=m_2/m_1\leq1$).}
    \label{sec4fig:zmerger_q}
\end{figure*}

\begin{figure}
    \centering
    \includegraphics[width=\columnwidth]{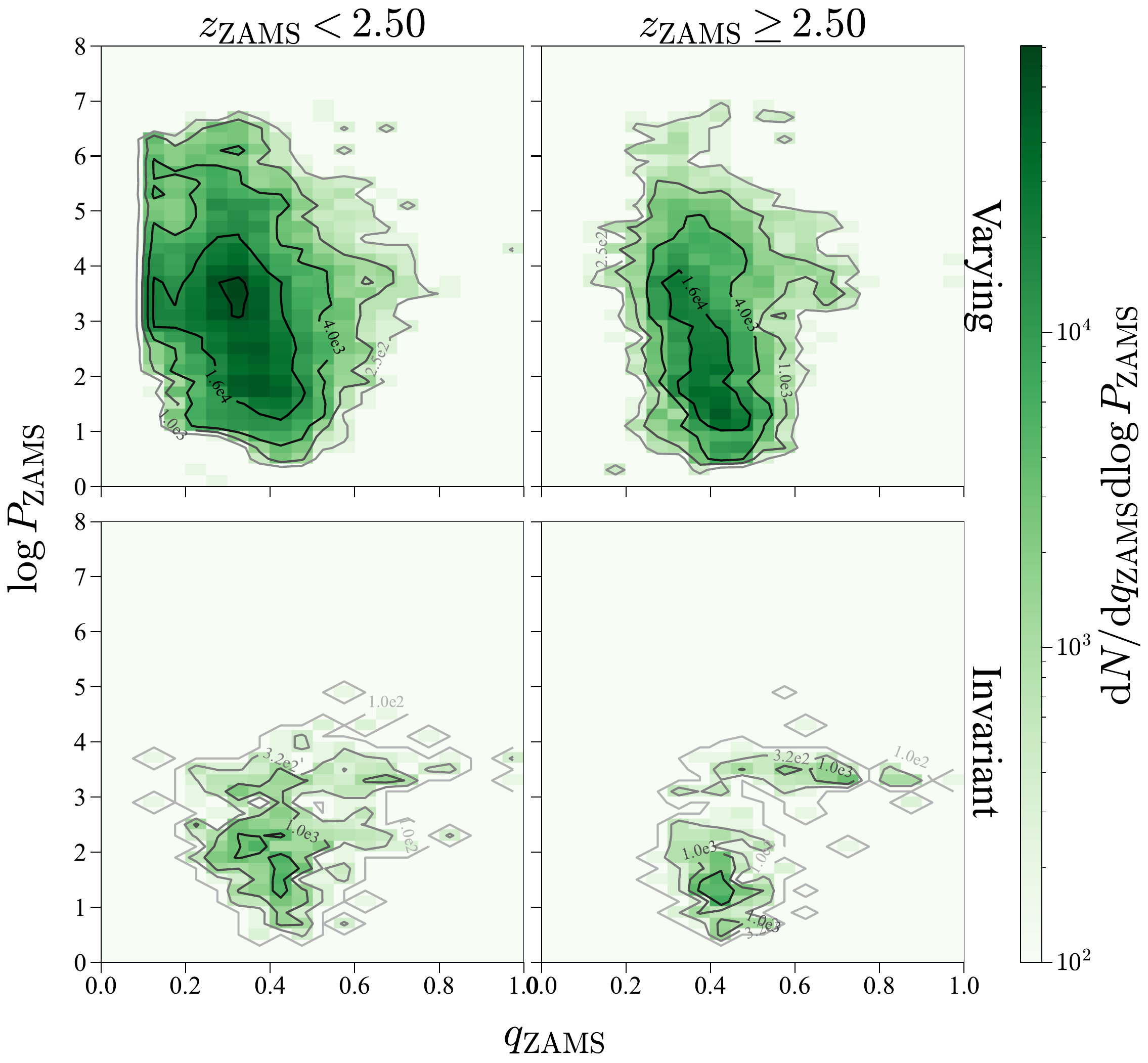}
    \caption{Initial (progenitor) mass ratio ($q_\ZAMS$)--initial orbital period ($P_\ZAMS$) distribution of BHNSs as a two-dimensional histogram, for $\zzams<2.5$ (left column) and $\zzams\geq2.5$ (right column), from the Varying (top row) and Invariant (bottom row) models. Each two-dimensional bin is assigned the frequency of binaries per mass ratio--orbital period decade bin ($\d N/\d q_\ZAMS\d\log P_\ZAMS$), which is encoded both by the right color bar and the gray scale solid contours. The histogram is defined over 30 uniform bins within $0\leq\log P_\ZAMS\leq8$, and 20 uniform bins over $0\leq q_\ZAMS\leq1$. The lightly colored background is not populated. The extended $0.4\lesssim q_\ZAMS\lesssim1$ component within $3\lesssim\log P_\ZAMS$, present only in the Invariant model, and can be associated to a particular BHNS formation channel involving a single-core CE episode as the \textit{first} MT channel.}
    \label{sec4fig:bhns_qzams_pzams}
\end{figure}

\begin{figure}
    \centering
    \includegraphics[width=\columnwidth]{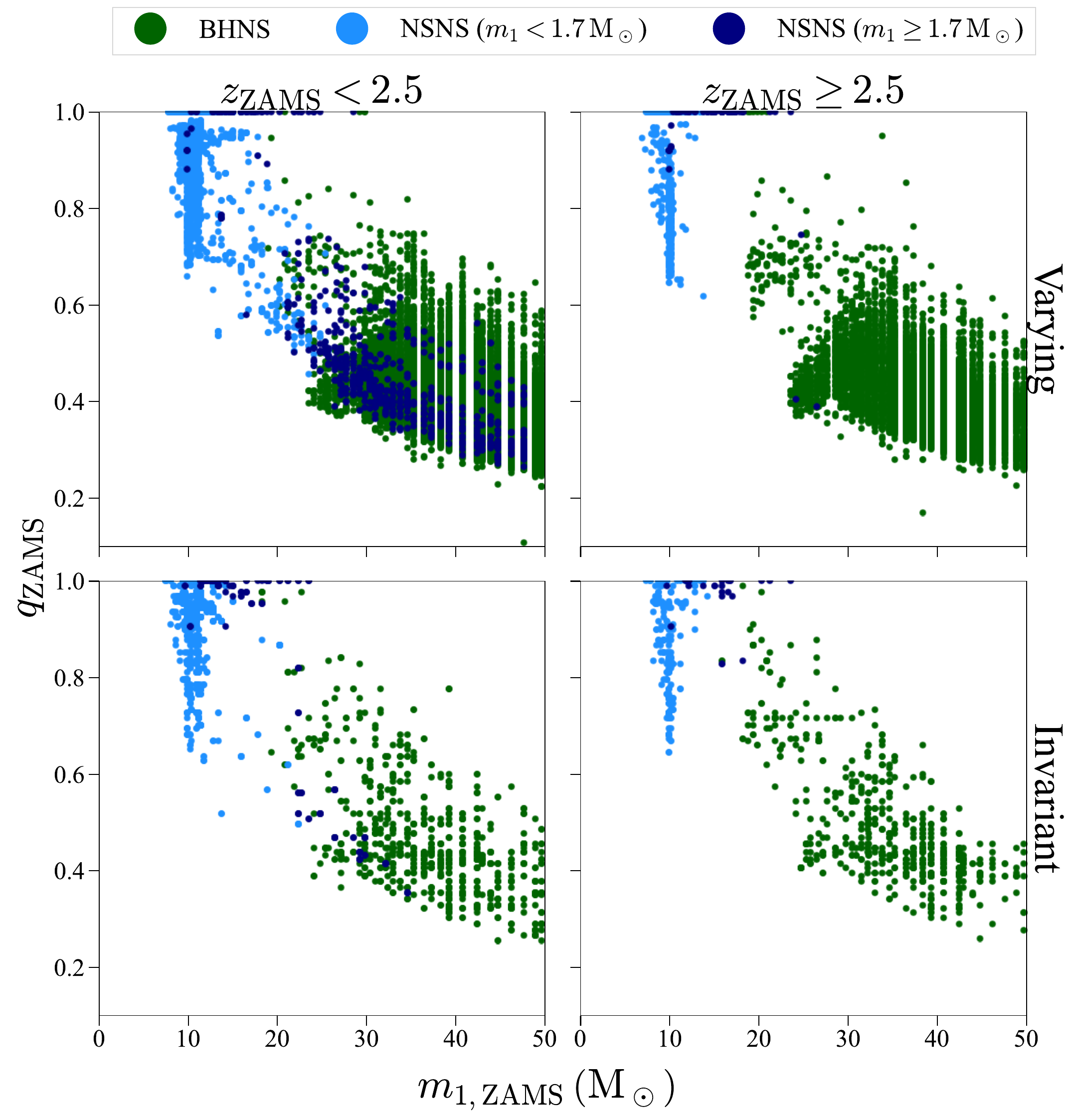}
    \caption{Scatter of BHNSs (excluding NSBHs) and NSNSs over the initial (progenitor) primary mass ($m_{1,\ZAMS}$)--initial mass ratio ($q_\ZAMS$) plane, for $m_{1,\ZAMS}\leq50\,\Msun$, for $\zzams<2.5$ (left column) and $\zzams\geq2.5$ (right column), from the Varying (top row) and Invariant (bottom row) models. BHNSs are plotted in green, and NSNSs distinguished between those with $m_1<1.7\,\Msun$ (light blue) and $m_1\geq1.7\,\Msun$ (dark blue). The vertical gaps in the scatter are an artifact of our discretized sampling procedure (Sec. \ref{sec2sub:sampling}). The $m_1\geq1.7\,\Msun$ NSNSs share their region of the initial parameter space with BHNSs, and at the low metallicities of higher redshifts, have their formation suppressed by the reduced efficiency of wind mass loss. The regions of the initial parameter space that generate each class of merger remain fixed, but the initial model affects their degree of occupancy.}
    \label{sec4fig:nsns_mzams1_qzams_scatter}
\end{figure}

We are able to tentatively establish an analogous dynamics in our sample by examining our BHNS sample in the $q_\ZAMS\en P_\ZAMS$ plane, in a lower ($\zzams<2.5$) and upper ($\zzams\geq2.5$) redshift range, in Fig. \ref{sec4fig:bhns_qzams_pzams}. We determine, in the Invariant model at low redshift (high metallicity), a dominant contribution within $q_\ZAMS\sim0.2\en0.6$ and $\log P_\ZAMS\sim1\en4$, which retreats to $P_\mathrm{ZAMS}\lesssim3$ at higher redshifts (lower metallicity); and two secondary contributions, one more concentrated within $q_\ZAMS\sim0.3\en0.6$ and $\log P_\ZAMS\sim0.2\en2$, and another within $q_\ZAMS\sim0.4\en0.9$ and  $\log P_\ZAMS\sim3\en4$. The two first components overlap around $\log P_\ZAMS\sim2$. These three components can be naturally associated to channels I, II and III from \citet{broekgaardenImpactMassiveBinary2021a}, respectively. In the Varying model, then, we notice two major differences, the first being an extension of the overlapped channel I and II to lower $q_\ZAMS$, and of channel II to longer $P_\ZAMS$. The second, a \textit{suppression} of channel III: nearly no $q_\ZAMS\gtrsim0.7$ BHNS are produced, in spite of a larger sample. Although further channels might play a role, within this three-channel picture we may conclude that the Varying model suppresses the "single-core CE as first MT" channel (channel III), and amplifies the CE channel (channel I), in relation to the Invariant, losing the significant contribution of channel III to $m_1\sim5\,\Msun$ at low metallicities. Accordingly, as we had noted, the Varying model loses the $\sim5\,\Msun$ feature for $\zzams\geq3.05$, when compared to the Invariant (Fig. \ref{sec4fig:zzams_m1}). This might be understood as a consequence of the correlation between initial orbital period, primary mass and mass ratio: the "single-core CE as first MT" channel prefers relatively broader initial separations, which in the Varying model strongly favor asymmetrical progenitors ($q_\ZAMS<0.3$). The same channel, however, also favors more symmetric binaries ($q_\ZAMS>0.4$). In short, the Varying model induces a dearth of channel III BHNS progenitors.

Like for the BHBH merger population in Fig. \ref{sec4fig:bhbh_mco1_pzams}, Fig. \ref{sec4fig:bhns_qzams_pzams} shows an unexpected $\log P_\ZAMS>6$ component. This is again a contribution from initially extremely eccentric ($e_\ZAMS>0.9$) binaries, which are driven to closer orbits upon the first episode of interaction. 

The mass ratio peaks at $\sim0.15$ and $\sim0.3$ correspond to $\sim6.7$ and $\sim3.3$ in terms of $Q=m_1/m_2$, which were found by \citet{broekgaardenImpactMassiveBinary2021a}, in addition to a $Q\approx12$ feature. They note that both channels II and III characteristically produce $Q\approx3$ ($q\sim0.3$) binaries. They also find an increasing contribution of $Q\approx3$ relative to $Q\approx6$ with decreasing metallicity. We may compare this with the trend here observed in the Invariant model of an increasing contribution from $q\sim0.3$ relative to $q\sim0.15$ binaries up to $\zzams=2.49$, where the BHNS progenitor population peaks. In our results, however, the significance of this upper feature gradually decreases as the redshift increases. We do not observe a $q\sim0.08$ feature correspondent to $Q\approx12$. However, we do not allow $q_\ZAMS\leq0.1$ in our initial sample, and this might be the reason for the lack of extremely asymmetrical BHNSs. If we interpret the $q\sim0.3$ feature as hosting most of the contributions from channels II and III, then its relatively lesser contribution in the Varying model could be explained by the suppression of channel III discussed above.

For NSBHs, the  general shift from $m_1\sim5$ to $\sim15\,\Msun$ between the Invariant and Varying models can be attributed to two main factors. First, the top-heaviness of the IMF, which is visible in the Varying model as a slight upward shift in the distribution peak with increasing redshift, up to $\zzams=3.05$. Second, and perhaps more importantly, the preference for closer initial orbits for more massive progenitors, which increases the chances of a MT episode occurring while the initially more massive component (the NS progenitor) is still in the main sequence (case A MT), allowing the BH progenitor to accrete a more mass. This would also allow for NSBHs to be generally formed from more massive progenitors, in particular those that would have formed BHBHs at greater initial separation. A "case A MT" channel was in fact expected to be a minor contributor to BH+NS mergers by \citet{broekgaardenImpactMassiveBinary2021a}, but a significant pathway for NSBH merger formation. The strong NSBH shift to $\lesssim5\,\Msun$ at the highest redshifts, however, is present in both models, and we suggest it to be an effect of low metallicities: wind mass loss would be an important way of accessing more massive progenitors for NSBH formation and, once it becomes ineffective at very low metallicities, only relatively light binaries may lead to mass ratio inversion and NSBH formation. In terms of mass ratio, the NSBH distribution (Fig.\ref{sec4fig:zzams_q}, third column left-to-right) is limited to $0.1\lesssim q\lesssim0.8$ but is not sampled densely enough to have its shape appropriately constrained, and at this resolution we may at most say that it generally allows for any NS mass to be paired to the BH. For $\zzams\leq4.64$, the Invariant model shows an approximately flat distribution, while the Varying model is shifted to extreme mass ratios, both simply reflecting the relative lighter and more massive BHs in the respective models. In both cases the distribution shifts to $q\gtrsim0.4$ at $\zzams=10$, which is necessary to keep the predominantly lighter BHs paired to NSs.

The $\zmerger$-evolution of the primary mass and mass ratio distributions largely follow from the $\zzams$-evolution. For BHNSs, the $m_1$ distribution (Fig. \ref{sec4fig:zmerger_m1}, second column left to right) shifts from a lower $\sim5\,\Msun$ to an upper $\sim12\,\Msun$ feature with increasing $\zmerger$ in both models, and differ mainly by a relative shift of the lower-$\zmerger$ distributions toward greater masses in the Varying model. Another important shift is that the two intervals within $0.9<\zmerger\leq2.5$ dominate the population in the Invariant model, while $0\leq\zmerger\leq0.9$ is dominant in the Varying model. We connect this to the coalescence times in Sec. \ref{sec4sub:bhns_ctimes}. The mass ratio distribution (Fig. \ref{sec4fig:zmerger_q}, second column left to right) maintains the $q\sim0.15$ and $q\sim0.3$ features, with the latter being characteristic of $\zmerger\leq1.6$ in the Invariant model and $\zmerger\leq0.9$ in the Varying model. We connect this loss of $0.9<\zmerger\leq1.6$, $q\sim0.3$ mergers to a loss of long coalescence time BHNSs in Sec. \ref{sec4sub:bhns_ctimes}. 

For NSBHs (Fig. \ref{sec4fig:zmerger_m1}, third column left to right), the distribution peaks in the Invariant model within $\sim2.5\en10\,\Msun$ in all $\zmerger$ bins, with a secondary peak within $\sim10\en30\,\Msun$ that shifts to higher masses for increasing $\zmerger$. Most of the NSBH binaries are located in the lower bin for the Invariant model, and it might be interpreted as an indication that old binaries in this range tend to have longer coalescence times. However, as we note in Sec. \ref{sec4sub:bhns_ctimes}, the present sample does not allow for a clear picture of the $\zzams$-evolution of typical NSBH coalescence times. The trend for $m_1\gtrsim10\,\Msun$ NSBHs to merge earlier the more massive they are is clearer, but it might be a simple consequence of more massive binaries having shorter coalescence times, all else being equal. In the Varying model we find a stronger trend for more massive binaries to have merged earlier. This can in part be connected to the trend of $m_1$ to increase with $\zzams$ (Fig. \ref{sec4fig:zmerger_m1}), but that trend is much less significant, and the oldest binaries ($\zzams=10$) in fact tend to produce the lightest primaries. More massive progenitors are, however, shifted to lower periods, a possible origin for this behavior. The $\zmerger$-evolution of NSBHs (Fig, \ref{sec4fig:zmerger_q}, third column left to right) reveals little more than the constraint of NS masses at this sample size. Both models show an upper and lower feature which connect to the $m_1\lesssim10\,\Msun$ and $m_1\gtrsim10\,\Msun$ features of Fig. \ref{sec4fig:zmerger_m1}, respectively, and $\zmerger$ bins with more relatively prominent high $m_1$ features result in more prominent low $q$ features.

\subsubsection{Neutron star+neutron star mergers}
\label{sec4sub:nsns_masses}

The NSNS primary mass distributions are generally dominated by two features: a low-mass, $\sim1.2\en1.3\,\Msun$ peak which we always find to be dominant, with an extended tail up to the adopted maximum NS mass of $2.5\,\Msun$, and a break around $1.7\,\Msun$. These two features arise directly from the default models implemented in COMPAS for ECSNe and CCSNe \citep{compasRapidStellarBinary2022}. The low-mass peak is set by both: the default ECSNe prescription maps all progenitors with helium core masses $1.6\en2.25\,\Msun$ \citep{hurleyEvolutionBinaryStars2002a} to $m_\mathrm{NS}=1.26\,\Msun$, as an approximation of the model by \citet{timmesNeutronStarBlack1996}, while the delayed \citet{fryerCOMPACTREMNANTMASS2012} prescription for CCSNe maps all progenitors with $\lesssim2.5\,\Msun$ carbon-oxygen cores at supernova to $m_\mathrm{NS}\approx1.28\,\Msun$. The break is an actual discontinuity around $1.7\,\Msun$ in the NS mass distribution, which is a consequence of a discontinuity in the relation employed by the delayed CCSNe model \citep{fryerCOMPACTREMNANTMASS2012} for the proto-compact object mass as a function of carbon-oxygen core mass, which jumps from $1.2$ to $1.3\,\Msun$ at $3.5\,\Msun$. 

We show the $\zzams$-evolution of the NSNS $m_1$ distribution in the last (left to right) column of Fig. \ref{sec4fig:zzams_m1}. In both models, and at all $\zzams$, the $\sim1.3\,\Msun$ feature is dominant. The $\zzams=0.01$ sample is expressive in this case, but, as shown in Sec. \ref{sec4sub:nsns_ctimes}, it shows hints of having produced a significant $\tc\gtrsim0.1\,\Gyr$ component cutoff by the time available for merger, and we therefore continue to focus on $\zzams\geq0.5$.

Similar to BHNS mergers, the NSNS merger population is much less sensitive to changes in initial conditions compared to the BHBH merger population. Analogously, this may be attributed to the the dominance of the CE channel (channel I) in producing NSNS mergers, while the stable RLOF channel (channel II) is unable to produce NSNS mergers to any significant degree \citep{dominikDOUBLECOMPACTOBJECTS2012,giacobboProgenitorsCompactobjectBinaries2018,vigna-gomezFormationHistoryGalactic2018a,kruckowProgenitorsGravitationalWave2018,gallegos-garciaEvolutionaryOriginsBinary2023,iorioCompactObjectMergers2023a}.

However, we do note a degree of variation of the distribution with metallicity/$\zzams$. In the Invariant model, the distribution tends to shift toward greater masses with increasing redshift, but only up to $\zzams=2.49$, beyond which NSNSs with $m_1\gtrsim1.7\,\Msun$ become increasingly uncommon. Although NSNSs are increasingly disfavored in relation to BHNSs and BHBHs at high redshift (Sec. \ref{sec4sub:formation_efficiency}), this is not a case of undersampling, as all $\zzams\geq1.8$ produce $\gtrsim200$ NSNSs ($\gtrsim300$ in the Varying model). It also does not fit well within the two main formation channels determined by \citet{vigna-gomezFormationHistoryGalactic2018a} with COMPAS, which generate a mass ratio distribution peaking at $q\approx1$ (characteristic of the double-core CE channel) and $q\approx0.88$ (characteristic of the single-core CE channel), smoothly falling toward lower $q$.

We suggest that this behavior is linked to the decreasing efficiency of wind mass loss with decreasing metallicity, which up to a point generally favors increasingly massive remnants, including NSs, but increasingly disfavors massive NSs as massive progenitors retain enough mass to produce BHs instead. We verify in Fig. \ref{sec4fig:nsns_mzams1_qzams_scatter} that, for $\zzams<2.5$, the $m_1\gtrsim1.7\,\Msun$ NSNSs indeed occupy the same region of the initial parameter space as BHNSs do, within $m_{1,\ZAMS}\sim10\en50\,\Msun$ and $q_\ZAMS\sim0.3\en0.7$, but are no longer able to reach it for $\zzams>2.5$. The effect of the Varying model (top-right panel in Fig. \ref{sec4fig:zzams_m1}) is straightforward: except at the lowest redshift, a top-heavy IMF favors more massive progenitors, which in turn increases the contribution of $m_1\gtrsim1.7\,\Msun$ NSNSs. However, the highest $\zzams$ still disfavor the formation of massive NSNSs, although in this case they do extend further to higher $m_1$ than their Invariant counterparts. As a result, unlike for BHBH and BHNSs, where to good approximation $m_1$ tends to increase with $\zzams$, it is difficult to establish a simple picture of the correlation between NSNS primary masses and $\zzams$, which might be particularly sensitive to model choices. Finally, we recall that the $\zzams=0.01$ NSNS sample is particularly affected by undersampling at high metallicities in the Varying model (see Fig. \ref{sec4fig:metal_fractions} and discussion), which are also the sub-samples for which one might expect formation of massive primary NSs following the above discussion. Therefore, we do not discount that the Varying model might produce $m_1>1.7\,\Msun$ NSNSs at $\zzams=0.01$ which were missed from the sampling. 

The narrow range of NS masses and tendency of the SNe prescriptions to produce $\sim1.3\,\Msun$ NSs makes the NSNS mass ratio distributions strongly shifted to symmetric masses, as seen for both models in the rightmost column in Fig. \ref{sec4fig:zzams_q}. The lighter the primary, more symmetric a NSNS must be, and so the highest redshifts are more strongly concentrated around $q=1$. The Varying model extends further toward asymmetric binaries than the Invariant model, down to $q\sim0.5$, with $0.5\leq\zzams\leq2.49$ in particular displaying a secondary peak within $q\sim0.5\en0.6$ that is prominent relative to $q=1$ for lower redshift, and which is thus associated to the $m_1\gtrsim1.7\,\Msun$ component (also characterized by asymmetric progenitors, see Fig. \ref{sec4fig:nsns_mzams1_qzams_scatter}). 

The result of this more complicated picture of $\zzams$-evolution is that $m_1\gtrsim1.7\,\Msun$ NSNSs become more common at \textit{lower} $\zmerger$, as seen in the rightmost column of Fig. \ref{sec4fig:zmerger_m1}. While this trend is stronger in the Varying model, both display it: in the Invariant model all $\zmerger\leq2.5$ are approximately equally favored over $\zmerger>2.5$, while in the Varying the two $\zmerger\leq1.6$ ranges are comparable and individually favored over the two higher ranges. The mass ratio distribution in terms of $\zmerger$ (rightmost column Fig. \ref{sec4fig:zmerger_q}) follows directly from the patterns discussed before: $\zmerger$ ranges with greater $m_1\gtrsim1.7\,\Msun$ contribution show an increased concentration of $q\sim0.5\en0.6$ NSNSs, which are consequently more common in the Varying model, but the $q=1$ peak remains dominant in all cases.

\subsection{Coalescence times}
\label{sec4sub:ctimes}

\begin{figure*}
    \centering
    \includegraphics[width=\textwidth]{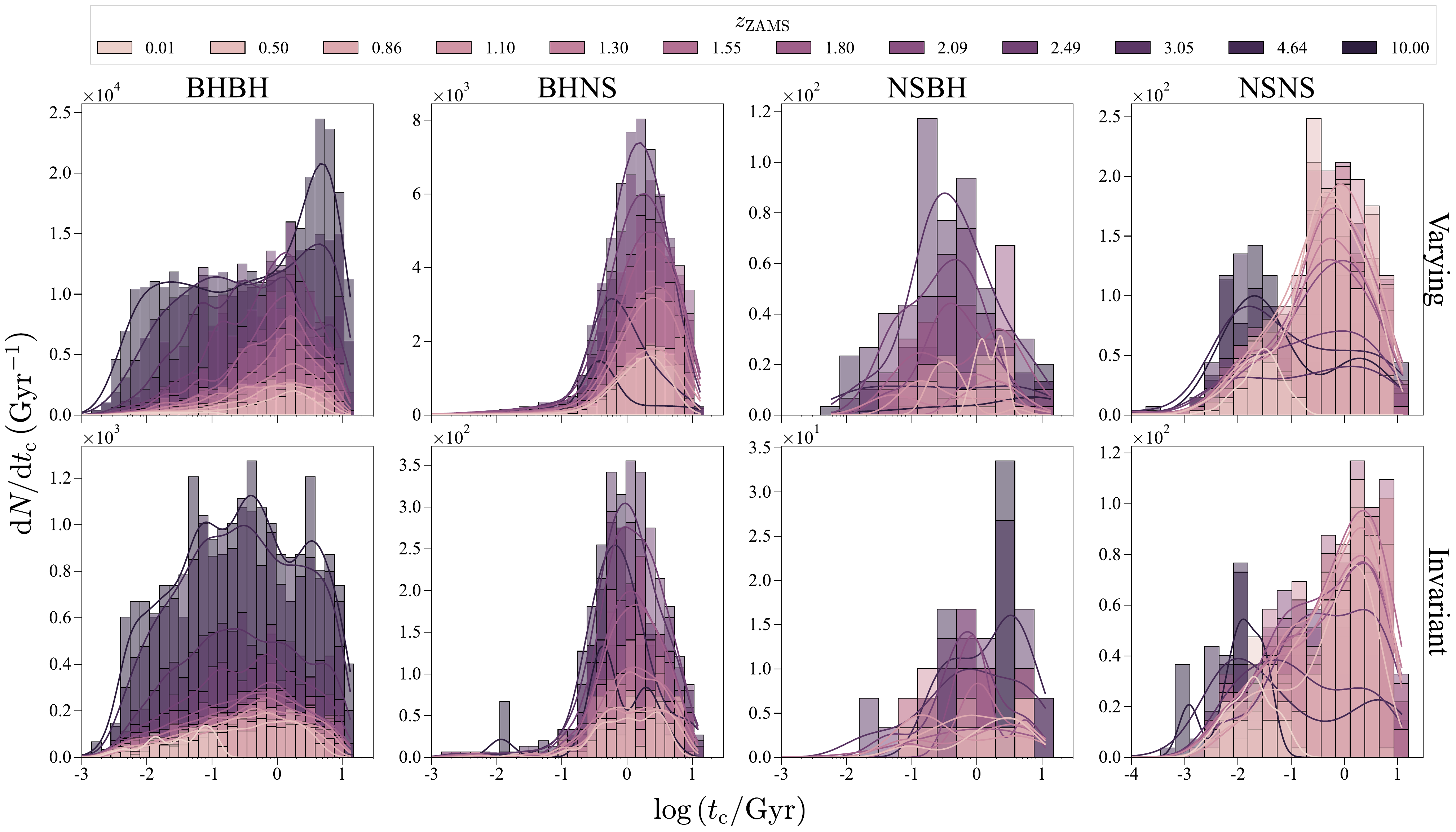}
    \caption{The same as Fig. \ref{sec4fig:zzams_m1}, but for coalescence times ($t_\mathrm{c}$).}
    \label{sec4fig:zzams_tc}
\end{figure*}

\begin{figure*}
    \centering
    \includegraphics[width=\textwidth]{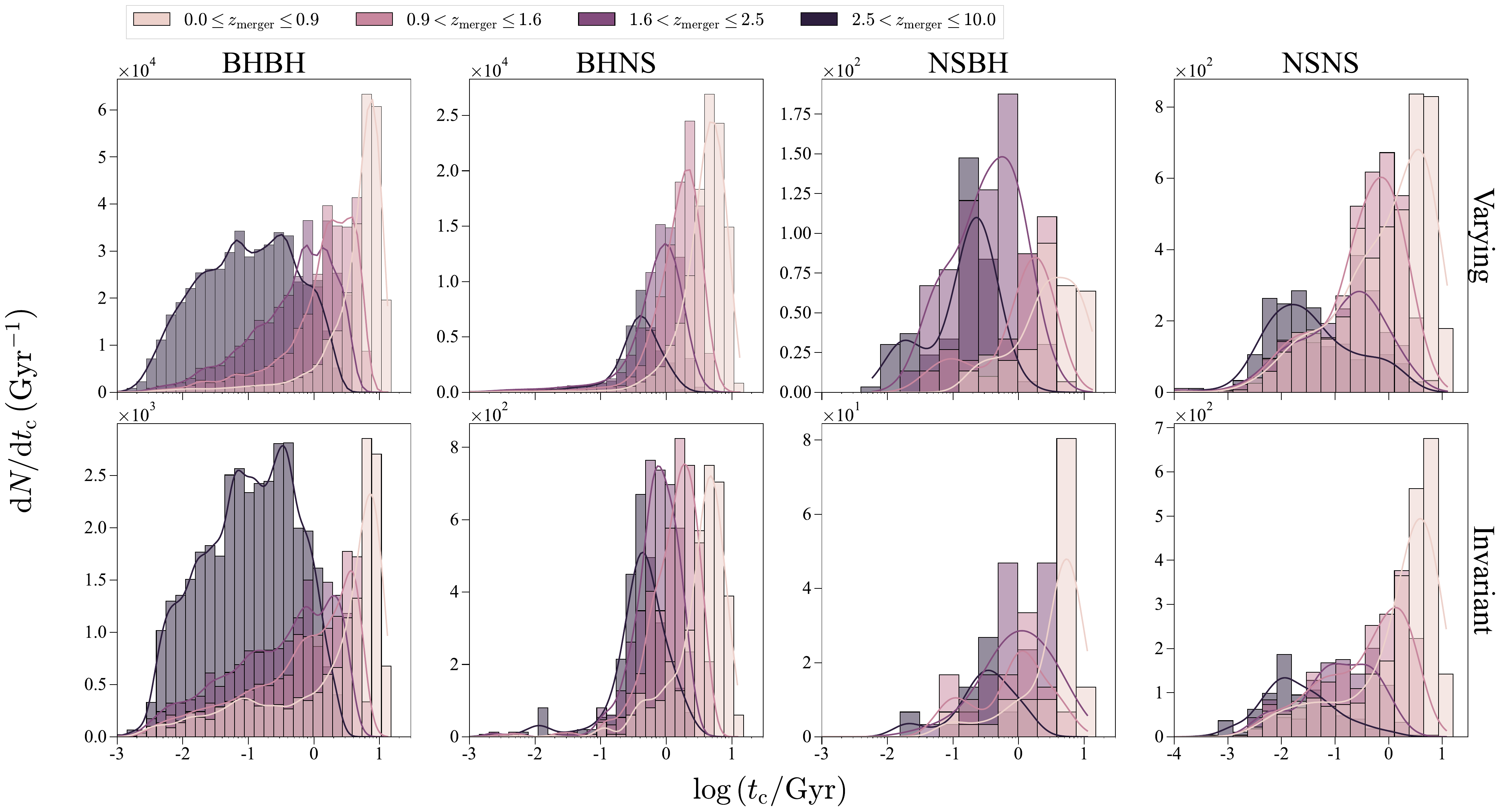}
    \caption{The same as Fig. \ref{sec4fig:zmerger_m1}, but for coalescence times ($t_\mathrm{c}$).}
    \label{sec4fig:zmerger_tc}
\end{figure*}

\begin{figure}
    \centering
    \includegraphics[width=\columnwidth]{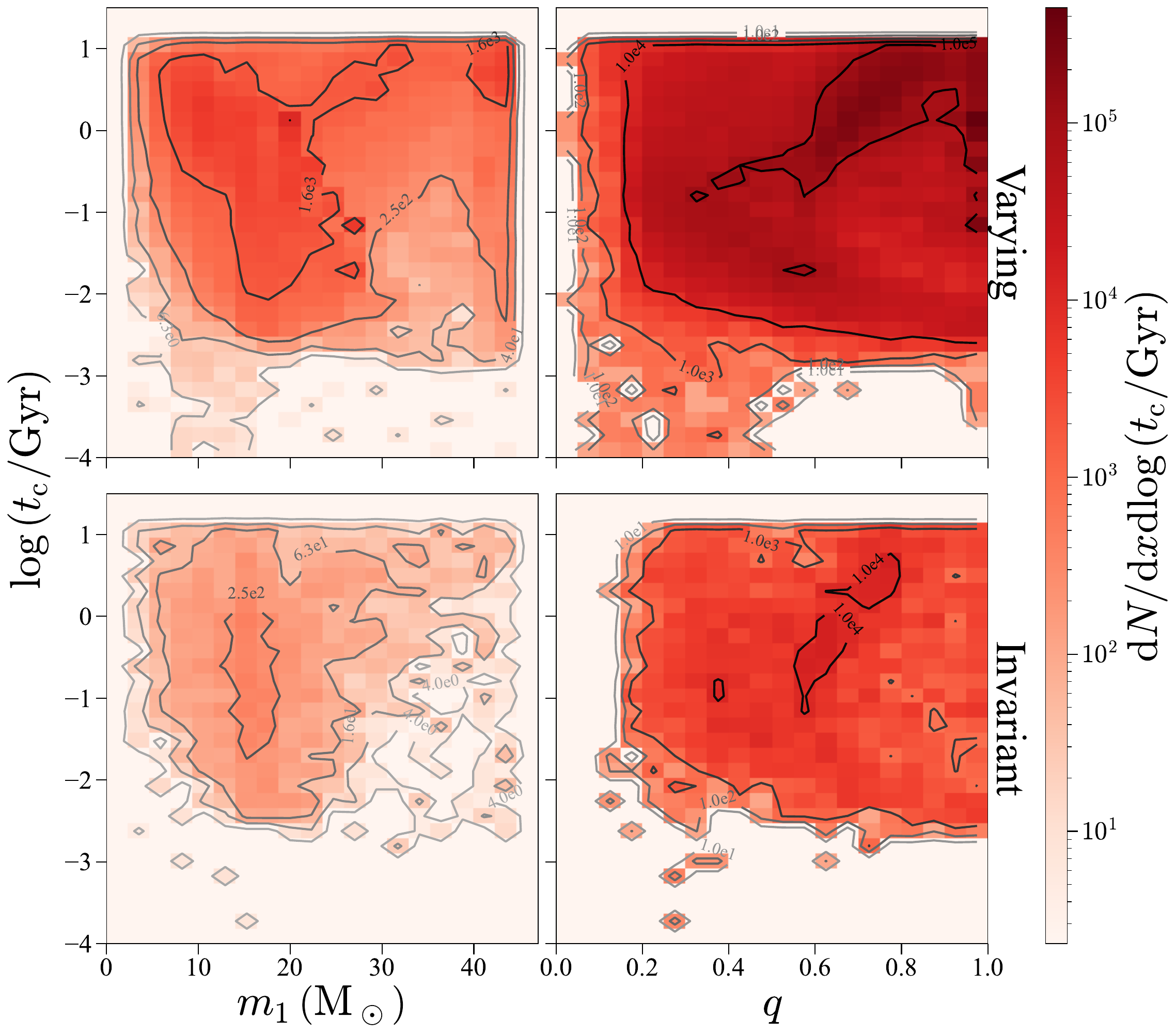}
    \caption{Primary mass ($m_1$)--coalescence time ($\tc$) (left column) and mass ratio ($q$)-coalescence time (right column) distributions of BHBHs as two-dimensional histograms, from the Varying (top row) and Invariant (bottom row) models. Each two-dimensional bin is assigned the frequency of binaries per $x$-axis variable--orbital period decade bin ($\d N/\d x\d\log\left(\tc/\Gyr\right)$, according to the column, which is encoded both by the right color bar and the grayscale solid contours. The histogram is defined over 30 uniform bins within $-4\leq\log\left(\tc/\Gyr\right)\leq1.5$ and 20 uniform bins within either $0\leq q\leq1$ or $0\leq m_1/\Msun\leq47$, according ot the column. The ligtly colored background is not populated. The extension to very short $\tc$ in the Varying model comes from the strong preference for very short $P_\ZAMS$ in our extrapolation of the companion frequency fit by \citet{moeMindYourPs2017} to $m_1\gtrsim40\,\Msun$. PPSINe pileup binaries ($m_1\sim45\,\Msun$) prefer longer $\tc$ but have a long tail, and the preferred $q$ shifts from $\sim0.6$ to $\sim0.8$ with increasing $\tc$.}
    \label{sec4fig:bhbh_qmco1_tc}
\end{figure}

\subsubsection{Black hole+black hole mergers}
\label{sec4sub:bhbh_ctimes}

In the first column (left to right) of Fig. \ref{sec4fig:zzams_tc} we compare the coalescence time ($\tc$) distributions of BHBH mergers as a function of $\zzams$ for both models, which differ considerably. The Invariant model peaks within the $\tc\sim0.1\en1\,\Gyr$ interval at all redshifts (again, disregarding $\zzams=0.01$), albeit with expressive contributions from the entire range. The distributions tend to broaden to both longer and shorter $\tc$ at greater $\zzams$, but particularly toward $\sim10\,\Gyr$. This behavior reinforces the tenability of the two-channel picture, as, in \citet{sonRedshiftEvolutionBinary2022}, the CE channel was characterized by a broad delay time ($t_\mathrm{d}$) distribution, concentrated within $t_\mathrm{d}\sim0.1\en1\,\Gyr$ at high metallicities, and relatively evenly spread across $t_\mathrm{d}\sim0.001\en10\,\Gyr$ at low metallicities, and the RLOF channel by a distribution concentrated above $t_\mathrm{d}\sim1\,\Gyr$ with a peak at $t_\mathrm{d}\sim10\,\Gyr$, mainly significant for low metallicities. As supernovae typically occur within $\sim0.01\,\Gyr$ of ZAMS for massive progenitors, we consider the order of magnitude comparison between $\tc$ and $t_\mathrm{d}$ adequate for identifying common distribution features. We also note the clear cutoff of the $\zzams=0.01$ population just above $\tc=0.1\,\Gyr$ due to the lookback time of $\approx0.14\,\Gyr$ at that redshift.

In the Varying model we look first at $\zzams=10$, for which the $\tc$ distribution is well-approximated by a peak at $\tc\sim5\,\Gyr$ overlaid on a flat distribution across $\tc\sim0.01\en10\,\Gyr$, which sharply drops for short $\tc$. We associate the high $\tc$ peak to the stable RLOF channel, and the flat component to the CE channel, based on the previous discussion. Below $\tc=1\,\Gyr$, the distribution is much flatter than in the Invariant model, and we attribute this increased relative production of short $\tc$ binaries to the shift toward shorter $P_\ZAMS$ for the massive primaries that dominates at high redshifts. The $\tc\sim1\en10\,\Gyr$ peak is linked to the overall increase in the contribution of the stable RLOF channel, which we noted in Sec. \ref{sec4sub:bhbh_masses} to manifest as a similarly intense growth of the high-$m_1$ tail and PPSINe peak at the highest $\zzams$. For $\zzams\leq3.05$, the distribution shifts to a single peak at about $\sim2\,\Gyr$, which still remains within the stable RLOF range. The broad feature we have associated with the CE channel, however, is suppressed with decreasing redshift, following the behavior seen in Fig. \ref{sec4fig:zzams_m1} for the $m_1$ CE component.

We verify consistency with our previous characterization of the two channels in our results by plotting $m_1$ and $q$ against $\tc$ in Fig. \ref{sec4fig:bhbh_qmco1_tc}.  In the Varying model we clearly identify an overdensity about $\sim9\,\Msun$ favoring $\tc\sim1\en10\,\Gyr$, associated to a longer high mass tail, consistent with the stable RLOF channel, as well as an overdensity about $\sim16\,\Msun$ favoring $\tc\sim0.01\en5\,\Gyr$, consistent with the CE channel. We clearly distinguish the PPISNe pileup at $\sim45\,\Msun$ overlaid on the two components, with a broad $\tc$ distribution biased toward $\sim10\,\Gyr$. The Invariant model then shows the lack of a strong PPISNe pileup, and the entire distribution dominated by the CE channel, with the stable RLOF component distinguishable chiefly by the high mass tail at long coalescence times. The mass ratio plots show that the $\sim1\en10\,\Gyr$ component favors $q\sim0.6\en0.8$, consistent with the stable RLOF channel. Within $\sim0.01\en5\,\Gyr$  the distribution is nearly flat above $q=0.1$, but we identify the $q\sim0.3$ peak identified with the CE channel in Fig \ref{sec4fig:zzams_q} at $\tc\approx0.1\,\Gyr$.

The $\zmerger$-evolution of coalescence times (Fig. \ref{sec4fig:zmerger_tc}, first column left to right) is dominated by the natural bias of longer $\tc$ toward later $\zmerger$. Between the Varying and Invariant models, we note that the latter has a sharper contribution of $\zmerger>2.5$ than the former, which closely follows the shape of the $\zzams=10$ and $4.64$ distributions in Fig. \ref{sec4fig:zzams_tc}. The high $\zzams$ distributions in the Varying model might be looked at as the corresponding Invariant distributions with an added long $\tc$ peak, but most binaries in this peak fall out of the $2.5<\zmerger\leq10$ interval. 

Finally, we verify that two trends discussed in Sec. \ref{sec4sub:bhbh_masses} are connected with coalescence times. With regard to the PPISNe pileup around $m_1\sim45\,\Msun$, we notice in Fig. \ref{sec4fig:bhbh_qmco1_tc} that, while BHBHs in this region have predominantly longer coalescence times, $\sim1\en10\,\mathrm{Gyr}$, their massive progenitors are also shifted toward short initial periods in the Varying model, which increases the fraction of these systems with coalescence times as short as $\sim1\,\mathrm{Myr}$. These two factors together contribute to the pileup for $\zmerger>2.5$, which predominantly consists of $\tc\sim10^{-3}\en1\,\mathrm{Gyr}$, and the pileup for $\zmerger<2.5$, which predominantly consists of $\tc\sim1\en10\,\mathrm{Gyr}$. As for the shift of the mass ratio peak from $q\sim0.6$ to $q\sim0.8$ with decreasing $\zmerger$ (Fig. \ref{sec4fig:zmerger_q}), we see clearly in Fig. \ref{sec4fig:bhbh_qmco1_tc} that it is related to a simultaneous shift from $\tc\sim1\,\Gyr$ to $\tc\sim10\,\Gyr$. The latter case will naturally contribute to mergers only at a lower $\zmerger$, leading to the shift in Fig. \ref{sec4fig:zmerger_q}.

\subsubsection{Black hole+neutron star mergers}
\label{sec4sub:bhns_ctimes}

We show in the second (left to right) column of Fig. \ref{sec4fig:zzams_tc} the $\tc$ distributions for BHNSs as a function of $\zzams$. For both models they are largely restricted to $\tc\sim0.1\en10\,\Gyr$ at all $\zzams$, but have a peak that shifts from $\sim3\,\Gyr$ to $\sim0.3\,\Gyr$ between $\zzams=0.5$ and $\zzams=10$. As with the $m_1\en\zzams$ evolution (Fig. \ref{sec4sub:bhns_masses}), the varying IMF in the Varying model makes this shift smoother than in the Invariant case, through the short ZAMS orbital period bias for increasingly massive primaries. The main difference between the two models is the permanence of a feature at $\sim3\,\Gyr$ in the Invariant model, even as the distribution shifts to $\sim0.3\,\Gyr$ with increasing $\zzams$. Although we cannot make a definite connection, the first would be compatible with channel III for BHNS formation, which produces coalescence times centered at $\sim1\,\Gyr$, but with a significant $\sim0.1\en10\,\Gyr$ spread, and is associated to wider ZAMS orbits \citep[see][but also our own tentative conclusions in Sec. \ref{sec4sub:bhns_masses} associated to Fig. \ref{sec4fig:bhns_qzams_pzams}]{broekgaardenImpactMassiveBinary2021a}. The outlier at $\sim0.01\,\Gyr$ is an artifact of the initial sampling having sampled the same pair of initial masses $10$ times at $\zzams=10$, and which does not affect our results.

\begin{figure*}
    \centering
    \includegraphics[width=\textwidth]{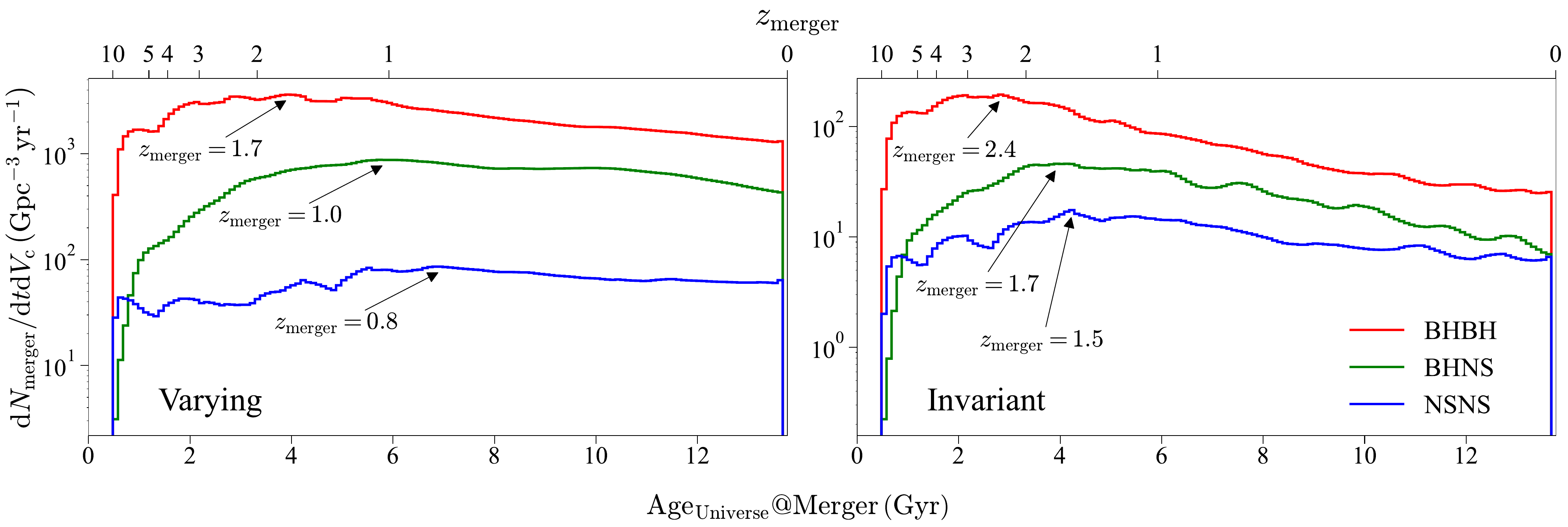}
    \caption{Merger rate densities over time ($\zmerger$) for BHBHs (red lines), BHNSs (green lines) and NSNSs (blue lines), in both the Varying (left) and Invariant (right) models. No distinction is made between BHNSs and NSBHs. The merger rates are calculated over bins of constant width $\Delta t=0.1\,\Gyr$. The "bumpy" aspect of the curves is an artifact of the relative low resolution on $\zzams$; see Sec. \ref{sec:3computation} for discussion. Peak merger rates are anottated with the corresponding $\zmerger$. Peak merger rates are concentrated around peak SFRs ($z\sim1\en3$), followed by a smooth decrease down to $\zmerger=0$, the rate of which depends on the merger class, with NSNSs being the closest to a stable rate. Relative to the Invariant model, the Varying model results in BHBH and BHNS rates about one order of magnitude greater overall, as well as in later peaks, both features connected to the top-heaviness of the IMF at high SFR. As noted in Sec. \ref{sec4sub:formation_efficiency}, BH-hosting binaries are particularly affected. A significant consequence is that BHBH and BHNS mergers always remain dominant in the Varying model, NSNS mergers become as common as BHNS mergers locally.}
    \label{sec4fig:full_mrates}
\end{figure*}

In terms of $\zmerger$ (Fig. \ref{sec4fig:zmerger_tc}, second column left to right), we see the bias for binaries with longer $\tc$ to merge later and broadly similar distributions between the two models. The main deviations follow those we have already discussed: as seen in Figs. \ref{sec4fig:zmerger_m1} and \ref{sec4fig:zmerger_q}, BHNS mergers switch from favoring the two intervals between $0.9<\zmerger\leq2.5$ in the Invariant model to the $0\leq\zmerger\leq0.9$ interval in the Varying model. And, as seen in Fig. \ref{sec4fig:zzams_tc}, the Invariant model displays an underlying flattened feature within $\tc\sim0.1\en10\,\Gyr$, which here is visible for $\zmerger\leq0.9$, and the $\sim0.01\,\Gyr$ minor feature from $\zzams=10$, both of which the Varying model does not present. 
The shift in the favored $\zmerger$ interval results from a relative shift toward longer coalescence times across $\zzams$ in the Varying model, causing some systems to move from the upper to the lower $\zmerger$ intervals. The flattened feature we have linked to channel III. Only its lower part may contribute to early (high $\zmerger$) mergers, but these would blend with other short $\tc$ mergers contributing to high redshifts and would not appear distinct; therefore, it is only noticeable as a short $\tc$ contribution to the lowest $\zmerger$ interval, from relatively young systems. At this resolution, we do not find mass gap BHNSs ($m_1\lesssim5\,\Msun$) to be significantly distinct from the rest of the population on the basis of $\tc$.

As for NSBHs, in terms of $\zzams$ (Fig. \ref{sec4fig:zzams_tc}, third column left to right), the Invariant model displays a preference for the $\tc\sim0.1\en10\,\Gyr$ interval, as is the case with BHNSs. However, for a few redshifts there is a non-negligible contribution in the $\tc\sim0.01\en0.1\,\Gyr$ range, which we identify as stemming from $5\lesssim m_1/\Msun\lesssim10$ binaries. There are both high ($\zzams=4.64,3.05$) and low ($\zzams=0.86$) redshifts that contribute to this component, so it is difficult to establish a correlation between $\zzams$ and $\tc$ for NSBH. Since this population is in any case undersampled, we refrain from proposing specific mechanisms that set this distribution. The Varying model mainly broadens the $\tc$ distribution for NSBHs, as it did for $m_1$, and, due to the preference of shorter initial periods, suppresses $\tc\gtrsim1\,\Gyr$ binaries in relation to the Invariant model. In terms of $\zmerger$ (Fig. \ref{sec4fig:zmerger_tc}, third column left to right), there is little structure beyond the bias for longer $\tc$ systems to merge at lower $\zmerger$. Generally and in both models, the distribution contains a dominant longer $\tc$ peak, and a secondary shorter $\tc$ feature, either a peak or a flattening, but, as before, we do not characterize this short-$\tc$ contribution.

\subsubsection{Neutron star+neutron star mergers}
\label{sec4sub:nsns_ctimes}

\begin{table*}
 \caption{Local and peak merger rates, and peak locations, for the entire BHBH, BHNS and NSNS merger populations, as well as for a few different BHBH and BHNS primary mass bins, from the Varying and Invariant models. Peak rates are computed within $\Delta t=0.1\,\Gyr$ bins, the midpoint of which is reported as the location of the peak. Local rates are computed within $\zmerger<0.014$.}
 \label{sec4tab:peakrates}
 \begin{tabular}{lcccccccc}
  \hline
  {} & \multicolumn{4}{c}{Varying} & \multicolumn{4}{c}{Invariant} \\
  $m_1$ bin & $t_\mathrm{peak}$ & $z_\mathrm{peak}$ & Peak rate density & Local rate density & $t_\mathrm{peak}$ & $z_\mathrm{peak}$ & Peak rate density & Local rate density  \\
  $\Msun$ & $\Gyr$ & - & $\Gpc^{-3}\,\yr^{-1}$ & $\Gpc^{-3}\,\yr^{-1}$& $\Gyr$ & - & $\Gpc^{-3}\,\yr^{-1}$ & $\Gpc^{-3}\,\yr^{-1}$ \\
  \hline
  \multicolumn{9}{c}{BHBH} \\
  Full & 3.93 & 1.67 & 3631.5 & 1314.4 & 2.83 & 2.35 & 194.8 & 25.5 \\
  $(0,10]$ & 5.53 & 1.10 & 489.8 & 231.0 & 3.93 & 1.67 & 34.8 & 5.6 \\
  $(10,20]$ & 2.83 & 2.35 & 2276.4 & 694.7 & 2.13 & 3.06 & 145.7 & 13.8 \\
  $(20,30]$ & 4.03 & 1.63 & 1066.4 & 245.2 & 2.13 & 3.06 & 25.7 & 4.7 \\
  $(30,\infty)$ & 2.13 & 3.07 & 289.5 & 143.4 & 0.93 & 6.08 & 4.9 & 1.4 \\
  \hline
  \multicolumn{9}{c}{BHNS} \\
  Full & 5.83 & 1.02 & 878.0 & 432.5 & 3.93 & 1.67 & 46.0 & 7.0 \\
  $(0,3]$ & 5.73 & 1.05 & 25.6 & 7.3 & 3.73 & 1.77 & 3.4 & 0.8 \\
  $(3,5]$ & 6.03 & 0.97 & 89.6 & 62.4 & 4.53 & 1.42 & 8.8 & 1.4 \\
  $(5,10]$ & 6.33 & 0.91 & 458.6 & 169.7 & 4.23 & 1.54 & 27.7 & 2.6 \\
  $(10,20]$ & 10.53 & 0.28 & 338.0 & 188.2  & 3.23 & 2.06 & 18.1 & 2.0 \\
  $(20,\infty)$ & 3.7 & 1.83 & 14.6 & 3.7 & 4.23 & 1.54 & 1.0 & 0.0 \\
  \hline
  \multicolumn{9}{c}{NSNS} \\
  Full & 6.93 & 0.78 & 85.8 & 64.0 & 4.23 & 1.54 & 17.5 & 6.6
\\
  \hline
 \end{tabular}
\end{table*}

\begin{table}
 \caption{Resulting local merger rate densities for BHBHs, BHNSs and NSNSs from the Varying and Invariant models, and $90\%$ credibility intervals from GWTC-3 \citep{ligoscientificcollaborationPopulationMergingCompact2023}. No distinction is made between BHNSs and NSBHs. Local rates were computed within $\zmerger\leq0.014$.}
 \label{sec4tab:localrates}
 \begin{tabular}{lccr}
  \hline
  Source & \multicolumn{3}{r}{Local rate density ($\Gpc^{-3}\,\yr^{-1}$)} \\
  {} & BHBH & BHNS & NSNS \\
  \hline
  Varying & 1314.4 & 432.5 & 64.0 \\
  Invariant & 25.5 & 7.0 & 6.6\\
  GWTC-3 ($90\%$) & $16\en61$ & $7.8\en140$ & $10\en1700$ \\
  \hline
 \end{tabular}
\end{table}

NSNS mergers result in a wider range of coalescence times, with a small contribution from $\tc\sim0.1\en1\,\Myr$ in addition to the $\tc\sim0.001\en10\,\Gyr$ range shared by the other classes. As before, we may describe the general 
$\zzams$-evolution of the coalescence time distribution in terms of two features: one within $\sim1\en10\,\Gyr$ characteristic of low $\zzams$, and the other within $\sim0.01\en0.1\,\Gyr$, characteristic of high $\zzams$. As the shift from the upper to the lower feature happens with increasing $\zzams$, we note that nonetheless the other remains as a secondary peak or plateau in the distribution. This allows us to posit that the $\zmerger=0.01$ population, with a strong peak around $\tc\sim0.02\,\Gyr$, also chiefly produced long $\tc$ binaries, which have not yet had time to merge and were excluded from the sample. These behaviors are generally common to both models, but in the Invariant all $\zzams\geq3.05$ samples are shifted to shorter $\tc$ relative to the Varying model. The overall shift to shorter $\tc$ for high $\zzams$ in the Invariant relative to the Varying model is caused by the initial orbital period distribution, as the Varying model favors $\log P_\ZAMS\sim4$ for the $m_{1,\ZAMS}\lesssim20\,\Msun$ progenitors that are characteristic of NSNSs at $\zzams\geq2.5$ (Fig. \ref{sec4fig:nsns_mzams1_qzams_scatter}). 

The $\zzams=10$ Invariant distribution stands out due to its peculiar shape which seems to violate some of the trends described above, but this is also an artifact of the particular initial sampling run that generated this sample having resulted in $60$ systems distributed over only $3$ different $m_{1\,\ZAMS}$ (approximately $9$, $9.9$ and $12.5\,\Msun$), which, due to our initial sampling procedure (see Sec. \ref{sec2sub:sampling}), lead to only $4$ different $\log P_\ZAMS$, and finally to only $5$ different $\tc$ in this sample. The location of the sample, in any case, fits with the overall picture described above.

The location of the features themselves is connected to the characteristic $m_1$ of the different $\zzams$ ranges, because $m_1\lesssim1.7\,\Msun$ NSNSs populate the entire $\tc$ range, while those with $m_1\gtrsim1.7\,\Msun$ preferentially have $\tc\gtrsim0.1\,\Gyr$. As this trend is present in both models, it must be directly related to the formation channel of $m_1\gtrsim1.7\,\Msun$ NSNSs mergers, which remains to be investigated.

\subsection{Merger rates}

\subsubsection{Comparison between the Varying and Invariant models}

We show in Fig. \ref{sec4fig:full_mrates} the merger rate densities found for BHBH, BHNS and NSNS mergers in both the Varying and Invariant models between $\zmerger=10\en0$. We do not differentiate between BHNSs and NSBHs in this section. The general "bumpy" aspect of the curves is an artifact of the discrete, relative low-resolution, sampling of $\zzams$ discussed in Sec. \ref{sec:3computation}. 

In all cases we find the curves to broadly follow the shape of the cSFH, which peaks within $z\sim1\en3$, with a sharp initial rise, a peak within or relatively shortly after the star-formation peak, followed by a smooth decrease down to $\zmerger=0$. The NSNS rate is the most stable, especially in its later portion, and the BHBH rate the most sharply peaked. Related to the pattern seen from the formation efficiencies (Sec. \ref{sec4sub:formation_efficiency}) of BHBH progenitor formation being favored at higher redshift and lower metallicities, NSNS progenitor formation at lower redshift and higher metallicities, and BHNS progenitor formation in an intermediate range, we see BHBH mergers peak first, and NSNS last. The relative stability of NSNS mergers is also in keeping with their progenitor formation efficiency being the least sensitive to environmental variations.

The main differences between both models are between peak locations and the overall magnitude of the merger rates, in particular those involving BHs. With regard to the position of peaks, indicated by the arrows on Fig. \ref{sec4fig:full_mrates} and the rows labeled "Full" in Table \ref{sec4tab:peakrates}, their position is always the same relative to each other (in redshift), but they are shifted down by $\Delta\zmerger\approx0.7$ in the Varying model. The immediate source is that the Varying IMF causes the SFR density to decrease at all except the lowest redshifts, resulting, for $\gtrsim2$, in about half of its value in the Invariant model \citep{chruslinskaEffectEnvironmentdependentIMF2020}. This simultaneously drags the peaks to later times and collaborates to make the post-peak evolution of the merger rates overall flatter in the Varying model than in the Invariant. From Sec. \ref{sec4sub:bhbh_ctimes} we also know that the oldest BHBHs are shifted to longer coalescence times by the Varying model, which could also play a role here. However, because the shift of the merger rate peak is common to all three classes, we conclude the cSFH to be the dominant factor.

The Varying model strongly amplifies the formation of COs, increasing all rates by about one order of magnitude (see the peak rates in Table \ref{sec4fig:full_mrates}, for example), but particularly those of BHBH and BHNS mergers. Because the Varying model is also flatter above the peak, the variation is even more extreme at low redshift, as exemplified by the local merger rates in Table \ref{sec4tab:localrates}, wherein the local BHBH and BHNS merger rates are greater by a factor of $\sim50\en60$, and that of NSNSs by $\sim10$, in the Varying model. Here we quantify one feature already noted in Sec. \ref{sec4sub:formation_efficiency}: whereas in the Varying model BHNS mergers are always dominant over NSNS mergers, in the Invariant model they are comparable locally. However, neither of the models is fully compatible with the most recent $90\%$ credibility rates for local merger rates as inferred from GWTC-3 \citep{ligoscientificcollaborationPopulationMergingCompact2023}. This is not surprising, as we have so far only tested two of the 192 possible permutations of our initial conditions models, and merger rates estimated from isolated binary evolution tend to be very sensitive to model choices (see \cite{broekgaardenImpactMassiveBinary2022} for variations with regards to the evolution models in COMPAS, and \cite{mandelRatesCompactObject2022} for the full range of variations in the literature). Table \ref{sec4tab:localrates} does highlight that an overproduction of BHs is the most egregious feature of the Varying model here defined.

\begin{figure*}
    \centering
    \includegraphics[width=\textwidth]{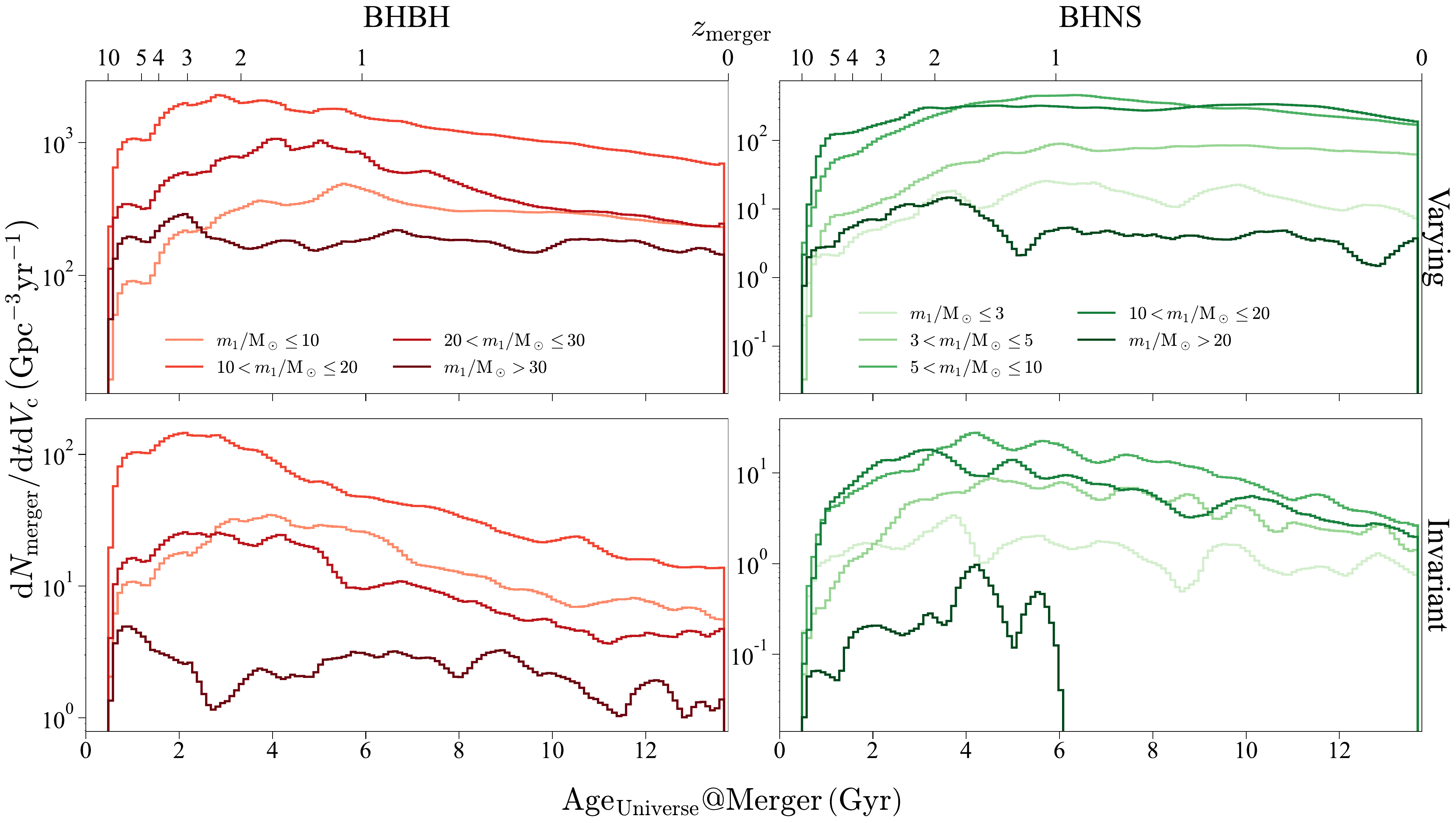}
    \caption{Merger rate densities over time ($\zmerger$) for BHBHs (left column) and BHNSs (right column), in both the Varying (top row) and Invariant (bottom row) models, for a few different primary mass ($m_1$) ranges motivated by Sec. \ref{sec4sub:masses}. No distinction is made between BHNSs and NSBHs. Besides trends in the position of the peaks already observed in Fig. \ref{sec4fig:full_mrates}, the top-heavy IMF in the Varying model has a strong effect of increasing the relative contribution of higher mass ranges. Some of the extreme mass ranges (the upper range for BHBHs; and both extremes for BHNSs) are based on too small samples to reliably predict the shape of the merger rate curve of that range, and instead should be taken as representing the order of their contribution at each time.}
    \label{sec4fig:massbin_mrates}
\end{figure*}

For the two largest and most varied populations --- BHBHs and BHNSs --- we also break the population into different $m_1$ bins, motivated by the discussion in Secs. \ref{sec4sub:bhbh_masses} and \ref{sec4sub:bhns_masses}, and consider their individual contribution to the merger rate over time in Fig. \ref{sec4fig:massbin_mrates}. The least populated bins are "noisy" and cannot be used to reliably predict the shape of the contribution from that bin, only the magnitude of its contribution. The peaks and local rates for each mass interval are identified in Table \ref{sec4tab:peakrates}, alongside those of the full sample for reference. For BHBHs, in the Invariant model the dominant contribution is from $10<m_1/\,\Msun\leq20$, followed by the $m_1/\,\Msun\leq10$ and $20<m_1/\,\Msun\leq30$, and a minor contribution from $m_1/\,\Msun>30$. While all components are shifted up in the Varying model, the higher mass bins are relatively favored, with the $20<m_1/\,\Msun\leq30$ component surpassing $m_1/\,\Msun\leq10$, and $m_1/\,\Msun>30$ becoming comparable to it. For the three lower bins, the shift of the peak to lower redshift is evident. The $m_1/\,\Msun>30$ component is closer to being constant, especially in the Varying case. There is a slight trend for it to decay over time in the Invariant model, but the sample in this bin is too small to firmly establish this trend.

For BHNSs, the Invariant model results in the $5<m_1/\,\Msun\leq10$ component being dominant, but remains comparable to $10<m_1/\,\Msun\leq20$ and, to a lesser extent, $3<m_1/\,\Msun\leq5$ components throughout. For $\zzams\gtrsim2$ the $3<m_1/\,\Msun\leq5$  contribution is much less significant, but $10<m_1/\,\Msun\leq20$ mergers are slightly more frequent than $5<m_1/\,\Msun\leq10$. The extreme $m_1/\,\Msun\leq3$ mergers contribute with $\sim1\,\Gpc^{-3}\,\yr$ throughout; $m_1/\,\Msun>20$ mergers, on the other hand, are rarer, contributing only down to $\zmerger\sim1$. As with BHBHs, the Varying model amplifies all components but privileges higher masses. The $5<m_1/\,\Msun\leq10$ and $10<m_1/\,\Msun\leq20$ populations dominate the sample with nearly equal contributions at all time, follower by the $3<m_1/\,\Msun<5$ component, and $m_1/\,\Msun\leq3$ mergers that become as frequent as $10\,\Gpc^{-3}\,\yr^{-1}$ below $\zzams\sim1$. The $m_1/\,\Msun>20$ remains the least common, but now contributes with a few mergers $\Gpc^{-3}\,\yr^{-1}$ down to $\zzams=0$.

The sharp artifacts on the merger rate curves make it difficult to decidedly ascertain a relation between mass and the position of the peaks, but we find evidence to support that the more massive contributions peak earlier, in line with expectations set by the evolution of metallicity (in both models) and the IMF (in the Varying model). This is clear in Table \ref{sec4tab:peakrates} for the Invariant model, in which the curves are steeper, except for the small $m_1/\Msun>20$ BHNS component. This trend is less apparent in the overall flatter Varying model. However, it becomes noticeable for BHNSs if the two extreme components are excluded, suggesting a flat evolution instead. For BHBHs, however, the $20<m_1/\Msun\leq30$ component has a clear later peak than the $10<m_1/\Msun\leq20$ component. From Sec. \ref{sec4sub:bhbh_masses} and Fig. \ref{sec4fig:bhbh_qmco1_tc}, we know that $20<m_1/\Msun\leq30$ BHBHs are generally older than $10<m_1/\Msun\leq20$ BHBHs and favor $\tc\gtrsim\,1\,\Gyr$, while the latter mass range shows coalescence times down to $\sim0.01\,\Gyr$, which might explain its later peak.

\subsubsection{Why does the Varying model overestimate the BH population?}
\label{sec4sub:varying_bhbh_overestimate}

Finally, it is important to make a point of the apparent conflict between the intense enhancement of BHBH and BHNS formation efficiencies, as well as merger rates, resulting from our Varying model, and previous conclusions on the impact of IMF variations for CBM population synthesis. As several additional factors are taken into account in the Varying model, it should not be considered to be directly at odds with previous work.

\citet{chruslinskaEffectEnvironmentdependentIMF2020} studied the influence of the non-Universal \citet{jerabkovaImpactMetallicityStar2018} IMF on the production of CO progenitors over time according to the corrected cSFH, and find that it has little effect on the production of BH progenitors. This occurs because BH progenitors are the main $\mathrm{H}\alpha$ emitters in a given galaxy, which they use as a SFR tracer. A top-heavy IMF implies then a smaller SFR, which counterbalances the greater fraction of BH progenitors to keep the absolute number of BH progenitors close to the one implied by the Invariant IMF. Our discussion of formation efficiencies in Sec. \ref{sec4sub:formation_efficiency} does not go immediately against this expectation because the formation efficiency is calculated with respect to the mass of the \textit{sample}, and not yet normalized to the SFH, and rather reflects an increased fraction of stars being CO progenitors, and BH progenitors in particular --- although there must also be a significant effect from orbital parameter and multiplicity models which we discuss shortly. Our discussion on merger parameter distributions in Secs. \ref{sec4sub:masses} and \ref{sec4sub:ctimes} stands on similar ground, as it is also not yet connected to the cSFH and reflects relative fractions rather than absolute numbers.

The merger rates in Sec. \ref{sec3sub:mrates}, however, are normalized by the SFH and continue to show that the Varying model results in $\sim10$ times more CBMs overall, with BHBHs being particularly affected, followed by BHNSs. We assume that, because \citet{chruslinskaEffectEnvironmentdependentIMF2020} have determined that the Varying IMF on its own has little effect on the production of BH progenitors, the particular permutation of models in our Varying model has introduced an exaggerated bias towards the formation of CBMs overall, and BHBHs in particular. It is not hard to imagine that this is the case. First, because the orbital parameter distributions by \citet{moeMindYourPs2017} introduce a bias towards short ZAMS orbital periods which is exacerbated by our choice of extrapolation of their fit for $m_1>40\,\Msun$. And, second, because we collapse all multiples into binaries, with a binary fraction that increases from $\sim0.7$ to $1$ with increasing $m_1$ in our range\footnote{An important caveat to this discussion is that, while we are concerned with binary populations, the spectral synthesis models used by \citet{chruslinskaEffectEnvironmentdependentIMF2020} do not account for potential effects of binary evolution on the UV part of the spectrum. While we do not perform spectral synthesis here, and thus suffer from the same uncertainties, we point the reader to Section 4.2 of \citet{chruslinskaEffectEnvironmentdependentIMF2020} for a discussion of the potential impact of binarity and rotation on the spectrum.}. In other words, it is important to consider that the particular biases of CBM formation do not readily translate isolated BH trends into trends for mergers.

Under these circumstances, the work by \citet{klenckiImpactIntercorrelatedInitial2018} provides an informative point of comparison as, employing the \texttt{StarTrack} population synthesis code, they compute the BHBH merger rate curve for the distributions from \citet{moeMindYourPs2017}, but by assuming that the orbital period distribution is invariant for $m_1>40\,\Msun$, instead of increasingly shifted toward shorter orbits, besides of a  modified version of the fit by \citet{marksEvidenceTopheavyStellar2012} of a metallicity-dependent IMF, which they set to be equal to Salpeter for $\FeHinline\gtrsim-0.55$. They find the \citet{moeMindYourPs2017} distributions to in fact \textit{decrease} the formation efficiency of BHBHs on their own, due mainly to a shift towards smaller $q_\mathrm{ZAMS}$, which disfavors formation of both BHBHs and NSNSs, producing a local BHBH merger rate of $88.8\,\Gpc^{-3}\,\yr^{-1}$, in contrast with the $203\,\Gpc^{-3}\,\yr^{-1}$ based on \citet{sanaBinaryInteractionDominates2012}. Coupling the \citet{moeMindYourPs2017} distributions to the varying IMF yields $181\,\Gpc^{-3}\,\yr^{-1}$. While this is not a controlled comparison of the effects of the models on their own due to differences in implementation, evolution models, SFH and multiplicity, it is notable that they find a much smaller range of variation in the local rates than we have. Their model also does not account for the possible dependence of the IMF on the SFR itself. On the other hand, they are careful to account for the counterbalancing between the SFR and the BH progenitor fraction from a top-heavy IMF, but still find the IMF to affect the BHBH merger rate by a factor of $\sim2$, which does reinforce the point that trends for isolated BH progenitors are not readily translatable to CBMs.

A similar approach is followed by \citet{deminkMergerRatesDouble2015}, \citet{kruckowProgenitorsGravitationalWave2018} and \citet{santoliquidoCosmicMergerRate2021} to investigate the effect of IMF variations, by manually varying the IMF slope within a given range \citep[in lieu of adopting a particular environment-dependent model, as done here and in][]{klenckiImpactIntercorrelatedInitial2018}. However, \citet{deminkMergerRatesDouble2015} vary the slope within $2.2\en3.2$, and \citet{santoliquidoCosmicMergerRate2020} compare slopes $2.0$ and $2.7$, while \citet{kruckowProgenitorsGravitationalWave2018} vary the slope within $1.5\en4$. This puts our work closer to the latter, due to the strong SFR-dependence of the \citet{jerabkovaImpactMetallicityStar2018} IMF, and helps explain why \citet{kruckowProgenitorsGravitationalWave2018} find one order of magnitude variations in the merger rates from varying the IMF, while \citet{deminkChemicallyHomogeneousEvolutionary2016} find rate variations by up to a factor $\sim6$, and \citet{santoliquidoCosmicMergerRate2021} by less than a factor of $\sim2$.

If we consider that it is clear that the overproduction of BHs in the Varying model must be connected not to a single component of the model permutation, but rather to a convenient combination of some components, this then highlights the need to evaluate the entire grid of model permutations, which should clearly reveal the presence of any correlated impact between multiple models. This is something already emphasized by \citet{stevensonConstraintsContributionsObserved2022}, who found local BHBH merger rates ranging from $0.01\,\mathrm{Gpc}^{-3}\,\mathrm{yr}^{-1}$ to $400\,\mathrm{Gpc}^{-3}\,\mathrm{yr}^{-1}$ by simultaneously varying four parameters within COMPAS, over a total of 2916 permutations. More refined techniques for dealing with the \textit{curse of dimensionality} associated to varying many parameters simultaneously have also been explored recently, such as by \citet{barrettAccuracyInferencePhysics2018}, who employ a Fisher matrix analysis; \citet{delfaveroIterativelyComparingGravitationalwave2023}, who use Gaussian process regression to interpolate the joint likelihood of a set of parameters from a small number of simulations; and by \citet{rileySurrogateForwardModels2023} who utilize artificial neural networks to interpolate from the results of running COMPAS on a reduced set of permutations to a much larger one. More recently, \citet{raufExploringBinaryBlack2023} estimated BHBH merger rates by joining COMPAS and the semi-analytic galaxy formation model \texttt{SHARK}, and \citet{raufTrifectaModellingTools2024} demonstrated that their results could be scaled to different COMPAS models without resampling from simulated galaxies.

Further variations of the IMF and initial orbital parameters can be expected to have the same kind of impact, with our results presenting an example of the interplay between the IMF and the orbital period distribution, and those by \citet{klenckiImpactIntercorrelatedInitial2018} of the interplay between the IMF and the mass ratio distribution. The low-mass slope, as well, could have an indirect effect on the BHBH yield by decreasing the fraction of BH progenitors in a given population, and how often these progenitors are paired to other BH progenitors might also be significantly influenced by how the IMF is sampled (e.g., whether for $m_1$ only, or also by $m_2$).
\section{Summary and conclusions}
\label{sec:5conclusions}

We have analyzed the scope of the impact of varying models for stellar/binary formation on the properties and time-evolution of CBM populations up to redshift $10$. Within a space of 192 possible permutations of initial condition models, we have chosen two representative permutations to contrast in depth: the Invariant model, with uncorrelated and universal IMF and orbital parameter distributions; and the Varying model, with an IMF that becomes top-heavy at high SFRs and low redshifts, as well as correlated orbital parameter distributions. We study the impact of these models on the primary mass, mass ratio and coalescence time of BHBH, BHNS, NSBH and NSNS mergers, in terms both of $\zmerger$, the redshift at merger, and $\zzams$, a proxy for the total age of the binary. Even though we do not track the individual formation channels in our sample, we find that we can consistently analyze our results based on previous work that has characterized these formation channels, within a picture where evolution models, which we did not vary, define the location of major features, and initial conditions set their relative weights. Finally, we also compute merger rate curves over time and compare them to the currently constrained local merger rates.

\textit{Formation efficiencies.} We track the evolution of the formation efficiencies (number of systems per unit-star forming mass) of BHBH, BHNS and NSNS progenitors as a function of metallicity and $\zzams$. For subsolar metallicities, BHBH progenitor formation dominates at all redshifts, followed by BHNSs, and then NSNSs. In both models their evolution is dominated by a sharp shift drop in BHBH and BHNS progenitor formation efficiency at solar metallicity, accompanied by a rise of the NSNS progenitors. In the Invariant model this results in NSNS progenitor formation becoming more efficient than that of BHNS progenitors for supersolar metallicity. We find that BHBH progenitor formation favors at lower metallicities, NSNS progenitor at higher and BHNS at an intermediate range, centered around $\FeHinline\sim-1$ in the Invariant model and $\FeHinline\sim-0.5$ in the Varying model. Finally, the NSNS progenitor formation efficiency is the least sensitive to metallicity variations. The Varying model increases all formation efficiencies by at least one order of magnitude, particularly that of BHBH progenitors, at all redshifts. In terms of $\zzams$ only, we find that BHBH progenitors remain dominant throughout. In the Varying model the BHNS progenitor formation efficiency is always greater than that of NSNSs progenitors (see Fig. \ref{sec4fig:redshift_frac}), while in the Invariant model NSNSs become more common than BHNSs for $\zzams\lesssim2$

\textit{BHBH merger parameters.} The BHBH merger populations has previously been characterized in terms of two main formation channels, the CE and stable RLOF channel, a picture which we find good agreement with. The CE channel is characterized by $m_1\sim10\en30\,\Msun$, a broad $q\sim0.2\en1$ distribution with an overdensity at $q\sim0.3$, and coalescence times within $\sim0.01\en10\,\Gyr$. The stable RLOF channel is characterized by a long high-mass tail for $m_1\gtrsim20\,\Msun$ and a $\sim9\,\Msun$ feature, $q\sim0.6\en0.8$ and $\tc\sim1\en10\,\Gyr$. In agreement with \cite{sonLocationsFeaturesMass2023}, our results suggest that, in the Invariant model, the CE channel is dominant for $m_1\lesssim20\,\Msun$ and the stable RLOF channel is dominant for $m_1\gtrsim20\,\Msun$, particularly for old (high $\zzams$) populations, with a smaller $\sim9\,\Msun$ feature from young populations (low $\zzams$). In association with typical coalescence times, we find that the contribution of the stable RLOF channel is most significant for mergers at lower redshifts, especially $\zmerger\leq0.9$. The Varying model strongly amplifies the stable RLOF channel, which dominates BHBH progenitors with $\zzams\leq2.49$ in this model, and for $\zzams\geq4.64$ produces a strong $\gtrsim20\,\Msun$ component alongside the CE peak, including a significant pileup at $\sim45\,\Msun$, in association with the PPSINe prescription by \citet{marchantPulsationalPairinstabilitySupernovae2019}. Consequently, the PPISNe pileup is present at all $\zmerger$ intervals in the Varying model, with $\zmerger\leq1.6$ being dominated by features of the stable RLOF channel. We are able to associate the favoring of the stable RLOF channel to an excess of short orbital periods at ZAMS in the Varying model, and the significant presence of the very old PPISNe pileup BHBHs in all $\zmerger$ bins to a broad $\tc\sim0.01\en10\,\Gyr$ distribution with a bias toward $\sim10\,\Gyr$. This signals the importance of considering initial conditions in investigating the balance between the two channels. From the evolutionary perspective, recent work has started to suggest a greater importance, or even prevalence, of the stable RLOF channel in forming BHBH mergers \citep{marchantRoleMassTransfer2021,gallegos-garciaBinaryBlackHole2021}, in contrast to the traditional view of the CE channel as the "standard" channel for BHBH formation \citep{vandenheuvelNatureXrayBinaries1973,taurisFormationEvolutionCompact2006,belczynskiRarityDoubleBlack2007,postnovEvolutionCompactBinary2014,belczynskiCompactBinaryMerger2016}.

We also find a small $m_1\sim35\,\Msun$ bump for $\zmerger\leq0.9$ in both models. In the Varying model, the primary mass distribution in $\zmerger\leq0.9$ is characterized by $\sim9\,\Msun$ and $\sim35\,\Msun$ overdensities, which are similar to the $35^{+1.7}_{-2.9}\,\Msun$ and $10^{+0.29}_{-0.59}\,\Msun$ overdensities found in BHBH mergers from GWTC-3 by \citet{ligoscientificcollaborationPopulationMergingCompact2023}. This feature, however, might be an unphysical artifact from the transition between the CCSNe and PPISNe prescriptions for massive stars \citep{sonRedshiftEvolutionBinary2022}. Its origin association with low merger redshfits remain to be verified. The Varying $\zmerger\leq0.9$ distribution also features the $\sim45\,\Msun$ pileup, not present in the observed sample, which might disfavor this combination of initial conditions and PPISNe model. However, because we have not implemented selection effects at this time, we refrain from further comparing our distributions directly to the known population.

\textit{BHNS merger parameters.} We further separated BH+NS mergers into BHNSs and NSBHs, depending on which CO formed first, and study their parameter distributions separately. We find the BHNS population to suffer less extreme variations with age than the BHBH population, but to display a clear shift between a $m_1\sim5\,\Msun$ peak for younger populations and a $m_1\sim12\,\Msun$ peak for older populations, in both models, with the Invariant model retaining a secondary $\sim5\,\Msun$ feature even at the highest $\zzams$. We find the mass ratio distributions to vary little with $\zzams$, except for a relative suppression of $q\sim0.3$ in relation to $q\sim0.15$ binaries in the BHNS model. By analyzing the sample in the $q_\ZAMS\en P_\ZAMS$, we identify in the Invariant model components analogous to the three most significant channels for BHNS formation identified by \citet{broekgaardenImpactMassiveBinary2021a}: channel I (CE phase as second mass transfer), channel II (stable RLOF only) and channel III (single-core CE as first mass transfer). We find channel III, here characterized by $\log P_\ZAMS\sim3\en4$ and $q\sim0.4\en0.9$, to be almost entirely suppressed in the Varying model. We attribute the relative loss of $m_1\sim5\,\Msun$, $q\sim0.3$ BHNSs to this suppression of channel III in the Varying model, which strongly disfavors high $q_\ZAMS$ for the $P_\ZAMS$ characteristic of channel III. The $\zmerger$-evolution of BHNS mergers closely follows their $\zzams$-evolution: both models show a shift from $\sim5\,\Msun$ to $\sim12\,\Msun$ with increasing $\zmerger$, the lower feature remaining even at high redshift in the Invariant model. Both models show that $q\sim0.15$ mergers are always the most common, but with an increasing contribution from $q\sim0.3$ for $\zmerger\leq1.6$. We show BHNSs to nearly exclusively populate $\tc\sim0.1\en10\,\Gyr$, with a clear shift from $\tc\gtrsim1\,\Gyr$ to $\tc\lesssim1\,\Gyr$ with increasing $\zzams$. While in the Invariant model both features remain present even as the dominant one shifts, the Varying model shows a much more well-defined, evolving, single $\tc$ peak.

Although our NSBH sample is too small to draw firm definitive conclusions, we may remark some of the resulting trends. We identify a slight trend for the distribution to shift toward larger masses with growing $\zzams$, but the preferred mass range varies strongly from model to model: the Invariant distribution is prefers the $m_1\sim5\en10\,\Msun$ range, while the Varying distribution $m_1\en10\en20\,\Msun$. The $\zzams=10$ sample is the most interesting case, as in both models it is entirely contained within the lower mass gap, $m_1\leq5\,\Msun$. We posit that this shift toward the lightest BHs at extremely high redshift is due to inefficient wind mass loss blocking NSBH production from more massive primaries. At lower $\zzams$, we associate the shift to larger masses in the Varying model to the bias toward short orbital periods for increasingly massive primaries, which increases the likelihood of mass transfer while the primary is still inside the Main Sequence. Although we do not verify the presence of the channel in this sample, such a "case A mass transfer" channel has been noted by \citet{broekgaardenImpactMassiveBinary2021a} to make a minor contribution to the BH+NS population as frequent channel for NSBH formation.

\textit{NS merger parameters.} We find NSNS primary masses to be strongly clustered around $\sim1.3\,\Msun$, a feature resulting from ECSNe and some CCSNe, with a tail up to the maximum $2.5\,\Msun$, and a discontinuity at $\sim1.7\,\Msun$ hailing from a discontinuity in the delayed prescription for CCSNe \citep{fryerCOMPACTREMNANTMASS2012}. Both models display the same trend with $\zzams$: a growth in the relative contribution of $m_1\gtrsim1.7\,\Msun$ mergers with redshift, up to a peak at $\zzams=1.49$. The high mass contributions are more significant overall in the Varying model, but the trend is the same. We note that the $m_1\gtrsim1.7\,\Msun$ NSNSs nearly exclusively originate from the same region of the initial parameter space as the least massive of the BHNSs, and that, for $\zzams\geq2.5$, the NSNS population ceases to access that region. We surmise that the observed $m_1$ trend is a consequence of the decreasing efficiency of wind mass loss at high redshift which increases the mean mass of COs. Up to an intermediate redshift this favors more massive NSs, but for even higher redshift they are lost, as their would-be progenitors become BHs instead. We find that the $q$ of NSNSs largely follow from the $m_1$ distributions and NS mass constraints. We also find a shift between a preferred $\tc\sim1\en10\,\Gyr$ to $\tc\sim0.01\en0.1\,\Gyr$ with increasing $\zzams$ in both models, and find that this is related to $m_1\gtrsim1.7\,\Msun$ preferring the $\tc\gtrsim0.1\,\Gyr$ range.

\textit{Merger rates.} Finally, we compute and compare the merger rate density over time for BHBH, BHNS and NSNSs up to $\zmerger=10$ in both models. We find that their evolution closely follows the SFH, with peaks within or slightly after the high SFR $\sim1\en3$ redshift range; BHBH mergers peak first, and NSNS mergers last, related to the evolution of formation efficiencies and coalescence times. Below the peak $\zmerger$, rates fall more slowly in the Varying model due to a flatter SFH evolution in that region. The shift in the SFH for the Varying IMF also shifts the peaks to later redshifts in that model. The NSNS merger rate is the most consistent with being stable after the peak, more so in the Varying model. Neither model results in local rates fully compatible with the $90\%$ credibility intervals of \citet{ligoscientificcollaborationPopulationMergingCompact2023} but, while the Invariant model predicts compatible BHBH rates ($\mathcal{R}_{z<0.01}=25.5\,\Gpc^{-3}\,\yr^{-1}$) and only slightly lower BHNS and NSNS rates ($\mathcal{R}_{z<0.01}=7.0\,\Gpc^{-3}\,\yr^{-1}$ and $\mathcal{R}_{z<0.01}=6.6\,\Gpc^{-3}\,.\yr^{-1}$, respectively), the Varying model far exceeds the observational constraints of BHBH and BHNS mergers, with their respective merger rates exceeding by factors of $\sim20$ and $\sim3$, ($\mathcal{R}_{z<0.01}=1314.4\,\Gpc^{-3}\,\yr^{-1}$ and $\mathcal{R}_{z<0.01}=432.5\,\Gpc^{-3}\,\yr^{-1}$, respectively) the reported upper limits. The NSNS merger rate predicted by the Varying model ($\mathcal{R}_{z<0.01}=64.0\,\Gpc^{-3}\,\yr^{-1}$), on the other hand, remains fully within the current constraints. Coupled with the formation efficiency, this signals that the Varying model significantly overestimates the frequency of BHBH formation. 

We are also able to separate the contributions of different BHBH and BHNS primary mass ranges to the merger rates over time, and find that the Varying model increases the contribution of all ranges, with a more significant relative increases for higher masses. We find that extreme mass ranges tend to have flatter contribution over time to the merger rate ($m_1/\Msun>30$ for BHBHs; $m_1/\Msun\leq3$ and $m_1/\Msun>20$ for BHNSs), while, for the intermediate mass ranges, there is a baseline trend for the rate of more massive intervals to peak at earlier times. Strong correlations between mass and coalescence time may nonetheless break this trend, as we find to be the case for BHBHs in the Varying model, for which the $10<m_1/\Msun\leq20$ contribution peaks at $\zmerger=2.35$, but the $20<m_1/\Msun\leq30$ at $\zmerger=1.63$.

\textit{Final remarks.} We cannot at this point draw a definitive conclusion with regard to the feasibility of any individual model considered in the Varying permutation, other than that the considered permutation itself is inadequate given current constraints on the local merger rate, in particular because it exaggerates the production of BHs overall. We conclude that this can only be explained by a convenient combination of initial condition models. Our results do, however, provide a strong case for the importance of evaluating the possible initial condition permutations extensively, and of minding the correlated impact of different models, as we find the Varying and Invariant models to result on significantly different primary mass, mass ratio and coalescence time distributions over time and, in the extreme case, differ by a factor $\sim50\en60$ on the local BHBH and BHNS merger rate densities.  The necessity for individually evaluating the Varying IMF and Varying orbital parameter distributions is clearly motivated, as well as the verification of the impact of our extrapolation of the orbital period distribution to very massive progenitors.

The breakdown of the final parameter distributions into the locations and relative weights of their key features, the first set by evolution models, and the second by initial conditions, proves to be particularly useful, although this picture still needs to be systematically validated for models with variations of the IMF and orbital parameters, as has been done for variations of the cSFH in \citet{sonLocationsFeaturesMass2023}. Uncertainties with regard to binary evolution are known to strongly affect population synthesis estimates of CBM populations, leading to variations of many order of magnitude in, e.g., merger rate estimates \citep{mandelRatesCompactObject2022}. Our results suggest that initial condition uncertainties are also significant and deserve a similar degree of attention. While here we have considered only a particular variation of the IMF and orbital parameters, it will also be important to simultaneously consider variations of the metallicity-specific cSFH \citep[as in, e.g.][]{neijsselEffectMetallicityspecificStar2019,broekgaardenImpactMassiveBinary2022}. Finer aspects of the mass sampling might also be worth considering, include that of multiplicity, and the possible dominance of higher-order multiples for very massive progenitors, where the treatment of triples by \citeauthor{klenckiImpactIntercorrelatedInitial2018}, \citeyear{klenckiImpactIntercorrelatedInitial2018} is instructive. Continuing work on constraining the IMF for binary components (see Paper I); and further constraining of the low-mass IMF slope, which could have an indirect effect on CBM populations \citep[][for example, suggest a linear instead of log-linear dependence on $\Z$]{yan2024gwIMFlowmass}, are possible further refinements. Beyond uncertainties in the predictions of the final parameters, population synthesis remains an important tool in discerning features of stellar formation over time from very old populations, such as CBMs. Including selection effects in our pipeline will be a fundamental step in furthering work in both directions, and in the future we expect to confront the full set of permutations collected here to already-established constraints.

\section*{Acknowledgements}

We thank Martyna Chru\'sli\'nska for providing the grid of IGIMF corrections for the SFR. We thank Lieke van Son for useful discussion and suggesting investigating the formation of mergers from very wide ZAMS binaries. We would like to thank the anonymous referee for several suggestions that contributed to the accuracy and clarity of the paper.

This paper made use of the BOSSA initial sampling code (version 1.0) (de S\'a et al., submitted). Simulations in this paper made use of the COMPAS rapid binary population synthesis code (version 02.32.03), which is freely available at \url{http://github.com/TeamCOMPAS/COMPAS}. This research was funded by S\~ao Paulo Research Foundation (FAPESP) grant number 2020/08518-2. L.M.S. acknowledges funding from the CNPq (Brazil), grant number 140794/2021-2. L.S.R acknowledges S\~ao Paulo Research Foundation (FAPESP) grant number 2023/08649-8. J.E.H. has been partially supported by the CNPq. 

This work made use of the following software packages: \texttt{Jupyter} \citep{2007CSE.....9c..21P, kluyver2016jupyter}, \texttt{matplotlib} \citep{Hunter:2007}, \texttt{numpy} \citep{numpy}, \texttt{pandas} \citep{mckinney-proc-scipy-2010}, \texttt{python} \citep{python}, \texttt{scipy} \citep{2020SciPy-NMeth}, \texttt{seaborn} \citep{Waskom2021} and \texttt{PyTables} \citep{pytables}. Cosmological calculations in this paper made use of the WMAP9 model of the \texttt{astropy} library \citep{astropy:2013,astropy:2018,astropy:2022}.

This research has made use of NASA's Astrophysics Data System.

Software citation information aggregated using \texttt{\href{https://www.tomwagg.com/software-citation-station/}{The Software Citation Station}} \citep{software-citation-station-paper, software-citation-station-zenodo}.

\section*{Data Availability}

All associated post-processing code, including for generating plots included here, will be made available on GitHub upon publication. Data products underlying this article will be made available on Zenodo upon publication.



\bibliographystyle{mnras}
\bibliography{bibliography}



\appendix


\bsp	
\label{lastpage}
\end{document}